\documentclass[nobibnotes,superscriptaddress,nofootinbib,twocolumn,floatfix]{revtex4-1}
\usepackage{float}
\usepackage{amsmath,amsfonts,amssymb}
\usepackage{multirow}
\usepackage{graphicx}
\usepackage{hhline}
\usepackage{bm}
\usepackage{tabulary}

\newcommand{\Yl}{\mathbf{Y}^\ell}
\newcommand{\Ynu}{\mathbf{Y}^\nu}
\newcommand{\Yu}{\mathbf{Y}^u}
\newcommand{\Yd}{\mathbf{Y}^d}
\newcommand{\Y}{\mathbf{Y}}
\newcommand{\Yx}{\mathbf{Y}^x}
\newcommand{\Mnu}{\mathbf{M}_\nu}
\newcommand{\Ml}{\mathbf{M}_\ell}
\newcommand{\Mx}{\mathbf{M}_x}
\newcommand{\Dnu}{\mathbf{D}_\nu}
\newcommand{\Dl}{\mathbf{D}_\ell}
\newcommand{\Hnu}{\mathbf{H}_\nu}
\newcommand{\Hl}{\mathbf{H}_\ell}
\newcommand{\URl}{\mathbf{U}_R^\ell}
\newcommand{\URnu}{\mathbf{U}_R^\nu}
\newcommand{\ULnu}{\mathbf{U}_L^\nu}
\newcommand{\ULl}{\mathbf{U}_L^\ell}
\newcommand{\ULnud}{\mathbf{U}_L^{\nu\,\dag}}
\newcommand{\ULld}{\mathbf{U}_L^{\ell\,\dag}}
\newcommand{\U}{\mathbf{U}}
\newcommand{\N}{\mathbf{N}}
\newcommand{\dmatm}{\Delta m^2_{31}}
\newcommand{\dmsol}{\Delta m^2_{21}}

\begin{document}

\title{Dirac neutrinos in the 2HDM with restrictive Abelian symmetries}

\author{S. S. Correia}
\affiliation{Departamento de F\'{\i}sica and CFTP, Instituto Superior T\'ecnico, Universidade de Lisboa, Lisboa, Portugal}
\author{R. G. Felipe}
\affiliation{ISEL - Instituto Superior de Engenharia de Lisboa, Instituto Polit\'ecnico de Lisboa, Rua Conselheiro Em\'{\i}dio Navarro, 1959-007 Lisboa, Portugal}
\affiliation{Departamento de F\'{\i}sica and CFTP, Instituto Superior T\'ecnico, Universidade de Lisboa, Lisboa, Portugal}
\author{F. R. Joaquim}
\affiliation{Departamento de F\'{\i}sica and CFTP, Instituto Superior T\'ecnico, Universidade de Lisboa, Lisboa, Portugal}

\begin{abstract}

Recently, there has been a growing interest in extensions of the Standard Model in which naturally small Dirac neutrino masses arise due to existence of a symmetry which protects neutrino's Diracness. Motivated by this, we consider an extension of the Standard Model with a second Higgs doublet (2HDM) and three right-handed neutrinos where lepton number is conserved and, thus, neutrinos are Dirac particles. In this framework, we identify the most restrictive texture-zero combinations for the Dirac-neutrino and charged-lepton mass matrices that lead to masses and mixings compatible with current experimental data. We then investigate, in a systematic way, which of these combinations can be realized by Abelian continuous U(1) or discrete $\mathbb{Z}_N$ symmetries. We conclude that, from the 28 initially possible sets of maximally-restricted lepton mass matrices, only 5 have a symmetry realization in the 2HDM. For these cases, one-to-one relations among the Yukawa couplings and the neutrino mass and mixing parameters are established, and the fermion interactions with the neutral and charged scalars of the 2HDM are also determined. Consequences for lepton universality in $\tau$ decays and rare lepton-flavor-violating processes are also discussed. 

\end{abstract}

\maketitle

\section{Introduction}
\label{Intro}

The discovery of neutrino masses and mixing through the observation of neutrino oscillations provided so far the only laboratory evidence for physics beyond the Standard Model (SM). In the last decades, several experiments using neutrinos from various sources (the Sun, the atmosphere, reactors, and accelerators, among others) have been measuring most of the parameters involved in their flavor oscillations with very good precision. In spite of this remarkable achievement, there remain several questions to be answered about neutrinos. Perhaps the most fundamental one is related with their nature, namely, whether neutrinos are Dirac or Majorana particles, which is of utmost importance when constructing SM extensions with massive neutrinos. Unfortunately, this question cannot be addressed by neutrino oscillation experiments and the experimental data presently available is compatible with the existence of either Dirac or Majorana massive neutrinos. Establishing the nature of neutrinos has proved very challenging and, among the several proposals to solve this riddle~\cite{Kayser:1981nw,Yoshimura:2006nd,Yoshimura:2011ri,Fukumi:2012rn,Dinh:2012qb,Berryman:2018qxn,Balantekin:2018ukw}, the most promising one seems to rely on neutrinoless double beta decays~\cite{Schechter:1981bd}. Meantime, in the absence of a solid evidence in favor (or against) the existence of Dirac and/or Majorana neutrinos, both scenarios should be equally considered.

Dirac neutrino masses require adding singlet right-handed (RH) neutrino fields $\nu_R$ to the SM field content. At the same time, Majorana mass terms $m_R \overline{\nu_R^c} \nu_R$ must be forbidden to ensure lepton number conservation. The main objection that may be raised against this simple Dirac-neutrino scenario is that extremely small Yukawa couplings $\sim\mathcal{O}(10^{-11})$ are required to generate sub-eV neutrino masses via the usual Higgs mechanism. This is in contrast with the Majorana neutrino case, in which neutrino masses are naturally suppressed by the existence of a large scale $\Lambda$ commonly related with the mass of some new (heavy) particles. From the effective theory viewpoint, integrating out these heavy mediators gives rise to the dimension-five Weinberg operator $\nu_L \nu_L \phi^0 \phi^0/\Lambda$~\cite{Weinberg:1979sa}, from which naturally small Majorana masses are generated after electroweak symmetry breaking (EWSB). The ultraviolet completions of the SM that realize this operator at tree level lead to the well-known type I~\cite{Minkowski:1977sc,GellMann:1980vs,Yanagida:1979as,Schechter:1980gr,Glashow:1979nm,Mohapatra:1979ia}, II~\cite{Barbieri:1979ag,Cheng:1980qt,Magg:1980ut,Lazarides:1980nt,Mohapatra:1980yp} and III~\cite{Foot:1988aq,Ma:1998dn,Brahmachari:2001bv} seesaw mechanisms. 

If neutrinos are indeed Dirac particles, the smallness of their masses should be natural~\cite{Ma:2016mwh} in the context of new symmetries beyond those of the SM.~\footnote{In models with large extra dimensions, bulk RH neutrinos couple very weekly with SM gauge fields localized on a 3-brane. Since couplings of bulk modes are suppressed by the volume of the extra dimensions~\cite{ArkaniHamed:1998vp,Dvali:1999cn}, small Dirac neutrino masses are natural.} To avoid dealing with very tiny couplings, several proposals have been put forward to explain Dirac neutrino mass suppression from an effective theory viewpoint~\cite{CentellesChulia:2018bkz,Calle:2018ovc,Enomoto:2019mzl}. Obviously, any of such solutions must contemplate a symmetry which forbids the dimension-four term $\overline{\ell_L} \nu_R \widetilde{\Phi}$. In this way, small Yukawa couplings may originate from higher-dimensional operators which can be realized as in the Majorana neutrino case through, for instance, a Dirac seesaw mechanism~\cite{Roncadelli:1983ty,Roy:1983be}. At lowest order in the effective theory, one can define the dimension-five operator
\begin{equation}
\label{dim5Dirac}
\begin{aligned}
-\mathcal{L}_5^{D} = \frac{\Y}{\Lambda}\overline{\ell_L} \nu_R \widetilde{\Phi} S + \text{H.c.},
\end{aligned}
\end{equation}
where $\ell_L$ and $\nu_R$ are a left-handed (LH) lepton doublet and RH neutrino singlet, respectively, and $\Phi$ is the SM Higgs doublet. Notice that, in this case, the generic couplings $\Y$ do not need to be very small. The scalar singlet $S$, after acquiring a vacuum expectation value (VEV) $\langle S\rangle=v_S$, leads to small effective couplings $\Y_\nu\overline{\ell_L} \nu_R \widetilde{\Phi}$ with $\Y_\nu \equiv \Y v_S/\Lambda \ll 1$ as long as $\Lambda \gg v_S$. 

In the above scenario, SM-like Dirac neutrino masses through a dimension-four term can be forbidden with a simple $\mathbb{Z}_2$ symmetry under which $\nu_R$ and $S$ are odd, while the SM fields are even.\footnote{All possible tree-level and one-loop realizations of $\mathcal{L}_5^{D}$ have been systematically studied in Ref.~\cite{Yao:2018ekp}.} To further prevent the appearance of the $\nu_R\nu_R$ Majorana mass term, at least a $\mathbb{Z}_3$ symmetry should be considered~\cite{Ma:2015mjd,Bonilla:2016diq}. For instance, the charge assignment $\nu_R \sim \omega$, with $\omega^3=1$ and all remaining fields transforming trivially, would guarantee the absence of $\Ynu$ couplings and $m_R$ masses at the renormalizable level. A more economical scenario relies on a ${\rm U(1)_{B-L}}$ symmetry with {\em unconventional} charges chosen in such a way that the gauge-invariant operators $\overline{\ell_L} \nu_R \widetilde{\Phi}$ and $\nu_R\nu_R$ are forbidden~\cite{Ma:2014qra}. 

The identification of new directions towards naturally small Dirac neutrino mass generation, has triggered a growing interest on this subject. Recently, several models have been proposed in the context of tree-level/radiative Dirac neutrino mass generation, and their classification with respect to the dimensionality of the corresponding generating operators has been put forward~\cite{Ma:2016mwh,Yao:2018ekp,CentellesChulia:2018gwr,Yao:2017vtm,CentellesChulia:2018bkz,CentellesChulia:2019xky}. Links to the dark matter problem have also been established within frameworks where the symmetry protecting neutrino Diracness play the additional role of stabilizing the dark matter particle~\cite{CentellesChulia:2019xky,Yao:2017vtm,Farzan:2012sa,Okada:2014vla,Chulia:2016ngi,Reig:2018mdk,Bonilla:2018ynb,Dasgupta:2019rmf,Ma:2019byo}. Moreover, mechanisms for the generation of the matter-antimatter asymmetry can also be envisaged within scenarios with Dirac neutrinos~\cite{Dick:1999je,Murayama:2002je} and, ultimately, both these problems can be related~\cite{Borah:2016zbd,Gu:2017bdw,Narendra:2017uxl}.

Independently of the mechanism for neutrino mass generation, one is always confronted with the problem of explaining the observed neutrino mass and mixing pattern. While the general frameworks described above provide an explanation for the smallness of (Majorana or Dirac) neutrino masses, in general they do not address the flavour problem per se. Thus, one is compelled to consider more sophisticated realizations of certain neutrino mass models in which flavour symmetries are considered. One of the approaches is to explore the existence of vanishing elements (texture zeros) in the Yukawa and mass matrices which reflect the violation of a symmetry by a certain interaction~\cite{Grimus:2004hf,Dighe:2009xj,Adhikary:2009kz,Dev:2011jc,Felipe:2014vka,Cebola:2015dwa,Samanta:2015oqa,Kobayashi:2018zpq,Rahat:2018sgs,Nath:2018xih,Barreiros:2018ndn,Barreiros:2018bju}. The simplest of these symmetries are those based on continuous U(1) and discrete $\mathbb{Z}_N$ transformations.

As it will be shown later, in the SM extended with RH neutrinos, U(1) or $Z_N$-motivated texture zeros turn out to be incompatible with data since, in general, they lead to massless neutrinos or charged-leptons and/or vanishing lepton mixing angles (which are already excluded by data). This is a direct consequence of the fact that all fermions in the SM couple to the same Higgs field. However, this is not the case in the 2HDM~\cite{Branco:2011iw}.

Abelian symmetries have been used in the context of the 2HDM with controlled flavor-changing neutral currents (FCNC). Namely, in the so-called Branco-Grimus-Lavoura (BGL) scenarios~\cite{Branco:1996bq,Botella:2009pq,Botella:2012ab}, flavor-changing couplings (FCC) depend only on the Cabibbo-Kobayashi-Maskawa (CKM) quark mixing matrix or on the lepton mixing matrix. In these cases, the symmetry realizes a minimal flavour violation scenario without addressing the question of reconstructing the parameters of the Lagrangian (e.g. Yukawa couplings) in terms of the observable fermion mass and mixing parameters. In the most economical generalized BGL realization for quarks, in which all Yukawa couplings are real and CP violation arises from the relative phase between the VEVs of the neutral components of the two Higgs fields~\cite{Nebot:2018nqn}, there are nineteen parameters to be confronted with ten physical quark mass and mixing observables (six masses, three mixing angles and one Dirac CP-violating phase). However, if CP is broken explicitly in the Yukawa interactions, that number increases.

In this work, we adopt a different approach by considering {\em restrictive} Abelian symmetries in the context of the 2HDM, extended with RH neutrino fields such that neutrinos are Dirac particles. We call these symmetries restrictive since the number of relevant flavor  (Yukawa coupling) parameters in the lepton sector is the same as the number of observables, i.e. nine (ten) in the case of two (three) massive Dirac neutrinos. 

The paper is organized as follows. After a brief introduction on the 2HDM and general considerations regarding Dirac neutrinos presented in Section~\ref{Dir2HDM}, some general aspects of Abelian symmetries in the 2HDM with Dirac neutrinos are described in Section~\ref{Abflvsym}. We then identify in Section~\ref{Max2HDM} the maximally-restrictive textures for the lepton mass matrices compatible with present neutrino oscillation data. Section~\ref{AbComp} is devoted to investigate whether those textures are realizable by Abelian flavor symmetries in the 2HDM. Here, we apply the procedure developed in Ref.~\cite{Serodio:2013gka}, together with the Smith normal form (SNF) and canonical methods for the identification of the U(1) flavor symmetry~\cite{Petersen:2009ip,Ivanov:2013bka}. We conclude that, among the 28 combinations of maximally-restrictive texture-zero mass matrices compatible with data, only 5 can be realized by Abelian symmetries in the 2HDM. In Section~\ref{Phenom}, we discuss the reconstruction of the Yukawa parameters in terms of lepton masses and mixing, paying special attention to leptonic CP violation. The constraints coming from lepton-flavour universality in leptonic $\tau$ decays, $\ell_\beta \rightarrow \ell_\alpha \gamma$ and $\ell_\alpha^-\to \ell_{\beta}^-\,\ell_{\gamma}^+\,\ell_{\delta}^-$ are also discussed. Finally, our concluding remarks are given in Section~\ref{Conclusions}.

\section{Dirac neutrinos in the 2HDM}
\label{Dir2HDM}

As in the SM, Dirac neutrino masses can be generated in the 2HDM by adding RH neutrino singlet fields $\nu_R$, which couple to the SM lepton doublets $\ell_L$ and the two Higgses $\Phi_{a}$ defined, as usual, by
\begin{align}
\Phi_{a}=\left(\begin{matrix}
\phi^+_{a}\smallskip\\
\phi_{a}^0 \end{matrix}
\right)\,,\,a=1,2\,,
\end{align}
being $\phi_{a}^+$ and $\phi_{a}^0$ the charged and neutral components of the scalar doublets. In this framework, the lepton Yukawa interactions can be written as
\begin{align}
\label{lagrangian}
\mathcal{L}_\text{Y} &= -\overline{\ell_L}\left(\Yl_1 \Phi_1 + \Yl_2 \Phi_2\right)e_R \nonumber\\
&- \overline{\ell_L}\left(\Ynu_1 \widetilde{\Phi}_1 + \Ynu_2 \widetilde{\Phi}_2\right)\nu_R + \text{H.c.}\,,
\end{align}
where $e_R$ are the charged-lepton RH singlets, and $\widetilde{\Phi}_{a}= i \sigma_2 \Phi_{a}^*$. The general $3\times 3$ complex matrices $\Yl_{a}$ and $\Ynu_{a}$ encode the charged-lepton and Dirac neutrino Yukawa interactions, respectively. In line with the discussion presented in Section~\ref{Intro}, very small Yukawa couplings $\Ynu_{a}$ may originate from dimension-five operators:
\begin{equation}
\label{dim5Dirac2}
\begin{aligned}
-\mathcal{L}_5^{D} = \frac{\Y_{a}}{\Lambda}\overline{\ell_L} \nu_R \widetilde{\Phi}_{a} S + \text{H.c.},
\end{aligned}
\end{equation}
such that $\Ynu_{a}\equiv \Y_{a} v_S/\Lambda$ are sufficiently suppressed to generate sub-eV Dirac neutrino masses upon EWSB, i.e. when $\phi_{a}^0$ acquire VEVs:
\begin{align}
\label{VEVs}
\langle \phi^0_{a}\rangle\equiv \frac{v_{a}}{\sqrt{2}}\;,\quad \tan\beta\equiv \frac{v_2}{v_1}\,,\quad v=\sqrt{v_1^2+v_2^2}\,,
\end{align}
with $v\simeq 246$~GeV. Thus, from now on we will consider that there is such a mechanism responsible for the smallness of $\Ynu_{a}$. 

The charged-lepton and Dirac-neutrino mass matrices
\begin{equation}
\label{mlmnu}
\begin{aligned}
\Ml = \Yl_1 \frac{v_1}{\sqrt{2}} + \Yl_2 \frac{v_2}{\sqrt{2}}\;,\quad
\Mnu = \Ynu_1 \frac{v_1}{\sqrt{2}} + \Ynu_2 \frac{v_2}{\sqrt{2}}\,,
\end{aligned}
\end{equation}
can be diagonalized by a set of appropriate unitary matrices $\U_{L,R}^{\ell,\nu}$ so that
\begin{equation}
\label{umlumnu}
\begin{aligned}
\ULld \Ml \URl = \Dl &= \text{diag}(m_e,m_\mu,m_\tau),\\
\ULnud \Mnu \URnu = \Dnu &= \text{diag}(m_1,m_2,m_3)\,.
\end{aligned}
\end{equation}
where $m_{e,\mu,\tau}$ and $m_{1,2,3}$ denote the charged-lepton and neutrino masses, respectively, all being real and positive. To extract the LH rotation matrices, one diagonalizes the Hermitian matrices
\begin{equation}
\label{Hdef}
\Hl = \Ml \Ml^\dagger, \quad \Hnu = \Mnu \Mnu^\dagger,
\end{equation}
in the following way:
\begin{equation}
\label{Hdiag}
\begin{aligned}
\ULld \Hl \ULl = \Dl^2 &= \text{diag}(m_e^2,m_\mu^2,m_\tau^2),\\
\ULnud \Hnu \ULnu = \Dnu^2 &= \text{diag}(m_1^2,m_2^2,m_3^2).
\end{aligned}
\end{equation}
The unitary transformations $\U_L^{\ell,\nu}$ define the lepton mixing matrix $\U$ appearing in lepton charged-current interactions as
\begin{equation}
\U= \ULld \ULnu\,.
\label{Udef2}
\end{equation}

In the case of massive Dirac neutrinos, $\U$ can be parametrized by three mixing angles $\theta_{ij}$ and a single CP-violating Dirac phase $\delta$, such that~\cite{Tanabashi:2018oca}
\begin{widetext}
\begin{gather}
\U=\begin{pmatrix}
c_{12}c_{13}&s_{12}c_{13}&s_{13}e^{-i\delta}\\
-s_{12}c_{23}-c_{12}s_{23}s_{13}e^{i\delta}&c_{12}c_{23}-s_{12}s_{23}s_{13}e^{i\delta}&s_{23}c_{13}\\
s_{12}s_{23}-c_{12}c_{23}s_{13}e^{i\delta}&-c_{12}s_{23}-s_{12}c_{23}s_{13}e^{i\delta}&c_{23}c_{13}
\end{pmatrix}\,,
\label{Uparam}
\end{gather}
\end{widetext}
with $c_{ij}\equiv\cos\theta_{ij}$ and $s_{ij}\equiv\sin\theta_{ij}$. Thus, the lepton sector is characterized by ten parameters: three charged-lepton and three neutrino masses, three mixing angles and one phase. 
\begin{table}[t]
\centering
\begin{tabular}{@{}c@{\hskip 0.25in}c@{\hskip 0.25in}c@{}}
\hline \hline\\[-0.2cm]
Parameter&Best fit $\pm1\sigma$&$3\sigma$ range\\[0.1cm]
\hline\\
$\theta_{12}\;(^{\circ})$&$34.5_{-1.0}^{+1.2}$&$31.5\rightarrow38.0$\\[0.15cm]
$\theta_{23}\;(^{\circ})$ [NO] &$47.7_{-1.7}^{+1.2}$&$41.8\rightarrow50.7$\\
$\theta_{23}\;(^{\circ})$ [IO] &$47.9_{-1.7}^{+1.0}$&$42.3\rightarrow50.7$\\[0.15cm]
$\theta_{13}\;(^{\circ})$ [NO] &$8.45_{-0.14}^{+0.16}$&$8.0\rightarrow8.9$\\
$\theta_{13}\;(^{\circ})$ [IO] &$8.53_{-0.15}^{+0.14}$&$8.1\rightarrow9.0$\\[0.15cm]
$\delta\;(^{\circ})$ [NO] &$218_{-27}^{+38}$&$157\rightarrow349$\\
$\delta\;(^{\circ})$ [IO] &$281_{-27}^{+23}$&$202\rightarrow349$\\[0.15cm]
$\Delta m_{21}^2\;(\times 10^{-5}\;\text{eV}^2)$&$7.55_{-0.16}^{+0.20}$&$7.05\rightarrow8.14$\\[0.15cm]
$|\Delta m_{31}^2|\;(\times 10^{-3}\;\text{eV}^2)$ [NO] &$2.50\pm0.03$&$2.41\rightarrow2.60$\\
$|\Delta m_{31}^2|\;(\times 10^{-3}\;\text{eV}^2)$ [IO] &$2.42_{-0.04}^{+0.03}$&$2.31\rightarrow2.51$\\[0.2cm]
\hline\hline
\end{tabular}
\caption{Neutrino oscillation parameters obtained from the global analysis of Ref.~\cite{deSalas:2017kay} for normal ordering (NO) and inverted ordering (IO) of the neutrino mass spectrum (see also Refs.~\cite{Esteban:2016qun,Capozzi:2017ipn}).}
\label{datatable}
\end{table}

Global analyses of all available neutrino oscillation data constrain the parameters of the matrix $\U$ as shown in Table~\ref{datatable} \cite{deSalas:2017kay,Esteban:2016qun,Capozzi:2017ipn}, for both normal ($m_1<m_2<m_3$) and inverted ($m_3<m_1<m_2$) ordering of neutrino masses. In the case of $m_1=0$ ($m_3=0$), $m_2$ and $m_3$ ($m_1$ and $m_2$) are fully determined by the mass-squared differences $\dmsol=m_2^2-m_1^2$ and $\dmatm=m_3^2-m_1^2$ as
\begin{equation}
\begin{aligned}
{\rm NO} &:\; m_2=\sqrt{\dmsol}\;,\; m_3=\sqrt{\dmatm}\,,\\
{\rm IO} &:\; m_1=\sqrt{|\dmatm}|\;,\; m_2=\sqrt{\dmsol+|\dmatm|}\,,
\end{aligned}
\label{NHIH}
\end{equation}
so that the number of measurable quantities is reduced to nine. The cosmological constraint on the sum of the neutrino masses, $\sum_i m_i < 0.12$~eV at 95\% CL~\cite{Palanque-Delabrouille:2015pga}, should also be taken into consideration. If one neutrino is massless, then $\sum_i m_i \simeq 0.059~ \text{eV}$ for NO and $\sum_i m_i \simeq 0.099~ \text{eV}$ for IO, thus being the cosmological limit automatically obeyed.

\section{Abelian symmetries in the 2HDM with Dirac neutrinos}
\label{Abflvsym}

As shown in Appendix~\ref{appdxA0}, the most constrained mass matrices $\Ml$ and $\Mnu$ (compatible with experiment) achievable by imposing Abelian symmetries in the SM with RH neutrinos contain 23 real parameters to be compared with 9 measurable quantities (lepton masses and mixings). This is not the case in models with extra scalar fields like the 2HDM, where Abelian symmetries may lead to interesting constraints in the flavor sector~\cite{Grimus:2004hf,Low:2005yc,Ferreira:2011}. Thus, in this section we will go through some general aspects on this subject as a setup for the forthcoming analysis presented in this work. 

Denoting $\Phi\equiv (\Phi_1\,\, \Phi_2)^T$, and requiring that the full Lagrangian is invariant under the field transformations
\begin{align}
\label{abelsym}
\begin{aligned}
\Phi &\rightarrow 
\mathbf{S}_\Phi \Phi\;,\quad \ell_L \rightarrow 
\mathbf{S}_\ell \ell_L\,,\\
e_R &\rightarrow 
\mathbf{S}_e e_R\;,\quad \nu_R \rightarrow 
\mathbf{S}_\nu \nu_R,
\end{aligned}
\end{align}
where $\mathbf{S}_\Phi \in$ U(2) and $\{\mathbf{S}_\ell,~\mathbf{S}_e,~\mathbf{S}_\nu\}$ $\in$ U(3) are unitary matrices, yields the following constraints on the Yukawa couplings
\begin{equation}
\label{yukawainvariance}
\begin{aligned}
\Yl_a = \mathbf{S}_\ell \Yl_b \mathbf{S}_e^\dagger (\mathbf{S}^\dagger_\Phi)_{ba}\;,\quad
\Ynu_a = \mathbf{S}_\ell \Ynu_b \mathbf{S}_\nu^\dagger (\mathbf{S}^T_\Phi)_{ba}\,,
\end{aligned}
\end{equation}
where a sum over $b=1,2$ is implicitly assumed. By performing basis transformations identical to those in Eq.~\eqref{abelsym}, with the appropriate choices of unitary matrices $\mathbf{V} \in$ U(2) and $\{\mathbf{V}_\ell, \mathbf{V}_e, \mathbf{V}_\nu\}\in$ U(3), one can bring the matrices $\mathbf{S}$ into the form~\cite{Ferreira:2011}:
\begin{align}
\label{symtransf}
\begin{aligned} 
\mathbf{S} &=  \text{diag}(e^{i \theta_1},e^{i \theta_2})\,,\quad
\mathbf{S}_\ell = \text{diag}(e^{i \alpha_1},e^{i \alpha_2},e^{i \alpha_3}),\\
\mathbf{S}_e &= \text{diag}(e^{i \beta_1},e^{i \beta_2},e^{i \beta_3})\,,\quad
\mathbf{S}_\nu = \text{diag}(e^{i \gamma_1},e^{i \gamma_2},e^{i \gamma_3})\,,
\end{aligned}
\end{align}
where $\theta_i$, $\alpha_i$, $\beta_i$ and $\gamma_i$ are continuous phases. 

Under the general transformation~\eqref{symtransf}, the invariance condition~\eqref{yukawainvariance} reads
\begin{equation}
\label{absyminv}
(\Yx_a)_{ij} = e^{i (\Theta_{a}^{x})_{ij}} (\Yx_a)_{ij},
\end{equation}
where $i,j=1,2,3$ are flavor indices and $x=\ell,\nu$. The phase matrices $\Theta_{a}^{x}$, which encode the transformation properties of each Yukawa interaction, are defined as
\begin{equation}
\label{chargphases}
\begin{aligned}
(\Theta_{a}^{\ell})_{ij} = \beta_j-\alpha_i + \theta_{a},\quad
(\Theta_{a}^{\nu})_{ij} =  \gamma_j-\alpha_i - \theta_{a}\,.
\end{aligned}
\end{equation}
These phases can be written in terms of charges ($\alpha^\prime, \beta^\prime, \gamma^\prime, \theta^\prime$) and a parameter $\varphi \in [0,~2 \pi[$ such that
\begin{equation}
\label{chargphases2}
\begin{aligned}
(\Theta_{a}^{\ell})_{ij} &= (\beta^\prime_j-\alpha^\prime_i + \theta^\prime_{a})\,\varphi,\\
(\Theta_{a}^{\nu})_{ij} &=  (\gamma^\prime_j-\alpha^\prime_i - \theta^\prime_{a})\,\varphi\,.
\end{aligned}
\end{equation}
The particular case of $\varphi = 2 \pi/N$, with $N=2,3,...$, corresponds to a discrete $\mathbb{Z}_N$ symmetry. With these redefinitions, Eq.~\eqref{yukawainvariance} can be interpreted in terms of charge relations. Namely, invariance of  $(\Yx_a)_{ij}$ under the U(1) symmetry implies $(\Theta_{a}^{x})_{ij} = 0~(\text{mod}~2\pi)$. It is also straightforward to conclude that, as a consequence of having $\theta_1 - \theta_2 \neq 0~(\text{mod}~2\pi)$,\footnote{Otherwise one recovers, in terms of a phase transformation like in Eq.~\eqref{chargphases}, the SM context, since the same texture would be enforced on the Yukawa matrices $\mathbf{Y}^x_1$ and $\mathbf{Y}^x_2$ and, hence, in the resulting mass matrix $\mathbf{M}_x$.} a non-zero entry in $\Yx_{1}$ will automatically imply a zero entry in $\Yx_{2}$, and vice versa. Moreover, one can see from Eq.~\eqref{yukawainvariance} that overall rephasings of the type
\begin{equation}
\label{rephasingfreedom}
\begin{aligned}
\mathbf{S} &\rightarrow e^{i \theta} \mathbf{S}, \quad
\mathbf{S}_\ell \rightarrow e^{i \alpha} \mathbf{S}_\ell, \\
\mathbf{S}_e &\rightarrow e^{i \beta} \mathbf{S}_e,\quad
\mathbf{S}_{\nu} \rightarrow e^{i \gamma} \mathbf{S}_\nu,
\end{aligned}
\end{equation}
with $e^{i (\alpha - \beta - \theta)} =e^{i (\alpha - \gamma + \theta)} = 1$, do not alter the invariance condition on the Lagrangian. Thus, without loss of generality, we can set one of the Higgs and fermion transformation phases to zero. Here, we choose $\theta_1=\alpha_1=0$.

The aforementioned U(1) charges determine the presence (or absence) of zero entries in the Yukawa and mass matrices $\Yx_a$ and $\Mx$, defined in Eqs.~\eqref{lagrangian} and \eqref{mlmnu}, respectively. In particular, 
\begin{equation}
\label{restrictions}
\begin{aligned}
 (\Mx)_{ij}=0 \quad \Leftrightarrow \quad &(\Theta_{1}^{x})_{ij} \neq 0~(\text{mod}~2\pi)\\
  \wedge\; &(\Theta_{2}^{x})_{ij} \neq 0~(\text{mod}~2\pi) ,\\
 (\Mx)_{ij} \neq 0 \quad \Leftrightarrow \quad &(\Theta_{1}^{x})_{ij} = 0~(\text{mod}~2\pi) \\
 \vee\; &(\Theta_{2}^{x})_{ij} = 0~(\text{mod}~2\pi)\,.
\end{aligned}
\end{equation}
Vanishing elements in a mass matrix or Yukawa interaction matrix are usually dubbed as ``texture zeros". In this work, whenever a general matrix structure contains texture zeros, we will refer to it as a ``texture".

Before concluding this section, it is worth comparing the lepton and quark sectors in the framework considered in this work, i.e. the 2HDM with RH neutrinos. The Yukawa interactions for quarks are the same as in Eq.~\eqref{lagrangian} after replacing $\ell_L$ by $Q_L$ (LH quark doublets), $e_R$ by $d_R$ (RH down-quark singlets) and $\nu_R$ by $u_R$ (RH up-quark singlets). The charged-lepton and Dirac neutrino Yukawa matrices must be also replaced by the down and up-quark ones, $\Yd_{a}$ and $\Yu_{a}$, respectively. Textures for quarks can, in principle, be implemented in the same way as for leptons by imposing Abelian symmetries as the ones discussed above. However, as explained in Appendix~\ref{appdxA0}, this cannot be done in the SM. As in the lepton sector, quark textures can be implemented by Abelian symmetries in the 2HDM, being the main difference in the fact that all six quarks must be massive. This is in contrast with leptons, for which the existence of a massless neutrino is allowed by current experimental data.

\section{Maximally-restrictive textures for leptons}
\label{Max2HDM}

In Ref.~\cite{Ludl:2014axa}, all possible textures for $\Ml$ and $\Mnu$ were identified and grouped into equivalence classes, considering both Majorana and Dirac massive neutrinos. 

For charged leptons, two textures $\Ml$ and $\Ml^\prime$ are equivalent if they can be transformed onto each other by performing permutations of the $\ell_L$ and $e_R$ fields, i.e. if
\begin{gather}
\Ml^\prime  = \mathbf{P}_\ell^{\dagger}\Ml \mathbf{P}_e,\label{weakbasis}
\end{gather}
where $\mathbf{P}_{\ell,e}$ can be any two matrices of the 3-dimensional representation of the $S_3$ permutation group, namely,
\begin{align}
\begin{aligned}
\mathcal{I}&=\begin{pmatrix} 1&0&0\\ 0&1&0\\ 0&0&1 \end{pmatrix}\;,
&\mathcal{P}_{12}&=\begin{pmatrix} 0&1&0\\ 1&0&0\\ 0&0&1 \end{pmatrix}\,,\\
\mathcal{P}_{13}&=\begin{pmatrix} 0&0&1\\ 0&1&0\\ 1&0&0 \end{pmatrix}\;,
&\mathcal{P}_{23}&=\begin{pmatrix} 1&0&0\\ 0&0&1\\ 0&1&0 \end{pmatrix}\,,\\
\mathcal{P}_{123}&=\begin{pmatrix} 0&0&1\\ 1&0&0\\ 0&1&0 \end{pmatrix}\;,
&\mathcal{P}_{321}&=\begin{pmatrix} 0&1&0\\ 0&0&1\\ 1&0&0 \end{pmatrix}\,.
\end{aligned}
\label{Pmat}
\end{align}
Said otherwise, two $\Ml$ textures are equivalent when they are equal up to permutations of rows and columns. 

In the case of Dirac neutrinos, two textures $\Mnu$ and $\Mnu^\prime$ are considered equivalent if 
\begin{gather}
\Mnu' = \Mnu \mathbf{P}_\nu\,,\label{weakbasis2}
\end{gather}
where $\mathbf{P}_\nu$ is also a permutation matrix. Thus, two neutrino mass matrix textures are equivalent if they can be transformed onto each other by column permutations.

We shall combine the above $\Ml$ and $\Mnu$ classes into all possible $(\Ml,\Mnu)$ pairs, keeping only one representative texture of each $\Ml$ and $\Mnu$ equivalence class. Pairs leading to the same leptonic mixing matrix $\U$ are equivalent and, thus, redundant. This is the case when the mass matrices can be related by
\begin{gather}
\Ml^\prime  = \mathbf{P}_\ell^{\dagger}\Ml \mathbf{P}_e\;,\quad {\Mnu^\prime} = \mathbf{P}_\ell^{\dagger} \Mnu \mathbf{P}_{\nu}\,,\label{weakbasis3}
\end{gather}
for any two texture pairs $(\Ml,\Mnu)$ and $(\Ml^\prime,\Mnu^\prime)$. Notice that, in order to leave $\U$ invariant, $\mathbf{P}_{\ell}$ must be the same in both transformations. Therefore, two texture pairs are equivalent if they can be obtained from each other through column and row permutations, being the row permutation identical for both mass matrices in the pair. This is why in Eq.~\eqref{weakbasis2} only column permutations are considered, avoiding the possibility of excluding relevant cases. The outlined procedure aims at eliminating redundant cases that lead to the same physics, i.e., that reproduce the same mass and mixing parameters.

Since in this work we are interested in the most predictive $\Ml$ and $\Mnu$, it is crucial to introduce the concept of maximally-restrictive textures~\cite{Ludl:2014axa,Felipe:2016sya}:

\vspace{2mm}
\noindent {\em A texture pair $(\Ml,\Mnu)$ is said to be maximally restrictive if the predicted values for the lepton masses, mixing angles and CP phase are compatible with the experimental data, and the addition of one more texture zero in either $\Ml$ or $\Mnu$ makes the pair incompatible with data}.
\vspace{2mm}

Essentially, these are the pairs with least parameters, which are viable when confronted with observations. In order to identify them, we shall perform an analysis similar to the one of Refs.~\cite{Ludl:2014axa,Felipe:2016sya}, considering the updated neutrino oscillation parameters and including the current ranges for the Dirac phase $\delta$ (see Table~\ref{datatable}). We will require compatibility at 3$\sigma$ confidence level (CL). 

\subsection{Compatibility with data}
\label{CompData}

In order to identify the maximally-restrictive texture pairs $(\Ml,\Mnu)$ we performed a standard $\chi^2$-analysis with the function
\begin{equation}
\chi^2(x) = \sum_i \frac{\left[\mathcal{P}_i(x)-\overline{\mathcal{O}}_i\right]^2}{\sigma_i^2},
\end{equation}
where $x$ denotes the matrix elements of $\Ml$ and $\Mnu$, $\mathcal{P}_i(x)$ is the model prediction for the observable $\mathcal{O}_i$, $\overline{\mathcal{O}}_i$ is the corresponding best-fit value, and $\sigma_i$ denotes its 1$\sigma$ error. 

In our search for viable pairs ($\Ml,\Mnu$), we require the charged-lepton masses to be at their central values~\cite{Tanabashi:2018oca}, so that the $\chi^2$-function is minimized only with respect to the six neutrino observables $\mathcal{O}_i$ (the two neutrino mass-squared differences $\Delta m^2_{21,31}$, the three mixing angles $\theta_{ij}$ and the Dirac phase $\delta$) following the numerical method presented in Refs.~\cite{Cebola:2015dwa,Felipe:2016sya}. If the deviation of each neutrino observable from its experimental value is at most 3$\sigma$ at the $\chi^2$ minimum for a given $(\Ml,\Mnu)$ pair, the corresponding lepton textures are said to be compatible with data. In such cases, we test compatibility at the 1$\sigma$ as well. 

Our results show that the maximally-restrictive pairs ($\Ml,\Mnu$) compatible with data are those presented in Table~\ref{texturepairs}, where the labeling follows the notation of Ref.~\cite{Ludl:2014axa}.\footnote{Note that our matrices $\Mnu$ correspond to  $\Mnu^\dagger$ in Ref.~\cite{Ludl:2014axa}, since in this reference the RH neutrino fields appear on the left in the Dirac neutrino mass term.} A representative texture of each equivalence class is presented in Tables~\ref{leptontextures} and \ref{neutrinotextures} for $\Ml$ and $\Mnu$, respectively. All pairs in Table~\ref{texturepairs} were found to be consistent with neutrino oscillation data at 1$\sigma$, for both NO and IO, except for the pair $(6_1^{\ell},4_{17}^{\nu})$ which is consistent with data only at 3$\sigma$ CL and for a NO mass spectrum. Also notice that, with the exception of texture $4_{17}^{\nu}$, any representative of $\Mnu$ given in Table~\ref{neutrinotextures} features a massless neutrino, since it contains a full column of zeros.
\begin{table}[t]
\centering
\begin{tabular}{@{}c@{\hskip 0.2in}ccccccccc@{}}
\hline \hline \\[-0.2cm]
$\Ml$ & \multicolumn{9}{c}{$\Mnu$} \\\hline\\[-0.2cm]
$3_2^{\ell}$ & $7_1^{\nu}$ &  $7_3^{\nu}$ & & & & & & & \\[0.3cm]
$4_1^{\ell}$ & $6_1^{\nu}$ &  $6_3^{\nu}$ & $6_4^{\nu}$ &  $6_5^{\nu}$ &  $6_6^{\nu}$ &  &  &  & \\[0.3cm]
$4_2^{\ell}$ & $6_1^{\nu}$ &  $6_2^{\nu}$ &  $6_3^{\nu}$ & $6_7^{\nu}$ &  $6_8^{\nu}$ &  &  & & \\[0.3cm]
$4_3^{\ell}$ & $6_1^{\nu}$ &  $6_2^{\nu}$ &  $6_3^{\nu}$ & $6_4^{\nu}$ &  $6_5^{\nu}$ &  $6_6^{\nu}$ & $6_7^{\nu}$ &  $6_8^{\nu}$ &  $6_9^{\nu}$ \\[0.3cm]
$5_1^{\ell}$ & $5_1^{\nu}$ & $5_4^{\nu}$  & $5_5^{\nu}$ &  $5_6^{\nu}$ &  $5_8^{\nu}$ & & & & \\[0.3cm]$6_1^{\ell}$ &  $4_1^{\nu}$ & $4_{17}^{\nu}$ &  & & &  & & & \\[0.1cm]
\hline \hline
\end{tabular}
\caption{Maximally-restrictive $(\Ml,\Mnu)$ texture pairs compatible with data at 1$\sigma$ CL for both NO and IO (see Table~\ref{datatable}). The pair $(6_1^{\ell},4_{17}^{\nu})$ was found to be consistent with experimental data only at 3$\sigma$ and for IO.}
\label{texturepairs}
\end{table}
\begin{table}[t]
\centering
\begin{tabular}{@{}c@{\hskip 0.15in}c@{\hskip 0.15in}c@{}}
\hline \hline \\
$3_2^{\ell} \sim \begin{pmatrix} 0&\times&\times\\0&\times&\times\\\times&0&\times \end{pmatrix}$ & & \\[0.6cm]
$4_1^{\ell} \sim  \begin{pmatrix} 0&0&\times\\0&\times&0\\\times&\times&\times \end{pmatrix}$  & $4_2^{\ell} \sim \begin{pmatrix} 0&0&\times\\0&\times&\times\\\times&0&\times \end{pmatrix}$ &  $4_3^{\ell} \sim  \begin{pmatrix} 0&0&\times\\0&\times&\times\\\times&\times&0 \end{pmatrix}$    \\[0.6cm]
$5_1^{\ell} \sim  \begin{pmatrix} 0&0&\times\\0&\times&0\\\times&0&\times \end{pmatrix}$  & & \\[0.6cm]
$6_1^{\ell} \sim   \begin{pmatrix} 0&0&\times\\0&\times&0\\\times&0&0 \end{pmatrix}$  & & \\[0.6cm] 
\hline \hline 
\end{tabular}
\caption{Representative textures of the $\Ml$ equivalence classes appearing in the maximally-restrictive $(\Ml,\Mnu)$ pairs shown in Table~\ref{texturepairs}. }
\label{leptontextures}
\end{table}
\begin{table}[!t]
\centering
\begin{tabular}{@{}c@{\hskip 0.15in}c@{\hskip 0.15in}ccc@{}}
\hline \hline\\
$4_1^{\nu} \sim  \begin{pmatrix} 0&0&\times\\0&\times&\times\\ 0&\times&\times \end{pmatrix}$ & $4_{17}^{\nu} \sim  \begin{pmatrix} 0&\times&\times\\\times&0&\times\\ 0&0&\times \end{pmatrix}$ & \\[0.9cm]

$5_1^{\nu} \sim  \begin{pmatrix} 0&0&\times\\0&0&\times\\0&\times&\times \end{pmatrix}$ &  $5_4^{\nu} \sim  \begin{pmatrix} 0&0&\times\\0&\times&0\\0&\times&\times \end{pmatrix}$ &  $5_5^{\nu} \sim  \begin{pmatrix} 0&0&\times\\0&\times&\times\\0&\times&0 \end{pmatrix}$\\  [0.6cm]
$5_6^{\nu} \sim  \begin{pmatrix} 0&\times&\times\\0&0&\times\\0&0&\times \end{pmatrix}$ &
$5_8^{\nu} \sim  \begin{pmatrix} 0&\times&\times\\0&0&\times\\0&\times&0 \end{pmatrix}$ & & \\[0.9cm]

$6_1^{\nu} \sim   \begin{pmatrix} 0&0&0\\ 0&0&\times\\ 0&\times&\times \end{pmatrix}$  &  $6_2^{\nu} \sim   \begin{pmatrix} 0&0&\times\\ 0&0&0\\ 0&\times&\times \end{pmatrix}$ &  $6_3^{\nu} \sim   \begin{pmatrix} 0&0&\times\\ 0&0&\times\\ 0&\times&0 \end{pmatrix}$  \\[0.6cm]

$6_4^{\nu} \sim   \begin{pmatrix} 0&0&0\\ 0&\times&\times\\ 0&0&\times \end{pmatrix}$    & 
$6_5^{\nu} \sim   \begin{pmatrix} 0&0&\times\\ 0&\times&0\\ 0&0&\times \end{pmatrix}$ & $6_6^{\nu} \sim   \begin{pmatrix} 0&0&\times\\ 0&\times&\times\\ 0&0&0 \end{pmatrix}$  \\[0.6cm]

$6_7^{\nu} \sim   \begin{pmatrix} 0&0&\times\\ 0&\times&0\\ 0&\times&0 \end{pmatrix}$ &  $6_8^{\nu} \sim   \begin{pmatrix} 0&\times&\times\\ 0&0&0\\ 0&0&\times \end{pmatrix}$   &
$6_9^{\nu} \sim   \begin{pmatrix} 0&\times&\times\\ 0&0&\times\\ 0&0&0 \end{pmatrix}$ \\[0.9cm]

        $7_1^{\nu} \sim \begin{pmatrix} 0&0&0\\0&0&\times\\0&\times&0 \end{pmatrix}$ &  $7_3^{\nu} \sim \begin{pmatrix} 0&0&\times\\0&\times&0\\0&0&0 \end{pmatrix}$ &  \\[0.6cm] \hline \hline
\end{tabular}
\caption{Representative textures of the $\Mnu$ equivalence classes appearing in the maximally-restrictive $(\Ml,\Mnu)$ pairs shown in Table~\ref{texturepairs}. }
\label{neutrinotextures}
\end{table}

We emphasize that these maximally-restrictive texture pairs cannot be implemented in the SM by imposing Abelian symmetries (see Appendix~\ref{appdxA0}). Hence, in the next section we will address the question of whether (or which of) the texture pairs in Table~\ref{texturepairs} can be implemented in the context of the 2HDM with Abelian flavor symmetries.

\section{Abelian symmetry realization of compatible textures}
\label{AbComp}

Having identified the maximally-restrictive texture pairs that are compatible with data, next we aim at ascertaining the pairs in Table~\ref{texturepairs} that can be realized in the 2HDM by imposing continuous or discrete Abelian symmetries. At the same time, we wish to identify the corresponding transformation properties of the various fields according to Eq.~\eqref{abelsym}. Keeping this in mind, two methods shall be employed, namely the canonical and SNF methods, which we briefly review below. 

In the canonical method, the phases in Eq.~\eqref{chargphases2} are considered as variables and, for each texture pair of Table~\ref{texturepairs}, a system of equations corresponding to the conditions in Eq.~\eqref{restrictions} is defined. Subsequently, one checks if the system has a non-trivial solution for the phases, considering all possible decompositions of $\Ml$ and $\Mnu$ into Yukawa matrices $\Yl_{1,2}$ and $\Ynu_{1,2}$. If no solution is found, the texture (or texture pair) is not realizable by means of an Abelian symmetry (see, for example, Refs.~\cite{Ferreira:2011,Serodio:2013gka} for more details). 

The SNF method is used to identify rephasing symmetries of the Lagrangian for specific decompositions of $(\Ml,\Mnu)$ texture pairs into Yukawa matrices. With $n_f$ complex fields, and in the absence of phase-sensitive terms, the Lagrangian is invariant under a [U(1)]$^{n_f}$ symmetry. In the presence of terms transforming non-trivially under rephasing, the symmetry group is altered and the SNF method can be used to identify it. With this purpose, one starts by defining a $k \times n_f$ integer matrix $\textbf{D}$ where each row $\textbf{d}_l$ corresponds to one of the $k$ interactions allowed in the Lagrangian, with
\begin{align}
\textbf{d}_l = (\textbf{d}_\Phi,\textbf{d}_{\ell_L},\textbf{d}_{e_R},\textbf{d}_{\nu_R}),
\end{align}
where the number of $\textbf{d}_f$ components is equal to the number of fields of type $f$, i.e. two for $\textbf{d}_\Phi$ and three for the remaining fields. Specifically, $\textbf{d}_{f_i}=n$ ($\textbf{d}_{f_i}=-n$) when the (conjugate of) field $f_i$ appears $n$ times in the interaction. For instance, $\textbf{d}_l=(0,1,-1,0,0,0,0,1,0,0,0)$ corresponds to the coupling $\overline{e_{L}}\tau_R \Phi_2$. 

The matrix \textbf{D} can then be brought to its unique SNF~\cite{Ivanov:2013} such that $\textbf{D}=\textbf{R}\,\textbf{D}_{\text{SNF}}\,\textbf{C}$ with
\begin{equation}
\textbf{D}_{\text{SNF}} = \text{diag}\{d_1,d_2,...,d_r,0,...,0\}\,,
\label{DSNFds}
\end{equation}
where $d_i$ is a positive integer, divisor of $d_{i+1}$, and $r$ is the rank of \textbf{D}. The matrices $\textbf{R}$ and $\textbf{C}$ encode the addition, sign flip and permutation operations on the rows and columns, respectively. 

The minimal rephasing symmetry of the Lagrangian can be identified as
\begin{equation}
[\text{U}(1)]^{n_F} \rightarrow G = \mathbb{Z}_{d_1} \times ... \times \mathbb{Z}_{d_r} \times [\text{U}(1)]^{n_f-r},
\end{equation}
where $d_i=1$ and $d_i=0$ correspond to a continuous U(1) symmetry and the trivial group, respectively. For the case of discrete groups, the symmetry charges can be determined from the columns of $\textbf{C}^{-1}$, although no information regarding the charges associated to continuous symmetries can be obtained. We should emphasize that the SNF method identifies symmetries under which all interactions included in $\textbf{D}$ are invariant, without guaranteeing that all remaining terms are absent from the Lagrangian. A more detailed discussion on this point can be found in Ref.~\cite{Ivanov:2013}.

We first apply the canonical method to find which of the $(\Ml,\Mnu)$ pairs given in Table~\ref{texturepairs} cannot be implemented with Abelian flavor symmetries in the 2HDM. If both $\Ml$ and $\Mnu$ textures can be implemented but not simultaneously, then the corresponding pair is not realizable by the symmetry. From this analysis, which is detailed in Appendix~\ref{appdxA}, we conclude that 23 of the 28 maximally-restrictive pairs $(\Ml,\Mnu)$ appearing in Table~\ref{texturepairs} cannot be realized through Abelian symmetries in the present context. Moreover, all redundant pairs (not presented in the table) which are equivalent to these 23 pairs cannot be implemented either.\footnote{This is clear since the operations (row and column permutations) that relate different texture pairs in an equivalence class do not alter the solution of the system of equations determined through the canonical method. Thus, if the system lacks a solution for a member of such class, it does so for all texture pairs in it.} 

In conclusion, only the 5 pairs $(4_3^{\ell},6_{1,3,7,9}^{\nu})$ and $(5_1^{\ell},5_{8}^{\nu})$ have no inconsistencies and can be implemented by an Abelian symmetry. In order to identify it, we now study all possible decompositions of these pairs into Yukawa matrices.

\subsection{Decomposition into Yukawa textures}
\label{decompyukawa}

In Ref.~\cite{Serodio:2013gka}, a thorough analysis on the realization of Yukawa textures through Abelian symmetries in multi-Higgs-doublet models was performed. In particular, it was established that any Yukawa texture $\Yx$ realizable by a transformation of the type \eqref{symtransf} can be expressed as
\begin{equation}
\mathbf{Y}^x=\mathcal{P}' A_k \mathcal{P}\,,
\end{equation}
where $A_k$ is one of the $3 \times 3$ texture matrices defined in Eq.~(31) of Ref.~\cite{Serodio:2013gka}, and $\mathcal{P}'$ and $\mathcal{P}$ are permutation matrices. The $A_k$ textures are classified according to the degeneracy of the LH and RH symmetry phases $\alpha_i$ and $\beta_j$ in Eq.~\eqref{symtransf}. Those $A_k$ that are realized through transformations with the same phase degeneracy are grouped together in classes ({\bf i},{\bf j}), where {\bf i} and {\bf j} indicates the number of non-degenerate $\alpha_i$ and $\beta_j$, respectively. Within each class ({\bf i},{\bf j}), all $\mathcal{P}' A_k \mathcal{P}$ which can be simultaneously realized without common non-zero elements are grouped into ``chains", being $C_n^{({\bf i},{\bf j})}$ the $n$th chain of the ({\bf i},{\bf j}) class.

In the 2HDM framework, a given decomposition of a mass matrix $\mathbf{M}_x$ into Yukawa textures is realizable through Abelian symmetries if there is at least one chain containing the corresponding $\mathbf{Y}^x_1$ and $\mathbf{Y}^x_2$. Since none of the textures belonging to the pairs $(4_3^{\ell},6_{1,3,7,9}^{\nu})$ and $(5_1^{\ell},5_{8}^{\nu})$ has identical columns, and neither $4_3^\ell$ nor $5_1^\ell$ has identical rows, the implementation of any of these textures require non-degenerate transformations, i.e three distinct $\alpha_i$, three distinct $\beta_i$ and three distinct $\gamma_i$ phases. Thus, the search for decompositions is limited to the class \textbf{(3,3)}, generated by 
\begin{gather}
A_{12} = \begin{pmatrix} 0&0&0\\0&0&0\\0&0&\times \end{pmatrix},
A_{13}= \begin{pmatrix} \times&0&0\\0&\times&0\\0&0&\times \end{pmatrix},\nonumber\\
A_{15}= \begin{pmatrix} 0&0&0\\0&\times&0\\0&0&\times \end{pmatrix}\,,
\end{gather}
with the corresponding chains $C_n^{({\bf 3},{\bf 3})}$ given in  Ref.~\cite{Serodio:2013gka}. By inspecting these chains, we determine the viable decompositions for each matrix of the $(4_3^{\ell},6_{1,3,7,9}^{\nu})$ and $(5_1^{\ell},5_{8}^{\nu})$ pairs (see  Table~\ref{leptontexturedecompositions}). The associated chains and their corresponding building matrices are shown in Tables~\ref{chargedleptondecomp} and~\ref{neutrinodecomp}, for the charged-lepton and neutrino textures, respectively. 

\begin{table}[!htb]
\centering
\begin{tabular}{@{}cc@{\hskip 0.25in}c@{}}
\hline \hline\\[-0.2cm]
\multicolumn{3}{c}{Charged leptons}\\
[0.25cm]
Texture decomposition & $\Yl_1$ & $\Yl_2$ \\\hline\\[-0.2cm]
$4_{3,\text{I}}^{\ell}$ 
& $\begin{pmatrix} 0&0&\times\\0&\times&0\\\times&0&0 \end{pmatrix}$ 
& $\begin{pmatrix} 0&0&0\\0&0&\times\\0&\times&0 \end{pmatrix}$ \\[0.6cm]

$5_{1,\text{I}}^{\ell}$ 
& $\begin{pmatrix} 0&0&\times\\0&\times&0\\\times&0&0 \end{pmatrix}$
& $\begin{pmatrix} 0&0&0\\0&0&0\\0&0&\times \end{pmatrix}$\\[0.6cm]

$5_{1,\text{II}}^{\ell}$ 
& $\begin{pmatrix} 0&0&\times\\0&0&0\\\times&0&0 \end{pmatrix}$
& $\begin{pmatrix} 0&0&0\\0&\times&0\\0&0&\times \end{pmatrix}$\\[0.5cm]
\hline \hline\\[-0.2cm]
\multicolumn{3}{c}{Dirac neutrinos}\\
[0.25cm]
Texture decomposition & $\Ynu_1$ & $\Ynu_2$ \\\hline\\[-0.2cm]
$5_{8,\text{I}}^{\nu}$ 
& $\begin{pmatrix} 0&0&\times\\0&0&0\\0&\times&0 \end{pmatrix}$ 
& $\begin{pmatrix} 0&\times&0\\ 0&0&\times \\ 0&0&0 \end{pmatrix}$ \\[0.6cm]

$6_{1,\text{I}}^{\nu}$ 
& $\begin{pmatrix} 0&0&0\\0&0&0\\0&0&\times \end{pmatrix}$ 
& $\begin{pmatrix} 0&0&0\\ 0&0&\times \\ 0&\times&0 \end{pmatrix}$ \\[0.6cm]

$6_{3,\text{I}}^{\nu}$ 
& $\begin{pmatrix} 0&0&0\\0&0&\times\\0&0&0 \end{pmatrix}$ 
& $\begin{pmatrix} 0&0&\times\\ 0&0&0 \\ 0&\times&0 \end{pmatrix}$ \\[0.6cm]

$6_{3,\text{II}}^{\nu}$ 
& $\begin{pmatrix} 0&0&\times\\0&0&0\\0&0&0 \end{pmatrix}$ 
& $\begin{pmatrix} 0&0&0\\ 0&0&\times \\ 0&\times&0 \end{pmatrix}$ \\[0.6cm]

$6_{7,\text{I}}^{\nu}$ 
& $\begin{pmatrix} 0&0&\times\\ 0&0&0 \\ 0&\times&0 \end{pmatrix}$ 
& $\begin{pmatrix} 0&0&0\\ 0&\times&0 \\ 0&0&0 \end{pmatrix}$ \\[0.6cm]

$6_{7,\text{II}}^{\nu}$ 
& $\begin{pmatrix} 0&0&\times\\ 0&\times&0 \\ 0&0&0 \end{pmatrix}$ 
& $\begin{pmatrix} 0&0&0\\ 0&0&0 \\ 0&\times&0 \end{pmatrix}$ \\[0.6cm]

$6_{9,\text{I}}^{\nu}$ 
& $\begin{pmatrix} 0&\times&0\\0&0&\times\\0&0&0 \end{pmatrix}$ 
& $\begin{pmatrix} 0&0&\times\\ 0&0&0 \\ 0&0&0 \end{pmatrix}$ \\[0.5cm]
\hline \hline 
\end{tabular}
\caption{Realizable decompositions into Yukawa matrices of the textures in the pairs $(4_3^{\ell},6_{1,3,7,9}^{\nu})$ and $(5_1^{\ell},5_{8}^{\nu})$. The subscripts I and II label different decompositions of the same mass matrix texture.}
\label{leptontexturedecompositions}
\end{table}

\begin{table*}[!htb]
\centering
\begin{tabular}{@{}c@{\hskip 0.4in}c@{\hskip 0.4in}c@{}}
\hline \hline\\[-0.2cm]
Texture decomposition & Building matrices & Associated chains\\[0.1cm] \hline\\[-0.2cm]
$4_{3,\text{I}}^{\ell}$ & \multirow{1}{*}{$A_{13}\mathcal{P}_{13} + A_{15}\mathcal{P}_{23}$} &  $C_2^{(3,3)}\mathcal{P}_{13}$\\[0.5cm]
\multirow{2}{*}{ $5_{1,\text{I}}^{\ell}$} & \multirow{1}{*}{$A_{13}\mathcal{P}_{13} + A_{12}$} &  $C_{3-5}^{(3,3)}\mathcal{P}_{13}$\\
& \multirow{1}{*}{$\mathcal{P}_{13}A_{13} + A_{12}$} &  $\mathcal{P}_{13}C_{4}^{(3,3)}$\\[0.5cm]
\multirow{4}{*}{ $5_{1,\text{II}}^{\ell}$} & \multirow{1}{*}{$\mathcal{P}_{123}A_{15}\mathcal{P}_{12}+A_{15}$} &  $C_{6,7}^{(3,3)}$\\
& \multirow{1}{*}{$\mathcal{P}_{12}A_{15}\mathcal{P}_{321}+A_{15}$} &  $\mathcal{P}_{13} C_{11}^{(3,3)}\mathcal{P}_{321}$\\
& \multirow{1}{*}{$\mathcal{P}_{123}A_{15}\mathcal{P}_{12}+\mathcal{P}_{23}A_{15}\mathcal{P}_{23}$} &  $\mathcal{P}_{23} C_{11,12}^{(3,3)} \mathcal{P}_{23}$~;~$\mathcal{P}_{123} C_{12}^{(3,3)} \mathcal{P}_{13}$\\
& \multirow{1}{*}{$\mathcal{P}_{12}A_{15}\mathcal{P}_{321}+\mathcal{P}_{23}A_{15}\mathcal{P}_{23}$} &  $\mathcal{P}_{12} C_{2,7}^{(3,3)} \mathcal{P}_{123}$~;~$\mathcal{P}_{23} C_{13}^{(3,3)} \mathcal{P}_{23}$\\[0.1cm]
\hline \hline
\end{tabular}
\caption{Building matrices and corresponding chains of Ref.~\cite{Serodio:2013gka} which contain the charged-lepton texture decompositions identified in Table~\ref{leptontexturedecompositions}.}
\label{chargedleptondecomp}
\end{table*}

\begin{table*}[!htb]
\centering
\begin{tabular}{@{}c@{\hskip 0.4in}c@{\hskip 0.4in}c@{}}
\hline \hline\\[-0.2cm]
Texture decomposition & Building matrices & Associated chains\\[0.1cm] \hline\\[-0.2cm]
\multirow{2}{*}{$5_{8,\text{I}}^{\nu}$} & \multirow{1}{*}{$\mathcal{P}_{123}A_{15} + \mathcal{P}_{321}A_{15}$} &  $C_{7}^{(3,3)}\mathcal{P}_{12}$~;~$C_{8}^{(3,3)}$~;~$\mathcal{P}_{123}C_{11}^{(3,3)}$\\ 
& \multirow{1}{*}{$\left(\mathcal{P}_{12}A_{15}+\mathcal{P}_{13}A_{15}\right)\mathcal{P}_{23}$} &  $\mathcal{P}_{13}C_{11}^{(3,3)}\mathcal{P}_{23}$~;~$\mathcal{P}_{12}C_{14}^{(3,3)}\mathcal{P}_{23}$\\ [0.5cm]

\multirow{2}{*}{$6_{1,\text{I}}^{\nu}$} & \multirow{1}{*}{$A_{12} + A_{15}\mathcal{P}_{23}$} 
&  $C_{3,15}^{(3,3)}\mathcal{P}_{13}$~;~$C_{9,12-14,17}^{(3,3)}\mathcal{P}_{23}$~;~$\mathcal{P}_{13}C_{11,12}^{(3,3)}\mathcal{P}_{13}$\\ 
& \multirow{1}{*}{$A_{12} + \mathcal{P}_{23}A_{15}$} 
&  $\mathcal{P}_{13}C_{14}^{(3,3)}$~;~$\mathcal{P}_{23}C_{9,11,13,15,17}^{(3,3)}$~;~$\mathcal{P}_{13}C_{6,13}^{(3,3)}\mathcal{P}_{13}$\\[0.5cm]

\multirow{7}{*}{$6_{3,\text{I}}^{\nu}$} & \multirow{1}{*}{$\mathcal{P}_{23}A_{12} + \mathcal{P}_{12}A_{15}\mathcal{P}_{23}$} 
&  $\mathcal{P}_{23}C_{14}^{(3,3)}\mathcal{P}_{23}$\\
& \multirow{1}{*}{$\mathcal{P}_{321}A_{12} + \mathcal{P}_{12}A_{15}\mathcal{P}_{23}$} 
&  $\mathcal{P}_{12}C_{7,10,15,16}^{(3,3)}\mathcal{P}_{23}$~;~$\mathcal{P}_{12}C_{10,15}^{(3,3)}\mathcal{P}_{13}$~;~$\mathcal{P}_{321}C_{11}^{(3,3)}\mathcal{P}_{13}$\\ 
& \multirow{1}{*}{$\mathcal{P}_{23}A_{12}\mathcal{P}_{12} + \mathcal{P}_{12}A_{15}\mathcal{P}_{23}$} 
&  $\mathcal{P}_{23}C_{4,13}^{(3,3)}\mathcal{P}_{321}$\\
& \multirow{1}{*}{$\mathcal{P}_{321}A_{12}\mathcal{P}_{12} + \mathcal{P}_{12}A_{15}\mathcal{P}_{23}$} 
&  $\mathcal{P}_{12}C_{10,12}^{(3,3)}\mathcal{P}_{12}$\\
& \multirow{1}{*}{$\mathcal{P}_{23}A_{12} + \mathcal{P}_{123}A_{15}$} 
&  $C_{11}^{(3,3)}$~;~$\mathcal{P}_{123}C_{12,13}^{(3,3)}$\\
& \multirow{1}{*}{$\mathcal{P}_{23}A_{12}\mathcal{P}_{12} + \mathcal{P}_{123}A_{15}$} 
&  $\mathcal{P}_{23}C_{12}^{(3,3)}\mathcal{P}_{321}$~;~$\mathcal{P}_{123}C_{12}^{(3,3)}\mathcal{P}_{123}$\\
& \multirow{1}{*}{$\mathcal{P}_{321}A_{12} + \mathcal{P}_{123}A_{15}$} 
&  $\mathcal{P}_{321}C_{13}^{(3,3)}\mathcal{P}_{13}$~;~$\mathcal{P}_{12}C_{16}^{(3,3)}$\\[0.5cm]

\multirow{6}{*}{$6_{3,\text{II}}^{\nu}$} &\multirow{1}{*}{$\mathcal{P}_{13}A_{12} + \mathcal{P}_{23}A_{15}$} 
&  $\mathcal{P}_{13}C_{4,13}^{(3,3)}\mathcal{P}_{13}$~;~$C_{16}^{(3,3)}$\\
&\multirow{1}{*}{$\mathcal{P}_{123}A_{12} + \mathcal{P}_{23}A_{15}$} 
&  $\mathcal{P}_{12}C_{11}^{(3,3)}$~;~$\mathcal{P}_{23}C_{12-15,17}^{(3,3)}$\\ 
&\multirow{1}{*}{$\mathcal{P}_{123}A_{12}\mathcal{P}_{12} + \mathcal{P}_{23}A_{15}$} 
&  $\mathcal{P}_{23}C_{12}^{(3,3)}\mathcal{P}_{123}$~;~$\mathcal{P}_{123}C_{12}^{(3,3)}\mathcal{P}_{321}$~;~$\mathcal{P}_{23}C_{15}^{(3,3)}\mathcal{P}_{321}$\\
&\multirow{1}{*}{$\mathcal{P}_{13}A_{12} + A_{15}\mathcal{P}_{23}$} 
&  $C_{7,10,15,16}^{(3,3)}\mathcal{P}_{23}$~;~$C_{10,15}^{(3,3)}\mathcal{P}_{13}$~;~$\mathcal{P}_{13}C_{11}^{(3,3)}\mathcal{P}_{13}$\\
&\multirow{1}{*}{$\mathcal{P}_{13}A_{12}\mathcal{P}_{12} + A_{15}\mathcal{P}_{23}$} 
&  $C_{10,12}^{(3,3)}\mathcal{P}_{12}$\\
&\multirow{1}{*}{$\mathcal{P}_{123}A_{12}\mathcal{P}_{12} + A_{15}\mathcal{P}_{23}$} 
&  $\mathcal{P}_{123}C_{13}^{(3,3)}\mathcal{P}_{321}$\\[0.5cm]

\multirow{7}{*}{$6_{7,\text{I}}^{\nu}$} &\multirow{1}{*}{$\left(\mathcal{P}_{12}A_{15} + \mathcal{P}_{23}A_{12}\right)\mathcal{P}_{23}$} 
&  $\mathcal{P}_{23}C_{4,13}^{(3,3)}\mathcal{P}_{321}$\\
&\multirow{1}{*}{$\mathcal{P}_{12}A_{15}\mathcal{P}_{23} + \mathcal{P}_{321}A_{12}\mathcal{P}_{123}$} 
&  $\mathcal{P}_{12}C_{7,14}^{(3,3)}\mathcal{P}_{13}$~;~$\mathcal{P}_{12}C_{10,12}^{(3,3)}\mathcal{P}_{12}$~;~$\mathcal{P}_{321}C_{12}^{(3,3)}\mathcal{P}_{13}$\\ 
&\multirow{1}{*}{$\left(\mathcal{P}_{12}A_{15} + \mathcal{P}_{321}A_{12}\right)\mathcal{P}_{23}$} 
&  $\mathcal{P}_{13}C_{11}^{(3,3)}\mathcal{P}_{23}$~;~$\mathcal{P}_{12}C_{12,14-17}^{(3,3)}\mathcal{P}_{23}$\\
&\multirow{1}{*}{$\mathcal{P}_{123}A_{15} + \mathcal{P}_{23}A_{12}\mathcal{P}_{23}$} 
&  $\mathcal{P}_{23}C_{11}^{(3,3)}\mathcal{P}_{321}$~;~$\mathcal{P}_{12}C_{13}^{(3,3)}\mathcal{P}_{23}$\\
&\multirow{1}{*}{$\mathcal{P}_{123}A_{15}  + \mathcal{P}_{23}A_{12}\mathcal{P}_{123}$} 
&  $\mathcal{P}_{123}C_{12}^{(3,3)}\mathcal{P}_{123}$\\
&\multirow{1}{*}{$\mathcal{P}_{123}A_{15}  + \mathcal{P}_{321}A_{12}\mathcal{P}_{123}$} 
&  $\mathcal{P}_{321}C_{13}^{(3,3)}\mathcal{P}_{13}$\\
&\multirow{1}{*}{$\mathcal{P}_{123}A_{15}  + \mathcal{P}_{321}A_{12}\mathcal{P}_{23}$} 
&  $\mathcal{P}_{12}C_{16}^{(3,3)}$\\[0.5cm]

\multirow{7}{*}{$6_{7,\text{II}}^{\nu}$} &\multirow{1}{*}{$\mathcal{P}_{13}A_{15} + A_{12}\mathcal{P}_{23}$} 
&  $\mathcal{P}_{13}C_{7,10,15,16}^{(3,3)}$~;~$\mathcal{P}_{13}C_{10,15}^{(3,3)}\mathcal{P}_{321}$~;~$C_{11}^{(3,3)}\mathcal{P}_{321}$\\
&\multirow{1}{*}{$\mathcal{P}_{13}A_{15} + A_{12}\mathcal{P}_{123}$} 
&  $\mathcal{P}_{13}C_{10,12}^{(3,3)}\mathcal{P}_{123}$\\
&\multirow{1}{*}{$\mathcal{P}_{13}A_{15} + \mathcal{P}_{12}A_{12}\mathcal{P}_{123}$} 
&  $\mathcal{P}_{12}C_{13}^{(3,3)}\mathcal{P}_{13}$\\
&\multirow{1}{*}{$\mathcal{P}_{13}A_{15} + \mathcal{P}_{12}A_{12}\mathcal{P}_{23}$} 
&  $\mathcal{P}_{12}C_{14}^{(3,3)}$\\
&\multirow{1}{*}{$\left(\mathcal{P}_{321}A_{15} + A_{12}\right)\mathcal{P}_{23}$} 
&  $C_{4,13}^{(3,3)}\mathcal{P}_{321}$~;~$\mathcal{P}_{13}C_{16}^{(3,3)}\mathcal{P}_{23}$\\
&\multirow{1}{*}{$\mathcal{P}_{321}A_{15}\mathcal{P}_{23} + \mathcal{P}_{12}A_{12}\mathcal{P}_{123}$} 
&  $\mathcal{P}_{123}C_{11}^{(3,3)}\mathcal{P}_{23}$~;~$\mathcal{P}_{321}C_{12,13}^{(3,3)}\mathcal{P}_{23}$\\
&\multirow{1}{*}{$\left(\mathcal{P}_{321}A_{15} + \mathcal{P}_{12}A_{12}\right)\mathcal{P}_{23}$} 
&  $\mathcal{P}_{12}C_{12}^{(3,3)}\mathcal{P}_{13}$~;~$\mathcal{P}_{321}C_{12}^{(3,3)}\mathcal{P}_{12}$\\[0.5cm]

\multirow{3}{*}{$6_{9,\text{I}}^{\nu}$} &\multirow{1}{*}{$\mathcal{P}_{321}A_{15} + \mathcal{P}_{13}A_{12}$} 
&  $C_{3,14}^{(3,3)}$~;~$C_{6,13}^{(3,3)}\mathcal{P}_{13}$~;~$\mathcal{P}_{321}C_{11,13}^{(3,3)}$\\ 
&\multirow{1}{*}{$\mathcal{P}_{13}\left(A_{15}\mathcal{P}_{23} + A_{12}\right)$} 
&  $C_{11,12}^{(3,3)}\mathcal{P}_{13}$~;~$\mathcal{P}_{13}C_{12-14,17}^{(3,3)}\mathcal{P}_{23}$~;~$\mathcal{P}_{13}C_{15}^{(3,3)}\mathcal{P}_{13}$\\ 
&\multirow{1}{*}{$\mathcal{P}_{13}\left(A_{15}\mathcal{P}_{23} + A_{12}\mathcal{P}_{12}\right)$} 
&  $C_{9}^{(3,3)}\mathcal{P}_{123}$\\[0.1cm]
\hline \hline
\end{tabular}
\caption{Building matrices and corresponding chains of Ref.~\cite{Serodio:2013gka} which contain the neutrino texture decompositions identified in Table~\ref{leptontexturedecompositions}.}
\label{neutrinodecomp}
\end{table*}

Based on Table~\ref{leptontexturedecompositions}, we generate all possible pairs of $(4_3^{\ell},6_{1,3,7,9}^{\nu})$ and $(5_1^{\ell},5_{8}^{\nu})$ decompositions. Notice that one must consider both Yukawa matrix orderings, since swapping $\Ynu_1$ and $\Ynu_2$, while maintaining the charged-lepton ordering of $\Yl_1$ and $\Yl_2$, will affect the corresponding implementation. In what follows, we choose to fix the charged-lepton Yukawa ordering and consider two possible orderings for the neutrino Yukawa textures.

At this point, it remains to determine which of the decomposition pairs are viable, i.e those in which all four Yukawa matrices can be simultaneously realized by the same symmetry. Obviously, at least one pair of decompositions must be realizable for at least one $\Ynu_{1,2}$ ordering since, otherwise, the corresponding mass matrix pair would have been already excluded by the canonical method. Given that we have reduced the number of possible combinations to only 16, including the two possible orderings (12 and 4 for the $4_3^\ell$ and $5_1^\ell$ textures, respectively), the SNF method becomes now more practical to identify the minimal Abelian symmetry realization. 

\subsection{Minimal Abelian symmetries}
\label{minabrealization}

We apply the SNF method to all the $(4_3^{\ell},6_{1,3,7,9}^{\nu})$ and $(5_1^{\ell},5_{8}^{\nu})$ decomposition pairs in Table~\ref{leptontexturedecompositions}, considering both orderings of $\Y_{1,2}^\nu$. We find that $\textbf{D}_{\text{SNF}}$ can only take one of the following two forms:

\setcounter{MaxMatrixCols}{20}
\begin{equation}
\textbf{D}_{\text{SNF}} = \begin{pmatrix} 
1 & 0 & 0 & 0 & 0 & 0 & 0 & 0 & 0 & 0 & 0 \\
0 & 1 & 0 & 0 & 0 & 0 & 0 & 0 & 0 & 0 & 0 \\
0 & 0 & 1 & 0 & 0 & 0 & 0 & 0 & 0 & 0 & 0 \\
0 & 0 & 0 & 1 & 0 & 0 & 0 & 0 & 0 & 0 & 0 \\
0 & 0 & 0 & 0 & 1 & 0 & 0 & 0 & 0 & 0 & 0 \\
0 & 0 & 0 & 0 & 0 & 1 & 0 & 0 & 0 & 0 & 0 \\
0 & 0 & 0 & 0 & 0 & 0 & 1 & 0 & 0 & 0 & 0 \\
0 & 0 & 0 & 0 & 0 & 0 & 0 & g & 0 & 0 & 0 \\
\end{pmatrix}\,,\,g=0,2\,,
\end{equation}
where, according to the discussion around Eq.~\eqref{DSNFds}, the two cases correspond to the Abelian symmetry groups $\textbf{G}$
\begin{equation}
\begin{aligned}
g = 0&:  \textbf{G} = [\text{U}(1)]^4, \\
g = 2&:  \textbf{G} = \mathbb{Z}_2 \times [\text{U}(1)]^3\,.
\end{aligned}
\end{equation}
This result can be interpreted as follows. Irrespective of the type of Yukawa textures, the Lagrangian is invariant under the global continuous symmetries ${\rm U(1)_Y}$ and ${\rm U(1)_L}$, where Y is the SM hypercharge and L is the lepton number. These global symmetries simply appear because the 2HDM is invariant under the global version of the ${\rm U(1)_Y}$ gauge symmetry, while $\text{U}(1)_L$ ensures lepton number conservation, which is required when Dirac neutrino Yukawa interactions are considered. Notice that neither of these symmetries imposes texture zeros in the mass matrices. As such, the flavor symmetry identified by the SNF method is, in fact, 
\begin{equation}
\begin{aligned}
g = 0&: \textbf{G}_F = \textbf{G}/\left[{\rm U(1)_Y}\times{\rm U(1)_L}\right]=[\text{U}(1)]^2,\\
g = 2&:\textbf{G}_F = \mathbb{Z}_2 \times{\rm U(1)}.
\end{aligned}
\end{equation}

For all Yukawa matrices in Table~\ref{leptontexturedecompositions}, couplings with the field $\nu_{R1}$ are forbidden, leading to the existence of a massless Weyl neutrino. The absence of phase-sensitive terms involving ${\nu_R}_1$ implies a continuous U(1) symmetry associated to that field, which is preserved from the original rephasing symmetry $[\text{U}(1)]^{n_F}$ of the Lagrangian without phase-sensitive terms. Therefore, the actual SNF symmetry which realizes the non-zero elements of the decomposition pairs is
\begin{equation}
\begin{aligned}
g = 0&:\textbf{G}_F = \text{U}(1)\times\text{U}(1)_{{\nu_{Ri}}}\,,\\
g = 2&:\textbf{G}_F = \mathbb{Z}_2\times\text{U}(1)_{{\nu_{Ri}}},
\label{G3}
\end{aligned}
\end{equation}
where $\text{U}(1)_{{\nu_{Ri}}}$ follows from the invariance under ${\nu_R}_i \rightarrow e^{i \alpha} {\nu_R}_i$, with all remaining fields transforming trivially. Notice that we take ${\nu_R}_i$ with $i=1,2,3$ since row permutations of $\Mnu$ generate an equivalent pair ($\Ml,\Mnu$), as seen from Eq.~\eqref{weakbasis3}. 

After applying the SNF we have concluded that the non-zero elements of all Yukawa matrices in each decomposition can be implemented (i.e. are compatible) by a U(1) or a $\mathbb{Z}_2$ symmetry. Moreover, a $\text{U}(1)_{{\nu_{Ri}}}$ is present due to the absence of $\nu_{Ri}$ interactions. It now remains to check whether the texture zeros can be implemented by the same symmetries. 

Using the canonical method, we determine the decompositions and orderings of the texture pairs $(4_3^{\ell},6_{1,3,7,9}^{\nu})$ and $(5_1^{\ell},5_{8}^{\nu})$ that are exactly realized by the symmetries~\eqref{G3} (see Table~\ref{decompositions}). We find that all 5 mass matrix pairs have one decomposition which can be implemented imposing a $\textbf{G}_F = \text{U}(1)\times\text{U}(1)_{{\nu_{Ri}}}$ symmetry, being $\text{U}(1)$ alone able to reproduce the entire corresponding texture structure. This means that $\text{U}(1)_{{\nu_{Ri}}}$ is, in fact, an accidental symmetry arising from the particular form of the $\Mnu$ texture. The viable ordering in each case corresponds to that indicated in Table~\ref{leptontexturedecompositions}. The three decompositions which cannot be reproduced (and are omitted in Table~\ref{decompositions}) exhibit, for one ordering each, a rephasing symmetry $\textbf{G}_F = \text{U}(1)\times\text{U}(1)_{{\nu_{Ri}}}$ that is unable to enforce one of their respective zero textures. On the other hand, the $\textbf{G}_F = \mathbb{Z}_2\times\text{U}(1)_{{\nu_{Ri}}}$ cases correspond to the alternative Yukawa orderings of all 8 decompositions, with the $\mathbb{Z}_2$ rephasing symmetry simply enforcing the corresponding ordering, while it fails in imposing any texture zero on the mass matrices.

From now on, we will only refer to the U(1) symmetry, corresponding to $g=0$ in Eq.~\eqref{G3}. The U(1) implementation of the realizable $(\Ml,\Mnu)$ pairs is summarized in Table~\ref{realization}, were the continuous phases and discrete charges are presented. Out of the 28 maximally-restrictive $(\Ml,\Mnu)$ texture pairs compatible with data given in Table~\ref{texturepairs}, only 5 can be realized through Abelian symmetries in a 2HDM, each admitting a single decomposition into Yukawa matrices, realizable by a U(1) flavor symmetry. As it turns out, in all cases, the minimal set of discrete charges corresponds to a $\mathbb{Z}_5$ symmetry.

\begin{table}[!htb]
\centering
\begin{tabular}{c@{\hskip 0.3in}c}
\hline \hline\\[-0.2cm]
$\Ml$ & $\Mnu$ \\[0.1cm]\hline\\[-0.2cm]
\multirow{4}{*}{
$4_{3,\text{I}}^{\ell}:\, \begin{pmatrix} 0&0&a_1\\0&a_2&b_1\\a_3&b_2 &0 \end{pmatrix}$}
& $6_{1,\text{I}}^{\nu}:\, \begin{pmatrix} 0&0&0\\0&0&x_2\\0&x_1&y e^{i\alpha} \end{pmatrix}$  \vspace{0.2cm}\\
& $6_{3,\text{I}}^{\nu}:\,\begin{pmatrix} 0&0&x_1\\0&0&y e^{i\alpha}\\0&x_2&0 \end{pmatrix}$  \vspace{0.2cm}\\
& $6_{7,\text{I}}^{\nu}:\,\begin{pmatrix} 0&0&x_2\\0&y e^{i\alpha}&0\\0&x_1&0 \end{pmatrix}$ \vspace{0.2cm}\\
& $6_{9,\text{I}}^{\nu}:\,\begin{pmatrix} 0&x_1&y e^{i\alpha}\\0&0&x_2\\0&0&0\end{pmatrix}$ \vspace{0.2cm}\\[0.1cm]\hline\\[-0.2cm]
$5_{1,\text{II}}^{\ell}:\, \begin{pmatrix} 0&0&a_1\\0&b_1&0\\ a_2&0&b_2 \end{pmatrix}$ & $5_{8,\text{I}}^{\nu}:\,\begin{pmatrix} 0&y_1&x_1\\ 0&0&y_2 \\ 0&x_2e^{i\alpha}&0 \end{pmatrix}$ \\ \\[-0.2cm]
\hline \hline
\end{tabular}
\caption{Parameter conventions for the maximally restrictive pairs $(\Ml,\Mnu)$ which are simultaneously compatible with charged-lepton and neutrino data and realizable in the 2HDM context through a U(1) flavor symmetry. The decompositions of $(\Ml,\Mnu)$ into $\Y_a^{\ell,\nu}$ and corresponding orderings are those of Table~\ref{leptontexturedecompositions}. We have took advantage of field-rephasing freedom to eliminate all unphysical phases, placing the only irremovable phase $\alpha$ in $\Mnu$. }
\label{decompositions}
\end{table}

\begin{table}[!htb]
\centering
\begin{tabular}{c@{\hskip 0.3in}c@{\hskip 0.1in}c@{\hskip 0.1in}c}
\hline \hline\\[-0.2cm]
$(\Ml,\Mnu)$  &{\rm U(1):} $(\alpha_1,\alpha_2,\alpha_3)$ & $(\beta_1,\beta_2,\beta_3)$ & $(\gamma_1,\gamma_2,\gamma_3)$\\[0.2cm]
&$\,\,\,\,\,\mathbb{Z}_5$: $(\alpha_1^\prime,\alpha_2^\prime,\alpha_3^\prime)$ & $(\beta_1^\prime,\beta_2^\prime,\beta_3^\prime)$ & $(\gamma_1^\prime,\gamma_2^\prime,\gamma_3^\prime)$ \\[0.1cm]\hline\\[-0.2cm]
$(4_{3,{\rm I}}^{\ell},6_{1,{\rm I}}^{\nu})$   & $(0,\theta,2\theta)$ & $(2\theta,\theta,0)$ & $(\eta,3\theta,2\theta)$ \\[0.2cm]
& $(0,1,2)$ & $(2,1,0)$ & $(4,3,2)$ \\[0.2cm]
$(4_{3,{\rm I}}^{\ell},6_{3,{\rm I}}^{\nu})$   & $(0,\theta,2\theta)$ & $(2\theta,\theta,0)$ & $(\eta,3\theta,\theta)$ \\[0.2cm]
& $(0,1,2)$ & $(2,1,0)$ & $(4,3,1)$ \\[0.2cm]
$(4_{3,{\rm I}}^{\ell},6_{7,{\rm I}}^{\nu})$   & $(0,\theta,2\theta)$ & $(2\theta,\theta,0)$ & $(\eta,2\theta,0)$ \\[0.2cm]
& $(0,1,2)$ & $(2,1,0)$ & $(4,2,0)$ \\[0.2cm]
$(4_{3,{\rm I}}^{\ell},6_{9,{\rm I}}^{\nu})$   & $(0,\theta,2\theta)$ & $(2\theta,\theta,0)$ & $(\eta,0,\theta)$ \\[0.2cm]
& $(0,1,2)$ & $(2,1,0)$ & $(4,0,1)$ \\[0.2cm]
$(5_{1,{\rm II}}^{\ell},5_{8,{\rm I}}^{\nu})$   & $(0,-\theta,\theta)$ & $(\theta,-2\theta,0)$ & $(\zeta,\theta,0)$ \\[0.2cm]
& $(0,4,1)$ & $(1,3,0)$ & $(3,1,0)$ \\[0.2cm]
\hline \hline
\end{tabular}
\caption{Implementation of the texture pairs from Table~\ref{decompositions} in a 2HDM. With $\theta \in \mathbb{R}$, $\eta \neq \{0,~\theta,~2\theta,~3\theta\}$ and $\zeta \neq \{-\theta,~0,~\theta,~2\theta\}$. The charges $(\alpha_i,\beta_i,\gamma_i)$ correspond to Eq.~\eqref{chargphases} with $(\theta_1,\theta_2)=(0,\theta)$, while the discrete charges $(\alpha_i',\beta_i',\gamma_i')$ are those of Eq.~\eqref{chargphases2} with $(\theta_1',\theta_2')=(0,1)$ and $\varphi=2\pi/5$.}
\label{realization}
\end{table}

\section{Phenomenology}
\label{Phenom}

In the previous section we have selected among all maximally restricted texture-zero pairs $(\Ml,\Mnu)$ those which are simultaneously compatible with neutrino data and realizable in the context of the 2HDM with a U(1) flavor symmetry. In the following, we will focus on their phenomenology in what concerns leptonic CP violation, rare lepton decays and lepton universality. More specifically, we will establish the relation between the only complex phase which remains in the Yukawa sector of the theory with that appearing in the lepton mixing matrix $\U$ of Eq.~\eqref{Uparam}, i.e. the Dirac phase $\delta$. We also apply the constraints coming from universality tests in $\tau$ decays and from rare (two- and three-body) lepton decay searches. 

\subsection{Leptonic CP violation}
\label{LCPV}

Although, in general, all elements of the Yukawa matrices $\Y_a^{\ell,\nu}$ in Eq.~\eqref{lagrangian} are complex, some phases have no physical significance and, thus, can be removed by rephasing the fermion fields as $\psi_j \rightarrow e^{i\varphi_j} \psi_j$. It is straightforward to show that, for the maximally-restricted pairs of matrices considered in this work, all elements of $\Ml$ and $\Mnu$ (or $\Y_{1,2}^{\ell,\nu}$) can be made real and positive, except one. This is due to the fact that a single unremovable phase $\alpha$, which will be necessarily correlated with the Dirac CP-violating phase $\delta$ in Eq.~\eqref{Uparam}, remains after exhausting the rephasing freedom. Our convention for the position of $\alpha$ is given in Table~\ref{decompositions}, where all parameters $a_i$, $b_i$, $x_i$ and $y_i$ are real and positive.
\begin{figure*}[!ht]
\centering
\begin{tabular}{cc}
\includegraphics[width=0.35\textwidth]{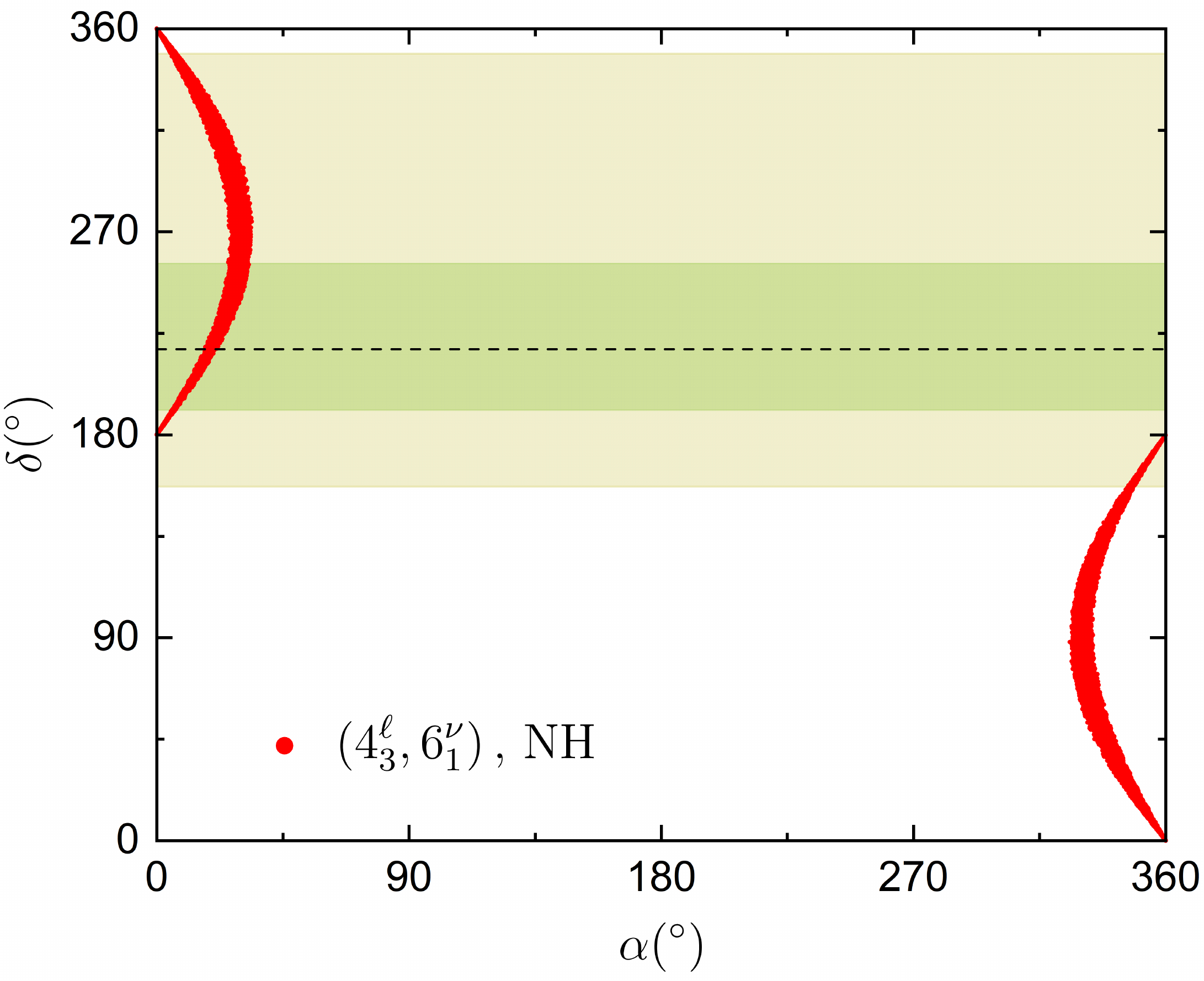} &
\includegraphics[width=0.35\textwidth]{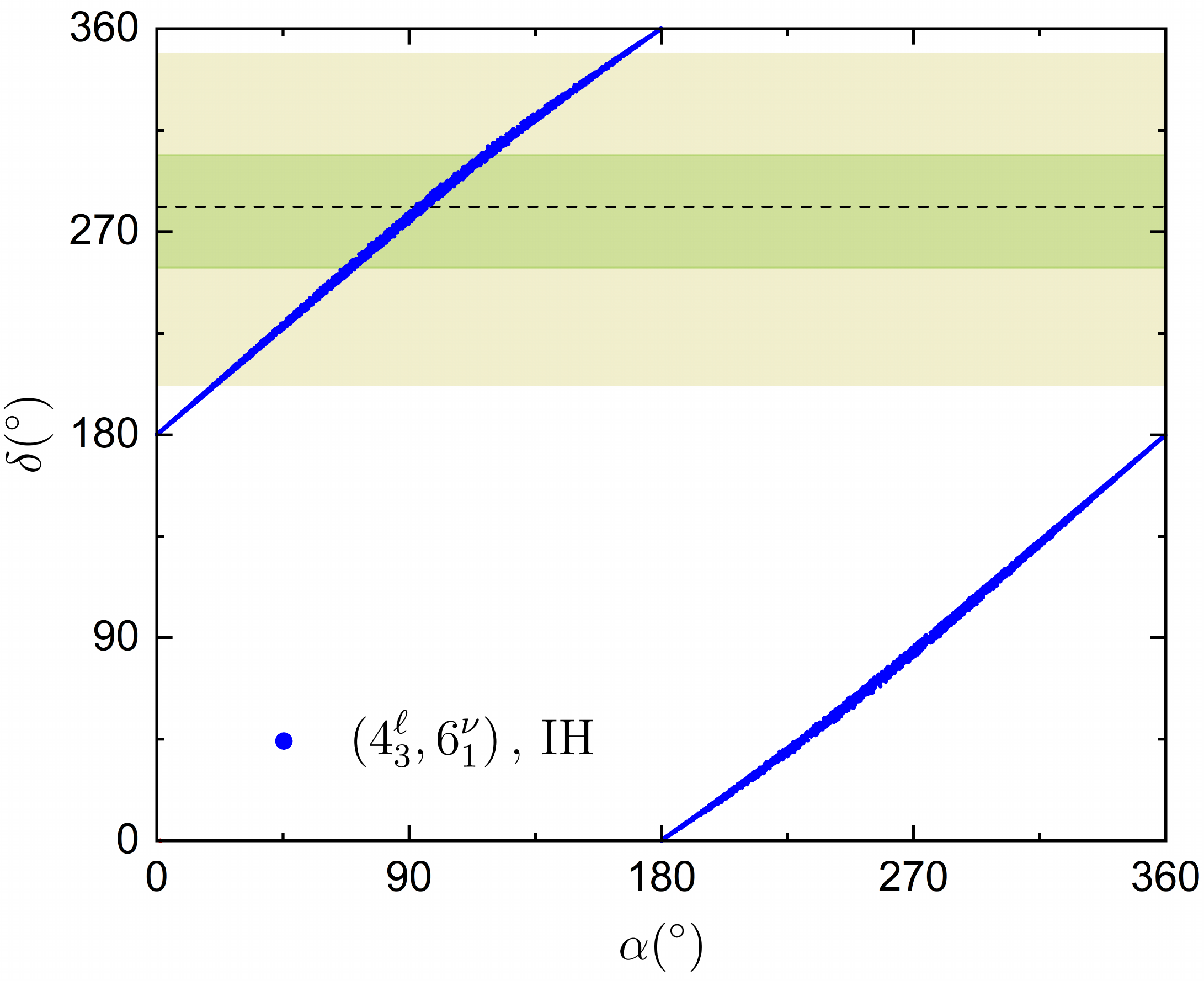} \\\includegraphics[width=0.35\textwidth]{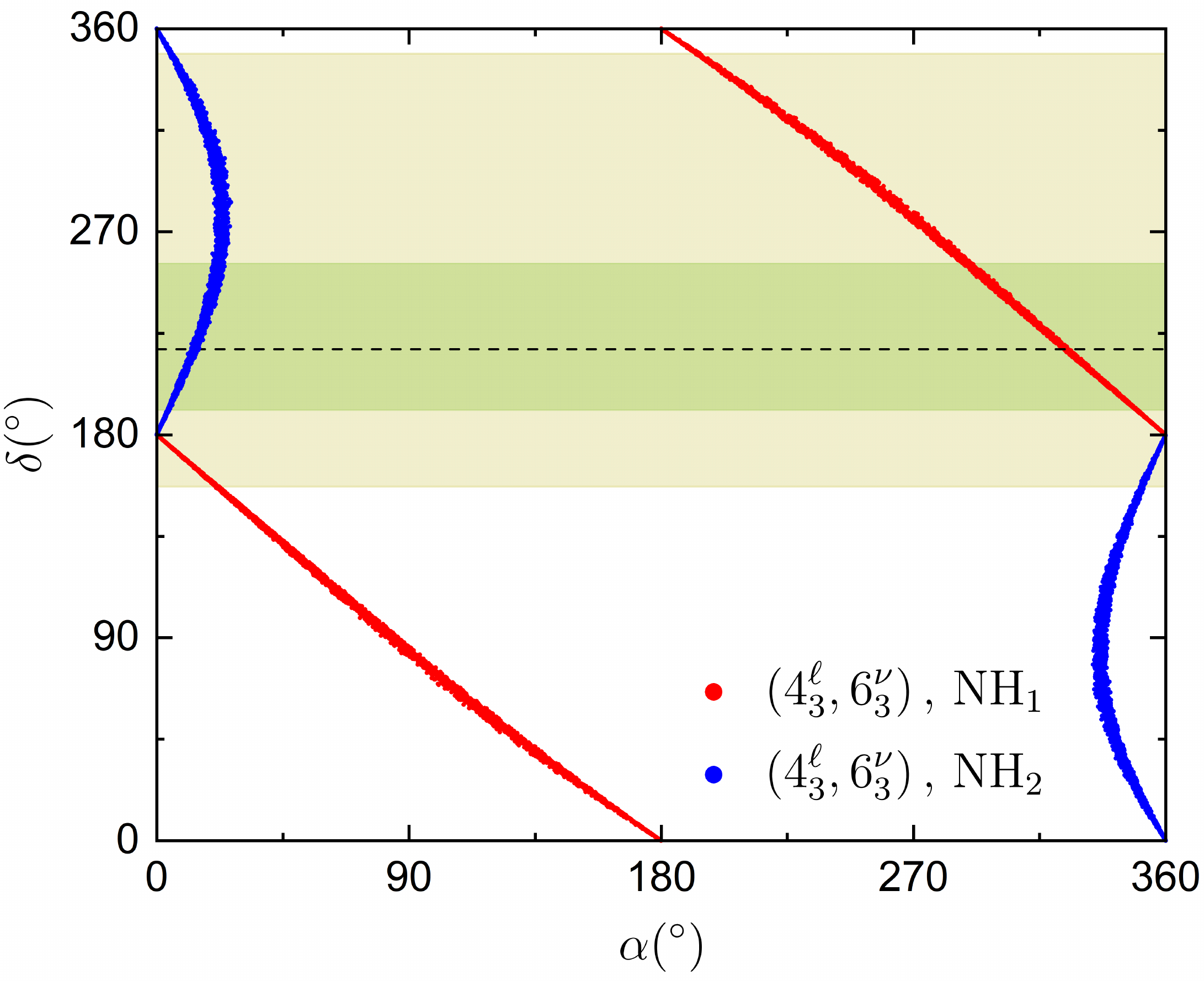} &
\includegraphics[width=0.35\textwidth]{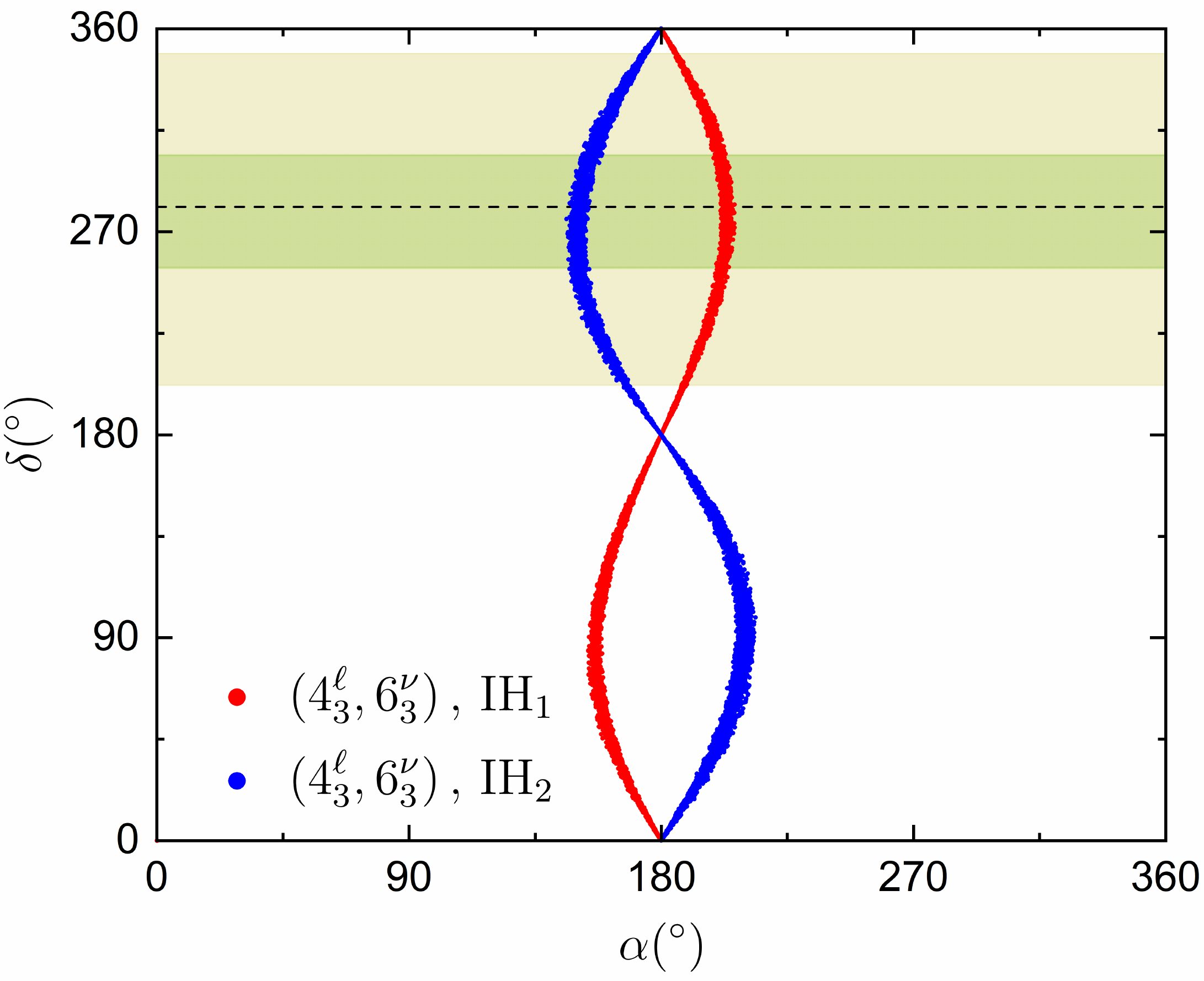} \\\includegraphics[width=0.35\textwidth]{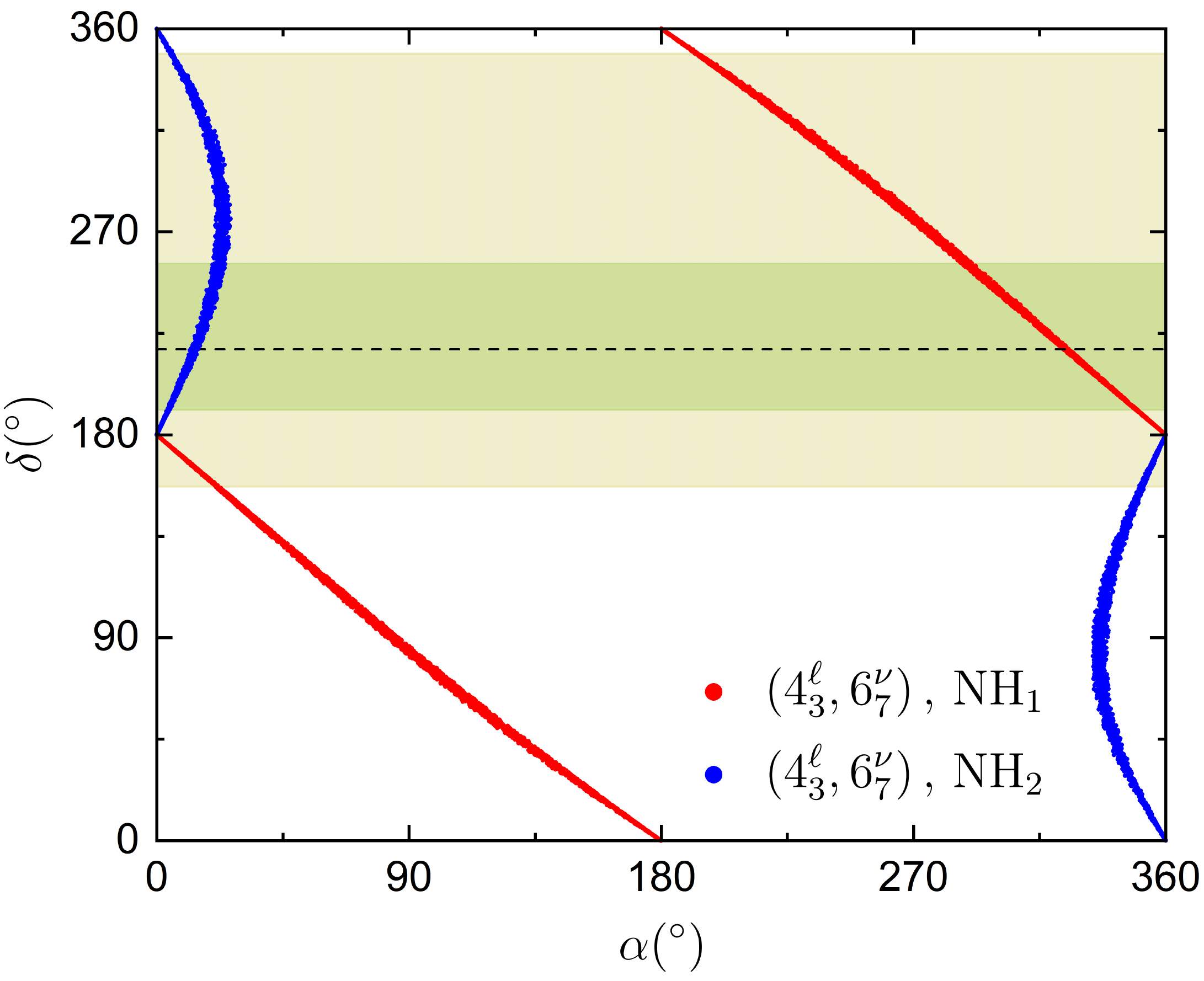} &
\includegraphics[width=0.35\textwidth]{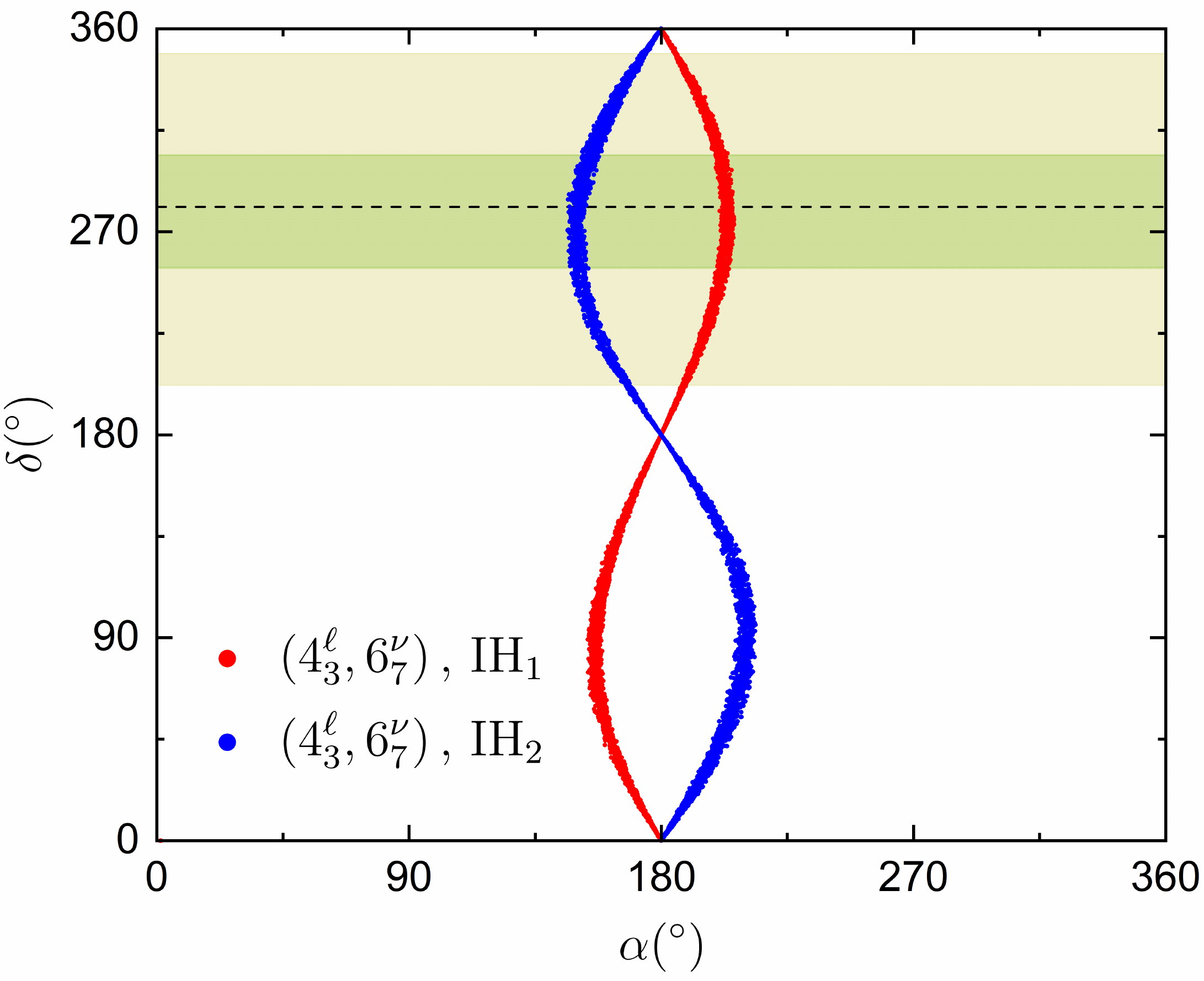} \\\includegraphics[width=0.35\textwidth]{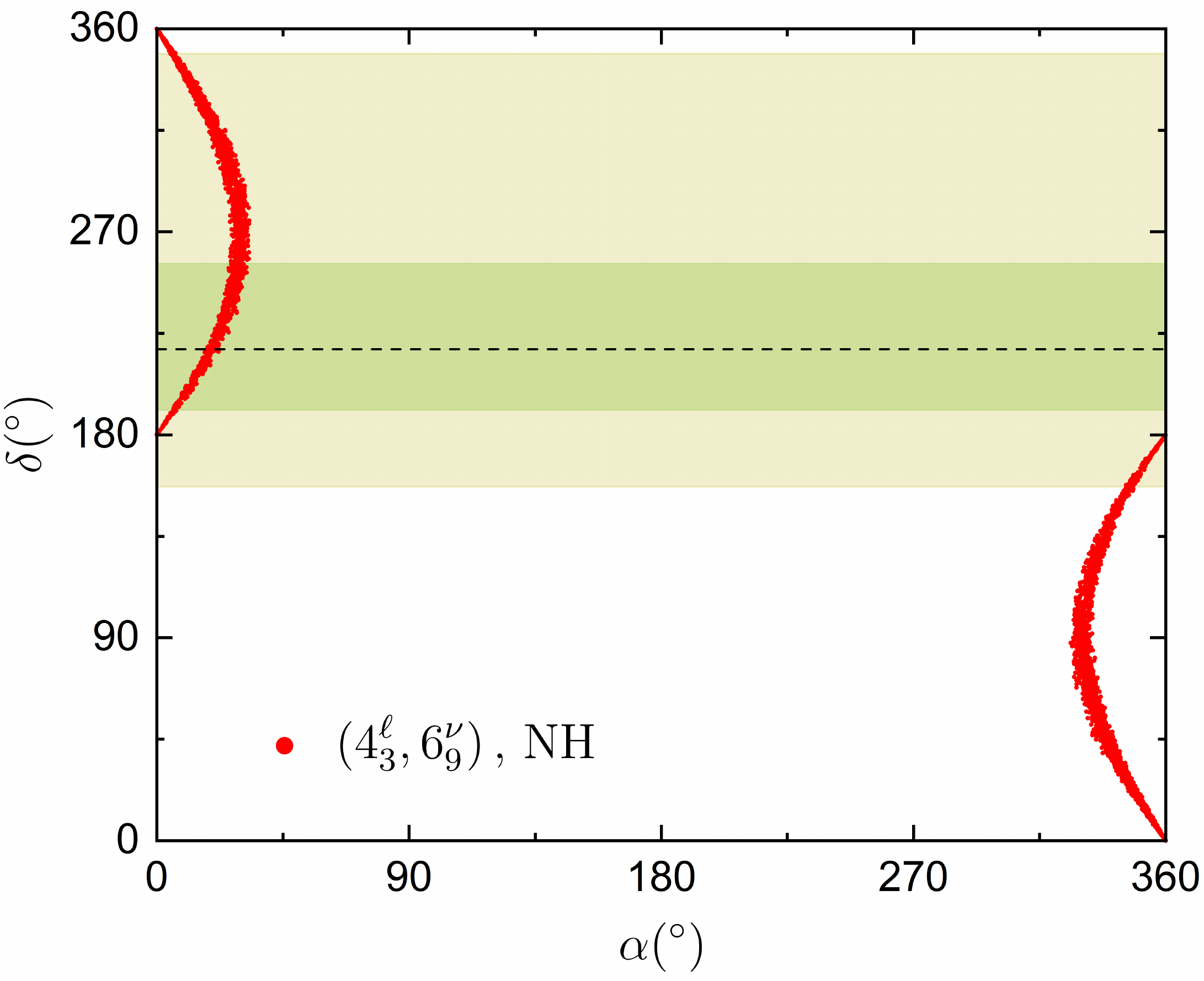} &
\includegraphics[width=0.35\textwidth]{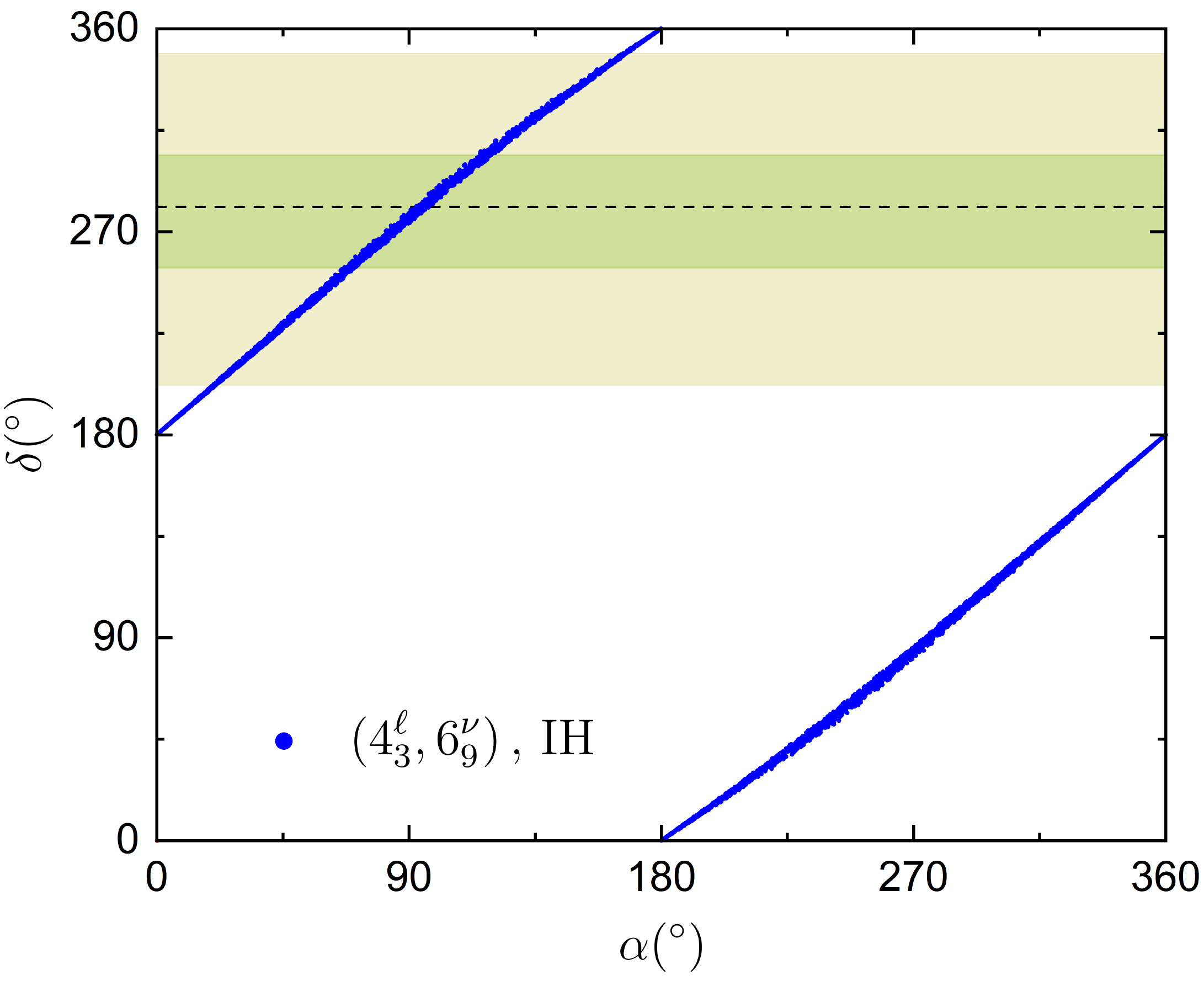} 
\end{tabular}
\caption{\label{fig1} Correlation between the phase $\alpha$ appearing in the Dirac neutrino mass matrix $\Mnu$ and the Dirac CP-violating phase $\delta$ of the lepton mixing matrix $\U$ for the $(4_3^\ell,6_i^\nu)$ texture pairs shown in Table~\ref{decompositions}. NH$_{1,2}$ and IH$_{1,2}$ solutions are classified according to which neutrino is considered to be the lightest (see Appendix~\ref{appdxB} for more details). In all points, the mixing angles $\theta_{ij}$ lie within the $3\sigma$ ranges given in Table~\ref{datatable}.}
\end{figure*}
\begin{figure*}[t]
\centering
\begin{tabular}{cc}
\includegraphics[width=0.4\textwidth]{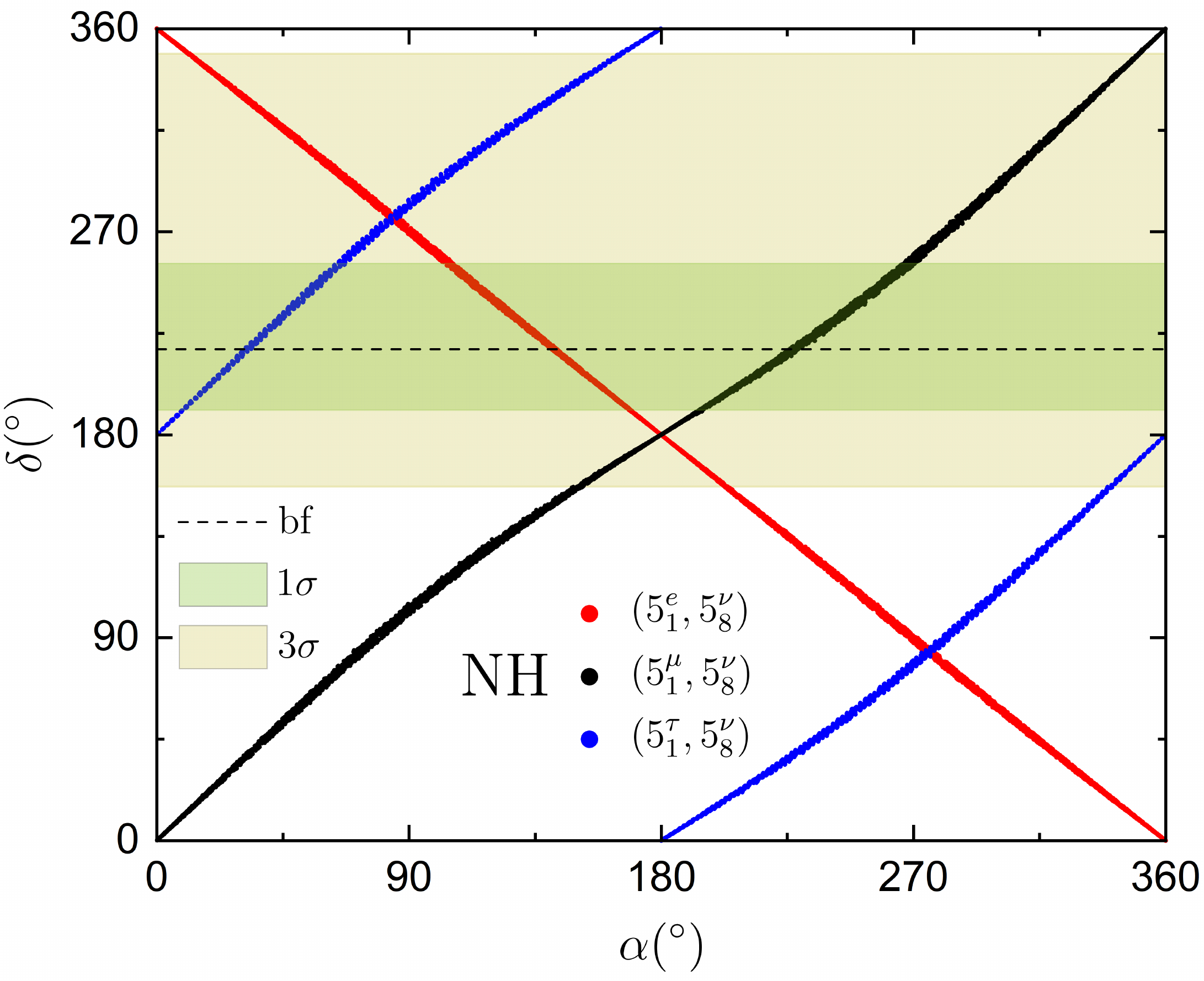} &
\includegraphics[width=0.4\textwidth]{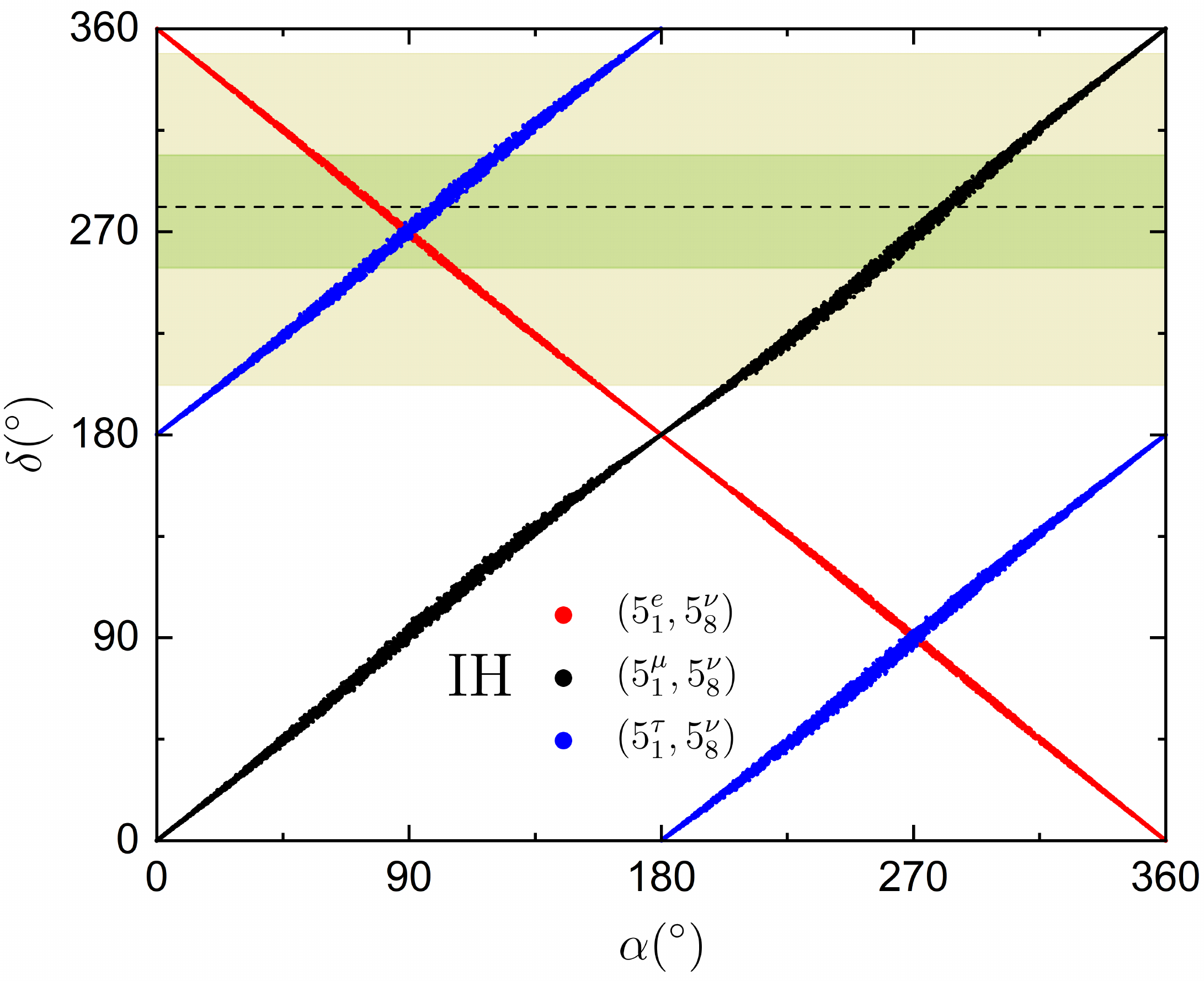} \\
\end{tabular}
\caption{\label{fig2}Correlation between the phase $\alpha$ appearing in $\Mnu$ and the Dirac CP-violating phase $\delta$ of the lepton mixing matrix $\U$ for the $(5_1^{e,\mu,\tau},5_8^\nu)$ texture pairs. The results shown in the left (right) panel correspond to a NH (IH) neutrino mass spectrum. In all points, the mixing angles $\theta_{ij}$ lie within the $3\sigma$ ranges given in Table~\ref{datatable}.}
\end{figure*}

Given that for all $(\Ml,\Mnu)$ pairs there are nine real parameters in total, four of them remain undefined after ensuring a mass spectrum that reproduces the observed charged-lepton masses $m_{e,\mu,\tau}$ and neutrino mass-squared differences $\Delta m_{21,31}^2$ (since one neutrino is massless, the two Dirac neutrino masses can be written in terms of $\Delta m_{21,31}^2$). In both $4_{3,\text{I}}^{\ell}$ and $6_{i,\text{I}}^{\nu}$ two free parameters remain, while in $5_{1,\text{II}}^{\ell}$ and $5_{8,\text{I}}^{\nu}$ we are left with one and three free parameters, respectively. For the texture $5_{1,\text{II}}^{\ell}$ ($4_{3,\text{I}}^{\ell}$) we choose to write $a_1$ and $b_{1,2}$ ($a_{1,2}$ and $b_1$) in terms of $a_2$ ($a_3$ and $b_2$) and $m_{e,\mu,\tau}$. As for $6_{i,\text{I}}^{\nu}$ ($5_{8,\text{I}}^{\nu}$), we express $x_2$ and $y$ ($x_1$ and $y_1$) in terms of $x_1$ ($x_2$, $y_2$) and $\Delta m_{21,31}^2$, as shown in Table~\ref{definingconditions} of Appendix~\ref{appdxB}. For the texture pairs in Table~\ref{decompositions}, we perform a scan of the free parameters in their validity ranges and, for each input set, $\Hl$ and $\Hnu$ are defined as in Eq.~\eqref{Hdef}. After diagonalizing these two matrices, $\U$ gets determined by Eq.~\eqref{Udef2}, and $\theta_{ij}$ and $\delta$ appearing in Eq.~\eqref{Uparam} can be extracted. Demanding agreement with the $3\sigma$ ranges given in Table~\ref{datatable}, we plot $\delta$ as a function of $\alpha$. Our results are shown in Figs.~\ref{fig1} and \ref{fig2} for the $(4_{3}^{\ell},6_{i}^{\nu})$ and $(5_{1}^{\ell},5_{8}^{\nu})$ pairs, respectively. In all cases, both the NH and IH mass spectra are considered.

By looking at Fig.~\ref{fig1} we see that the results are similar for different pairs of $(\Ml,\Mnu)$ textures. This could suggest that those pairs are equivalent in the sense that they can be transformed into each other by permutation transformations (see discussion at the beginning of Section~\ref{Max2HDM}). However, we advocated that only non-equivalent pairs of textures were kept. The similarity between some of the results have to do with the fact that up to a very small parameter, which does not have much impact in the results for the mixing angles and phases, those pairs are indeed equivalent. Let us illustrate this with one example, namely  $(4_{3}^{\ell},6_{3}^{\nu})$ and $(4_{3}^{\ell},6_{7}^{\nu})$. By looking at Table~\ref{decompositions} we notice that
\begin{align}
6_{3}^{\nu}=\mathcal{P}_{13}\,6_{7}^{\nu}\,\mathcal{P}_{23} \;,\; 4_{3}^{\ell} \underset{a_3 \rightarrow 0}{=} \mathcal{P}_{13}\,4_{3}^{\ell}\,\mathcal{P}_{23}\,.
\end{align}
The last relation simply indicates that $4_{3}^{\ell} =\mathcal{P}_{13}\,4_{3}^{\ell}\,\mathcal{P}_{23}$ in the limit $a_3\rightarrow 0$ and, thus, the two pairs would be equivalent in this case.\footnote{Regarding the mixing angles and CP-violating phase, the results are the same as long as the permutation on the left is the same since only this transformation affects $\U$.} Obviously, $a_3=0$ would lead to a massless charged lepton, which is not acceptable. However, it turns out that $a_3$ is very small and, indeed, $4_{3}^{\ell} \simeq \mathcal{P}_{13}\,4_{3}^{\ell}\,\mathcal{P}_{23}$, leading to a similar $\U$ for both pairs. A similar reasoning can be applied to the remaining pairs for which the results look the same. We emphasize that the existence in the mass matrices of such small parameters is a consequence of the hierarchical nature of the charged-lepton masses.

We now turn our attention to the results for the $(5_{1}^{\ell},5_{8}^{\nu})$ pair shown in Fig.~\ref{fig2}. As will become clear in the next section, this case is the most interesting one and deserves a more detailed attention. The first to notice is that in the matrix $5_{1}^{\ell}$ one of the charged-lepton states is decoupled. In particular, the Hermitian matrices $\Hl=\Ml \Ml^\dag$ and $\Hl^\prime=\Ml^\dag \Ml$ and their diagonalizing unitary transformations $\U_L$ and $\U_R$, are given by 
\begin{align}
\Hl&=\begin{pmatrix}
a_1^2 &0 &a_1 b_2\\
0 &b_1^2 &0\\
a_1 b_2 &0 & a_2^2+b_2^2
\end{pmatrix}, \U_L'=\begin{pmatrix}
c_L &0 & s_L\\
0 &1 &0\\
-s_L &0 & c_L
\end{pmatrix},\label{Hl5158a}\\
\Hl'&=\begin{pmatrix}
a_1^2 &0 &a_2 b_2\\
0 &b_1^2 &0\\
a_2 b_2 &0 & a_1^2+b_2^2
\end{pmatrix}, \U_R'=\begin{pmatrix}
c_R &0 & s_R\\
0 &1 &0\\
-s_R &0 & c_R
\end{pmatrix},
\label{Hl5158}
\end{align}
where $c_{L,R} \equiv \cos\theta_{L,R}$ and $s_{L,R} \equiv \sin\theta_{L,R}$ with
\begin{align}
\tan(2\theta_L)&=\frac{2\, m_{\ell_2}m_{\ell_3}\sqrt{(a_2^2-m_{\ell_2}^2)(m_{\ell_3}^2-a_2^2)}}{a_2^2\,(m_{\ell_2}^2+m_{\ell_3}^2)-2m_{\ell_2}^2m_{\ell_3}^2}\,,\label{tetaL}\\
\tan(2\theta_R)&=\frac{2\,\sqrt{(a_2^2-m_{\ell_2}^2)(m_{\ell_3}^2-a_2^2)}}{m_{\ell_2}^2+m_{\ell_3}^2-2a_2^2}\label{tg2R}\,.
\end{align}
As explained in Appendix~\ref{appdxA}, $m_{\ell_2,\ell_3}$ correspond to the masses of the two non-decoupled charged-lepton states. Thus, depending on which charged lepton $\ell_1$ is identified as decoupled, three different cases must be considered:
\begin{align}
\ell_1 = e,\mu,\tau\quad \rightarrow\quad 5_1^\ell \equiv 5_1^{e,\mu,\tau}\,.
\label{decid}
\end{align}
This explains the notation used in the plots of Fig.~\ref{fig1}. Taking into account that the unitary matrix $\ULl$ must be such that Eq.~\eqref{Hdiag} is verified with the correct charged-lepton mass ordering, we have
\begin{align}
5_1^e: &\;\U_{L,R}^\ell=\U_{L,R}'\mathcal{P}_{12}\,, \label{Ul51e}\\
5_1^\mu:&\;\U_{L,R}^\ell=\U_{L,R}'\label{Ul51mu}\,,\\
5_1^\tau:&\;\U_{L,R}^\ell=\U_{L,R}'\mathcal{P}_{23}\label{Ul51tau}\,.
\end{align}
When $\Mnu$ is of the type $5^\nu_8$, the unitary matrix which diagonalizes $\Hnu$ as in Eq.~\eqref{Hdiag} is presented in Eqs.~\eqref{UnupNH} and~\eqref{UnupIH} of Appendix~\ref{appdxB} for NH and IH, respectively. After some algebra, we conclude that the charged-lepton rotation set by $\theta_L$ given in Eq.~\eqref{Hl5158a} is crucial since compatibility with the measured neutrino mixing angles $\theta_{ij}$ would not be possible from $\ULnu$ alone.\footnote{Notice that neglecting the mixing coming from $\ULl$ would also lead to $\delta=0$ since $\alpha$ could be removed by rephasing the LH and RH charged-lepton fields. Thus, $\U$ can be considered real in the limit $\ULl=\openone$.} In particular, if one considers $\U=\ULnu$, then the following conditions must be satisfied when $\Ml$ is of the $5_1^{e}$ type:
\begin{align}
\text{NH:}&\;\,r\,\U_{12}\U_{32}+\U_{13}\U_{33}=0\,, \label{cond1}\\
\text{IH:}&\;\,\U_{11}\U_{31}+(1+r)\U_{12}\U_{32}=0\,,\label{cond2}
\end{align}
where $r \equiv \dmsol/\dmatm \simeq 0.03$, according to the data given in Table~\ref{datatable}. The above relations imply
\begin{align}
\tan^2\theta_{23}=\frac{4s_{13}^2(1\pm rs_{12}^2)^2}{r^2\sin^2(2\theta_{12})}\simeq \frac{4s_{13}^2}{r^2\sin^2(2\theta_{12})}\simeq 110 \label{t2351e}\,,
\end{align}
where the $-$(+) sign in $\pm$ is for the NH (IH) case, and in the final estimate the values of Table~\ref{datatable} have been considered. As for the $5_1^{\mu,\tau}$, relations \eqref{cond1} and \eqref{cond2} are replaced by
\begin{align}
\text{NH:}&\,\,r\,\U_{22}\U_{32}+\U_{23}\U_{33}=0\,, \\
\text{IH:}&\,\,\U_{21}\U_{31}+(1+r)\U_{22}\U_{32}=0\,,
\end{align}
leading to $\theta_{23}\simeq \theta,\pi/2-\theta$ with
\begin{align}
\tan^2\theta \simeq  \frac{r^2}{4}\frac{s_{13}^2\sin^2(2\theta_{12})}{c_{13}^4}\simeq 2\times 10^{-3}\,,
\end{align}
which implies $\theta_{23}\simeq 0,\pi/2$. From these results we conclude that the contribution to the mixing coming from the charged-lepton sector is crucial to get compatibility with data. Moreover, it can be shown that the Jarlskog invariant $\mathcal{J}_{CP}$, which signals Dirac-type CP violation, obeys  
\begin{align}
\mathcal{J}_{CP}={\rm Im}[\U_{11}\U_{22}\U_{12}^\ast\U_{21}^\ast] \propto \sin(2\theta_L)\sin\alpha\,,
\end{align}
confirming the fact that, for CP violation to occur in the lepton sector, $\theta_L\neq n\pi/2$ and $\alpha \neq n\pi$ (n is integer) must hold. In the following, we will obtain relations among the parameters in $\Ml$ and $\Mnu$ ($a_2,\,x_2\,,y_2$ and $\alpha$) and the three mixing angles $\theta_{ij}$ and the CP phase $\delta$.  Again, we focus on the  $(5_1^e,5_8^\nu)$ pair. 

From Eqs.~\eqref{Udef2}, \eqref{Hl5158}, \eqref{Ul51e}, \eqref{UnupNH} and \eqref{UnupIH}, the lepton mixing matrix $\U$ is computed and the mixing angles and the phase $\delta$ are extracted. Notice that, for the case $(5_1^e,5_8^\nu)$, one has $\U_{1j}=({\ULnu})_{2j}$. Therefore, given the parametrization \eqref{Uparam}, $x_2$ and $y_2$ in $\Mnu$ depend only on $\theta_{12}$ and $\theta_{13}$ through the relations
\begin{align}
{\rm NH:}\;x_2^2&= \frac{\dmsol c_{12}^2(rc_{13}^2s_{12}^2+s_{13}^2)}{rs_{12}^2(r-2s_{13}^2-rc_{13}^2s_{12}^2)+s_{13}^2}\,,\label{eq60}\nonumber\\
y_2^2&=\dmatm (s_{13}^2+rc_{13}^2s_{12}^2)\,,\\
{\rm IH:}\;x_2^2&= \frac{\dmatm(1+r)(1+rs_{12}^2)s_{13}^2}{rs_{12}^2(r-2s_{13}^2-rc_{13}^2s_{12}^2)+s_{13}^2}\,,\nonumber\\
y_2^2&=\dmatm c_{13}^2(1+rs_{12}^2)\,.
\end{align}
It now remains to express $\theta_L$ (or $b_2$) appearing in Eq.~\eqref{Hl5158} and the phase $\alpha$ in terms of the measurable neutrino parameters. Including the charged-lepton corrections to the mixing we have
\begin{align}
\tan^2\theta_{23} \simeq 
\frac{[r c_\alpha t_L\sin(2\theta_{12})-2s_{13}]^2+r^2t_L^2\sin^2(2\theta_{12})s_\alpha^2}{[r c_\alpha \sin(2\theta_{12})+2t_Ls_{13}]^2+r^2\sin^2(2\theta_{12})s_\alpha^2}\,,
\end{align}
where $t_L\equiv \tan\theta_L$, $c_\alpha \equiv \cos\alpha$ and $s_\alpha \equiv \sin\alpha$. In the limit $\theta_L,\alpha \rightarrow 0$, we recover the result \eqref{t2351e}, as expected. From the above equation, $\theta_L$ can be determined by the approximate relation
\begin{align}
\tan\theta_L \simeq \cot\theta_{23}\mp \frac{r c_\alpha \sin(2\theta_{12})}{2s_{13}s_{23}^2}\,,
\label{tgtL51e}
\end{align}
where the $-$ ($+$) sign corresponds to the NH (IH) case. The above relation provides a very good approximation for the behavior of the charged-lepton mixing angle $\theta_L$ in terms of $\theta_{ij}$, $r$ and $\alpha$. From Eqs.\eqref{tetaL}, \eqref{tgtL51e} and the defining conditions for $a_1^2$ and $b_{1,2}^2$ given in Table~\ref{definingconditions}, one can determine the parameters $a_{1,2}$ and $b_{1,2}$ in terms of $m_e$, $m_\mu$, $m_\tau$ and $\theta_{23}$, namely,
\begin{align}
a_2^2&\simeq\frac{2m_\mu^2 m_\tau^2 }{m_\mu^2+m_\tau^2\pm (m_\tau^2-m_\mu^2)\cos(2\theta_{23})} \,,\, \\
a_1^2&\simeq\frac{1}{2}\left[m_\mu^2+m_\tau^2\pm (m_\tau^2-m_\mu^2)\cos(2\theta_{23})\right]\,,\\
b_1^2&= m_e^2\,,\\
b_2^2&\simeq\frac{(m_\tau^2-m_\mu^2)^2\sin^2(2\theta_{23})}{2\left[m_\mu^2+m_\tau^2\pm (m_\tau^2-m_\mu^2)\cos(2\theta_{23})\right]}\,.
\label{eq69}
\end{align}

In order to relate $\delta$ with $\alpha$, we notice that 
\begin{align}
\arg(\U_{23})&\simeq \arctan\left[\frac{2s_{13}s_\alpha }{rt_L\sin(2\theta_{12})-2s_{13}c_{\alpha}}\right]\,,\\
\arg(\U_{33})&\simeq -\arctan\left[\frac{2s_{13}t_L s_\alpha }{r\sin(2\theta_{12})+2s_{13}t_L c_{\alpha}}\right]\,,
\end{align}
for both NH and IH. These relations imply
\begin{align}
\arg(\U_{23})\simeq\arg(\U_{33}) \simeq -\alpha\,,
\end{align}
from which, after performing some rephasing transformations to bring $\U$ to the form given in Eq.~\eqref{Uparam}, we obtain
\begin{align}
\delta=\arg(\U_{23})\simeq -\alpha\,,
\label{deltaa}
\end{align}
confirming the results plotted in Fig.~\ref{fig2}. Following the same procedure for the $5_1^\mu$ ($5_1^\tau$) we obtain $\delta\simeq \alpha$ ($\delta\simeq \pi+\alpha$), which also agrees with the numerical output shown in Fig.~\ref{fig2}. In conclusion, all parameters in the mass matrices $\Ml$ and $\Mnu$ can be determined in terms of the charged-lepton and neutrino masses and mixing angles through Eqs.~\eqref{eq60}-\eqref{eq69} and \eqref{deltaa}.

\subsection{Lepton universality and rare LFV decays}
\label{LFV}

In the 2HDM, Yukawa interactions may induce flavor-changing neutral currents (FCNC) at the tree and loop levels. Therefore, the viable maximally restrictive textures previously obtained (cf. Table~\ref{decompositions}) should be confronted with the current experimental bounds on such processes. In particular, the constraints on universality in purely leptonic decays, lepton-flavor-violating decays $\ell_\alpha^-\to \ell_{\beta}^-\,\ell_{\gamma}^+\,\ell_{\delta}^-$, and $\ell_\alpha\to \ell_{\beta}\, \gamma$ should be considered.\footnote{In Refs.~\cite{Botella:2014ska,Felipe:2016sya}, the implications of these processes have been analyzed in alternative 2HDM realizations.} With this purpose, we first briefly review the interactions among leptons and the neutral and charged scalars in the 2HDM. Since we are considering scenarios with a U(1) symmetry under which one of the Higgs doublets is charged, there is no CP violation in the scalar potential and, thus, no mixing between CP-even and CP-odd scalars. It is convenient to rotate ($\Phi_1,\Phi_2$) to the Higgs basis ($H_1,H_2$) through
\begin{align}
H_1&=\Phi_1\cos\beta+\Phi_2\sin\beta\nonumber\,,\\
H_2&=-\Phi_1\sin\beta+\Phi_2\cos\beta\,,
\label{H1H2}
\end{align}
such that one can write
\begin{align}
H_1=\frac{1}{\sqrt{2}}\begin{pmatrix}
\sqrt{2}\,G^+\\
v+H^0+i\,G^0
\end{pmatrix}\,,\,H_2=\frac{1}{\sqrt{2}}\begin{pmatrix}
\sqrt{2}\,H^+\\
R+i\,I\end{pmatrix}\,,
\label{H1H2def}
\end{align}
with $\langle H_2 \rangle=0$. The neutral (charged) Goldstone boson is denoted $G^0$ ($G^+$), while $I$ is the U(1) Goldstone boson, which is massless in the exact U(1) symmetric limit. In order to avoid this massless particle, a soft U(1) symmetry breaking term of the type $m_{12}^2 \Phi_1^\dagger\Phi_2$ can be included in the scalar potential originating a mass $m_I^2 \propto m_{12}^2$ for the decoupled CP-odd scalar. 

Throughout this work we will also assume that there is no mixing between $R$ and $H^0$ and identify $H^0$ with the SM Higgs boson discovered by the ATLAS and CMS collaborations at the LHC. Thus, we consider the limit where the physical mass state $h$ is identified with $H^0$, which implies $m_{H^0}\equiv m_h \simeq 125$~GeV. Under these premises, the CP-even scalars are $h$ and $R$, while $I$ is CP-odd. As already said, in our framework there is no CP-violation in the scalar sector and, thus, $I$ is decoupled from $h$ and $R$, which allows us to take $m_R$, $m_I$ and the mass of the charged scalar $m_{H^+}$ as independent parameters.

The relevant scalar-fermion interactions can be read off from the Lagrangian~\eqref{lagrangian} which, after appropriate transformations, takes the form
\begin{align}
-\mathcal{L}=&\frac{1}{v}\overline{e_L}\left[\Dl (v+h)+\N_{e}R+i\N_{e}I\right]e_R\nonumber\\
&+\frac{\sqrt{2}}{v}\overline{\nu_L}\,\U^\dagger \N_{e}  e_R H^++{\rm H.c.}\,,
\label{lagrangian2}
\end{align}
with
\begin{align}
\N_{e}=\U_L^{\ell\dagger} \N_{e}^0\URl\,,\,\N_{e}^0=\frac{v}{\sqrt{2}}(\Yl_1\sin\beta-\Yl_2\cos\beta)\,.
\label{Nedef}
\end{align}
In Eq.~\eqref{lagrangian2} all the fermion fields are mass eigenstates (see Section~\ref{Dir2HDM} for details and definitions regarding the unitary transformations and Yukawa matrix conventions).

Lepton universality tests aim at probing the SM prediction that all leptons couple with the same strength to the charged weak current. A relevant quantity to test universality in purely leptonic $\tau$ decays is
\begin{equation}
\label{eq:gmoverge}
\left|\frac{g_\mu}{g_e}\right|^2 \equiv \frac{\text{Br}\left(\tau\to \mu\nu\bar\nu\right)}{\text{Br}\left(\tau \to e\nu\bar\nu\right)}
\frac{f(x_{e\tau})}{f(x_{\mu\tau})}\,,
\end{equation}
where $x_{\alpha\beta}\equiv m^2_\alpha/m^2_\beta$\,. In the presence of scalar and vector interactions, the $\ell_\alpha\to \ell_{\beta}\, \nu\,\bar\nu$ branching ratio (BR) is~\cite{Botella:2014ska}
\begin{equation}
\label{eq:BRuniv}
\begin{split}
\text{Br}(\ell_\alpha\to \ell_{\beta}\, \nu\,\bar\nu)=
\Big(1+\frac{1}{4}\left|{g_{RR,\alpha \beta}^{S}}\right|^2\Big)
f(x_{\beta\alpha})\\
+2\,\text{Re}\left[g_{RR,\alpha \beta}^{S} \left(g_{LL,\alpha \beta}^{V}\right)^\ast\right]
x_{\beta\alpha}\, g(x_{\beta\alpha})\,,
\end{split}
\end{equation}
where 
\begin{align}
\begin{split}
f(x)&=1-8x+8x^3-x^4-12x^2 \ln x,\\
g(x)&=1+9x-9x^2-x^3+6x(1+x)\ln x,
\end{split}
\end{align}
are the phase space functions and 
\begin{align}
&\left|g_{RR,\alpha \beta}^{S}\right|^2\equiv\sum_{i,j=1}^3|\U_{\alpha i}|^2|\U_{\beta j}|^2 |g_{i\alpha j\beta}|^2\,,\\
&\left(g_{RR,\alpha\beta}^{S}\right) \left(g_{LL,\alpha \beta}^{V}\right)^\ast\equiv\sum_{i,j=1}^3|\U_{\alpha i}|^2|\U_{\beta j}|^2 g_{i\alpha j\beta}\,.
\end{align}
Finally, the flavor-dependent coefficients $g_{i\alpha j\beta}$ are model-specific, being in our framework given by
\begin{equation}
g_{i\alpha j\beta}=-\frac{(\U^\dagger \N_e)_{i \alpha} (\N_e^\dagger\, \U)_{\beta j}}{m_{H^+}^2 \U^\ast_{\alpha i} \U_{\beta j}},
\end{equation}
with $\mathbf{N}_e$ defined as in \eqref{Nedef}. Current experimental constraints yield~\cite{Amhis:2016xyh}
\begin{align}\label{eq:expuniv1}
\left|g_\mu/g_e\right|-1=0.0019\pm 0.0014
\end{align}
and
\begin{align}\label{eq:expuniv2}
\!\!\!\!\left|g_{RR,\mu e}^{S}\right|<0.035, \; \left|g_{RR,\tau e}^{S}\right|<0.70, \; \left|g_{RR,\tau \mu}^{S}\right|<0.72,
\end{align}
at 95\% CL~\cite{Tanabashi:2018oca}. 

In the present scenario, lepton-flavor violating decays $\ell_\alpha^-\to \ell_{\beta}^-\,\ell_{\gamma}^+\,\ell_{\delta}^-$ are mediated by the neutral scalars $R$ and $I$ at tree level. The BR for a generic $\ell_\alpha^-\to \ell_{\beta}^-\,\ell_{\gamma}^+\,\ell_{\delta}^-$ process (normalized to the BR of the flavor-conserving decay $\ell_\alpha \to \ell_\beta\nu_\alpha\bar{\nu}_\beta$) is~\cite{Botella:2014ska}
\begin{align}
&\frac{\text{Br}(\ell_\alpha^-\to \ell_{\beta}^-\,\ell_{\gamma}^+\,\ell_{\delta}^-)}{\text{Br}(\ell_\alpha \to \ell_\beta\nu_\alpha\bar{\nu}_\beta)}
=\frac{1}{16(1+\delta_{\beta\delta})}\times\nonumber\\
&\left[
\left|g_{LL}^{\alpha\beta,\gamma\delta}\right|^2+\left|g_{LL}^{\alpha\delta,\gamma\beta}\right|^2+\left|g_{RR}^{\alpha\beta,\gamma\delta}\right|^2+\left|g_{RR}^{\alpha\delta,\gamma\beta}\right|^2\right.\nonumber\\
&+\left|g_{LR}^{\alpha\beta,\gamma\delta}\right|^2+\left|g_{LR}^{\alpha\delta,\gamma\beta}\right|^2
+\left|g_{RL}^{\alpha\beta,\gamma\delta}\right|^2+\left|g_{RL}^{\alpha\delta,\gamma\beta}\right|^2\nonumber\\
&\left.-\text{Re}\left(g_{LL}^{\alpha\beta,\gamma\delta}{g_{LL}^{\alpha\delta,\gamma\beta}}^\ast+g_{RR}^{\alpha\beta,\gamma\delta}{g_{RR}^{\alpha\delta,\gamma\beta}}^\ast\right)
\right]\,,
\label{3body1}
\end{align}
where
\begin{align}
&g_{LL}^{\alpha\beta,\gamma\delta}=(\N_e^\dagger)_{\beta\alpha}(\N_e^\dagger)_{\delta\gamma}\left(\frac{1}{m_R^2}-\frac{1}{m_I^2}\right)\,,\nonumber\\ &
g_{RL}^{\alpha\beta,\gamma\delta}=(\N_e)_{\beta\alpha}(\N_e^\dagger)_{\delta\gamma}\left(\frac{1}{m_R^2}+\frac{1}{m_I^2}\right)\,,\nonumber\\
&g_{LR}^{\alpha\beta,\gamma\delta}=(\N_e^\dagger)_{\beta\alpha}(\N_e)_{\delta\gamma}\left(\frac{1}{m_R^2}+\frac{1}{m_I^2}\right)\,,\nonumber\\ &
g_{RR}^{\alpha\beta,\gamma\delta}=(\N_e)_{\beta\alpha}(\N_e)_{\delta\gamma}\left(\frac{1}{m_R^2}-\frac{1}{m_I^2}\right)\,,
\label{3body2}
\end{align}
and~\cite{Tanabashi:2018oca} 
\begin{align}
\begin{split}
\text{Br}(\mu \to e\nu_\mu\bar{\nu}_e)&\simeq 1.0\,,\\
\text{Br}(\tau \to \mu\nu_\tau\bar{\nu}_\mu)&\simeq 0.17\,,\\
\text{Br}(\tau \to e\nu_\tau\bar{\nu}_e)&\simeq 0.18\,.
\label{mugamma}
\end{split}
\end{align}
Currently, the experimental upper limits on the branching ratios of the 3-body LFV decays are~\cite{Tanabashi:2018oca} 
\begin{align}\label{eq:expltolll}
\begin{split}
&\text{Br}(\tau^- \to e^-e^+e^-)<2.7\times10^{-8},\\
&\text{Br}(\tau^- \to \mu^-\mu^+\mu^-)<2.1\times 10^{-8},\\ 
&\text{Br}(\tau^- \to e^-\mu^+e^-)<1.5\times 10^{-8},\\
&\text{Br}(\tau^- \to e^-e^+\mu^-)<1.8\times 10^{-8},\\ 
&\text{Br}(\tau^- \to \mu^-e^+\mu^-)<1.7\times10^{-8},\\
&\text{Br}(\tau^- \to \mu^-\mu^+e^-)<2.7\times10^{-8},\\ 
&\text{Br}(\mu^- \to e^-e^+e^-)<1.0\times 10^{-12},
\end{split}
\end{align}
at 90\% CL.

Finally, neglecting contributions proportional to the neutrino masses and sub-leading terms in $m_\ell^2/m_{R,I}^2$, the decay width of the radiative lepton-flavor violating process $\ell_\alpha\to \ell_{\beta}\, \gamma$ is given, up to one-loop level, by~\cite{Botella:2014ska}
\begin{align}
\begin{split}
\frac{\text{Br}(\ell_\alpha \to \ell_\beta\, \gamma)}{\text{Br}(\ell_\alpha \to \ell_\beta\nu_\alpha\bar{\nu}_\beta)}=\frac{3\alpha_{e}}{2\pi}
\left(\left|\mathcal{A}_L\right|^2+\left|\mathcal{A}_R\right|^2\right),
\end{split}
\label{2body1}
\end{align}
where $\alpha_e=e^2/(4\pi)$ and the amplitudes $\mathcal{A}_{L,R}$ are in the present framework given by
\begin{align}
&\mathcal{A}_L=
-\dfrac{(\N_e^{\dagger}\N_e)_{\beta \alpha}}{12\,m_{H^{+}}^2}
+\dfrac{(\N_e^{\dagger}\N_e)_{\beta \alpha}}{12}\left(\dfrac{1}{m_{R}^2}+\dfrac{1}{m_{I}^2}\right)\,, \nonumber\\
&\mathcal{A}_R=
\dfrac{(\N_e\N_e^{\dagger})_{\beta \alpha}}{12\, m_{R}^2}
-\dfrac{(\N_e)_{\beta i} (\N_e)_{i\alpha}}{2\,m_{R}^2\,m_{\alpha}/m_i}
\left[\dfrac{3}{2}+\ln\left(\dfrac{m_{i}^2}{m_{R}^2}\right)\right]\nonumber\\
&+\dfrac{(\N_e\N_e^{\dagger})_{\beta \alpha}}{12\, m_I^2}
+\dfrac{(\N_e)_{\beta i}(\N_e)_{i \alpha}}{2\,m_{I}^2\,m_{\alpha}/m_i} \left[\dfrac{3}{2}+\ln\left(\dfrac{m_{i}^2}{m_{I}^2}\right)\right]\,,
\label{eq:AR}
\end{align}
where a sum over $i=e,\mu,\tau$ is implicitly assumed and chirally suppressed terms proportional to $m_\beta/m_\alpha$ were neglected. Current experimental upper bounds at 90\% CL are~\cite{Tanabashi:2018oca}
\begin{align}\label{eq:expltolgamma}
\begin{split}
\text{Br}(\mu \to e\gamma)&<4.2\times 10^{-13},\\
\text{Br}(\tau \to e\gamma)&<3.3\times 10^{-8},\\
\text{Br}(\tau \to \mu\gamma)&<4.4\times 10^{-8}.
\end{split}
\end{align}

We now aim at studying the compatibility of the texture pairs given in Table~\ref{decompositions} with the constraints coming from lepton universality and rare LFV decays discussed above. Simultaneously to the analysis performed in the previous section, we randomly vary $\tan\beta$ in the range $0.01$ to $100$ (these values ensure that the Yukawa couplings are always $\lesssim 1$), the charged-Higgs scalar masses $m_{H^\pm}\gtrsim 80$~GeV~\cite{Abbiendi:2013hk}, and the neutral scalar masses $m_{R,I} \gtrsim 100$~GeV~\cite{Tanabashi:2018oca}. We limit our search to cases where the $m_{H^\pm}\lesssim 1$~TeV and $m_{R,I} \lesssim 10$~TeV. For each input parameter set compatible with neutrino data, we compute $|g_\mu/g_e|-1$, $g_{RR,ij}^S$, and the BRs of all LFV 3-body and radiative charged-lepton decays. In all cases, we keep only those points obeying Eqs.~\eqref{eq:expuniv2}, \eqref{eq:expltolll} and \eqref{eq:expltolgamma}. As for $|g_\mu/g_e|$, we demand $|g_\mu/g_e|-1 \ge 10^{-4}$, keeping in mind the result \eqref{eq:expuniv1}.

\begin{figure*}[t]
\centering
\begin{tabular}{ccc}
\includegraphics[width=0.31\textwidth]{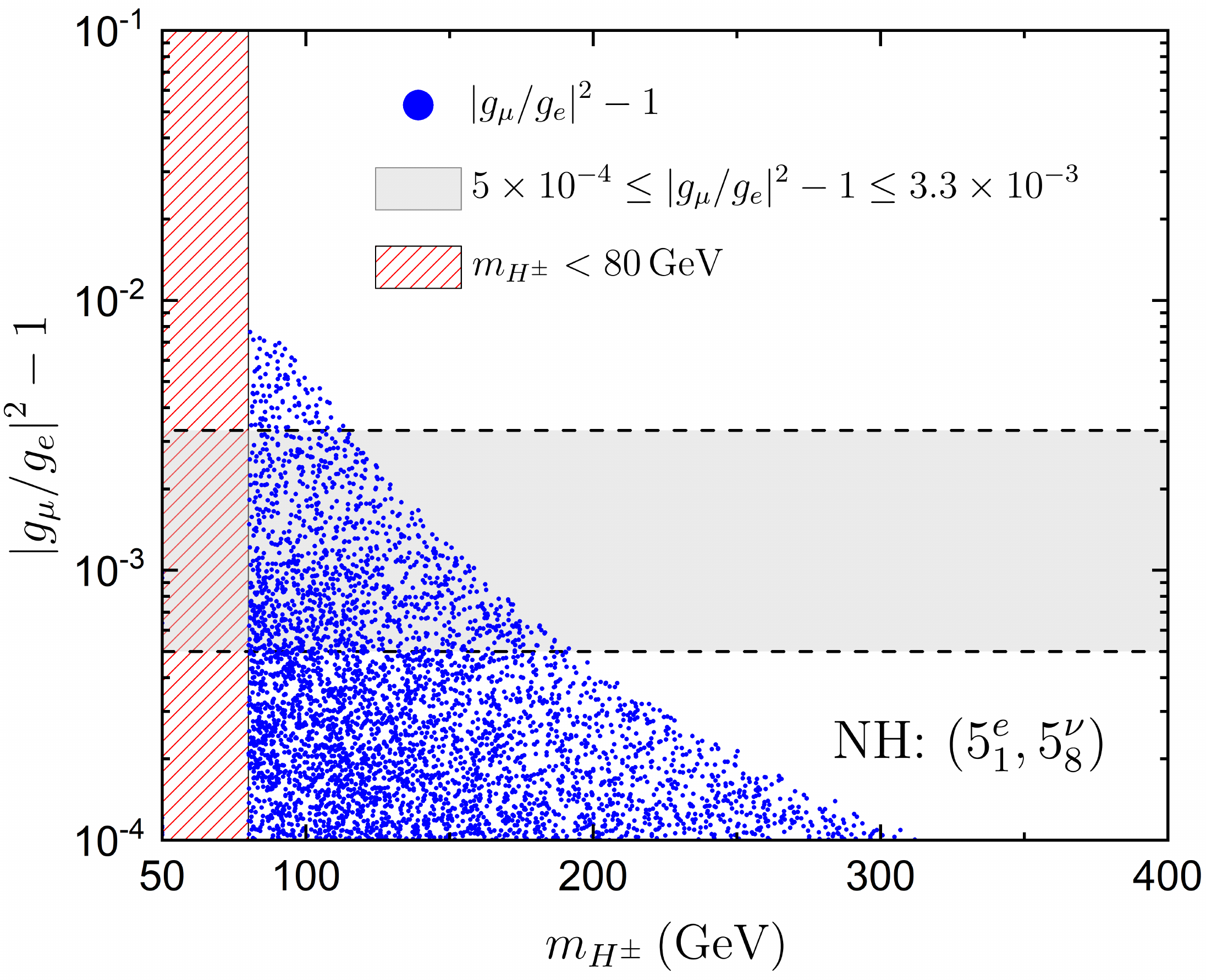} &
\includegraphics[width=0.31\textwidth]{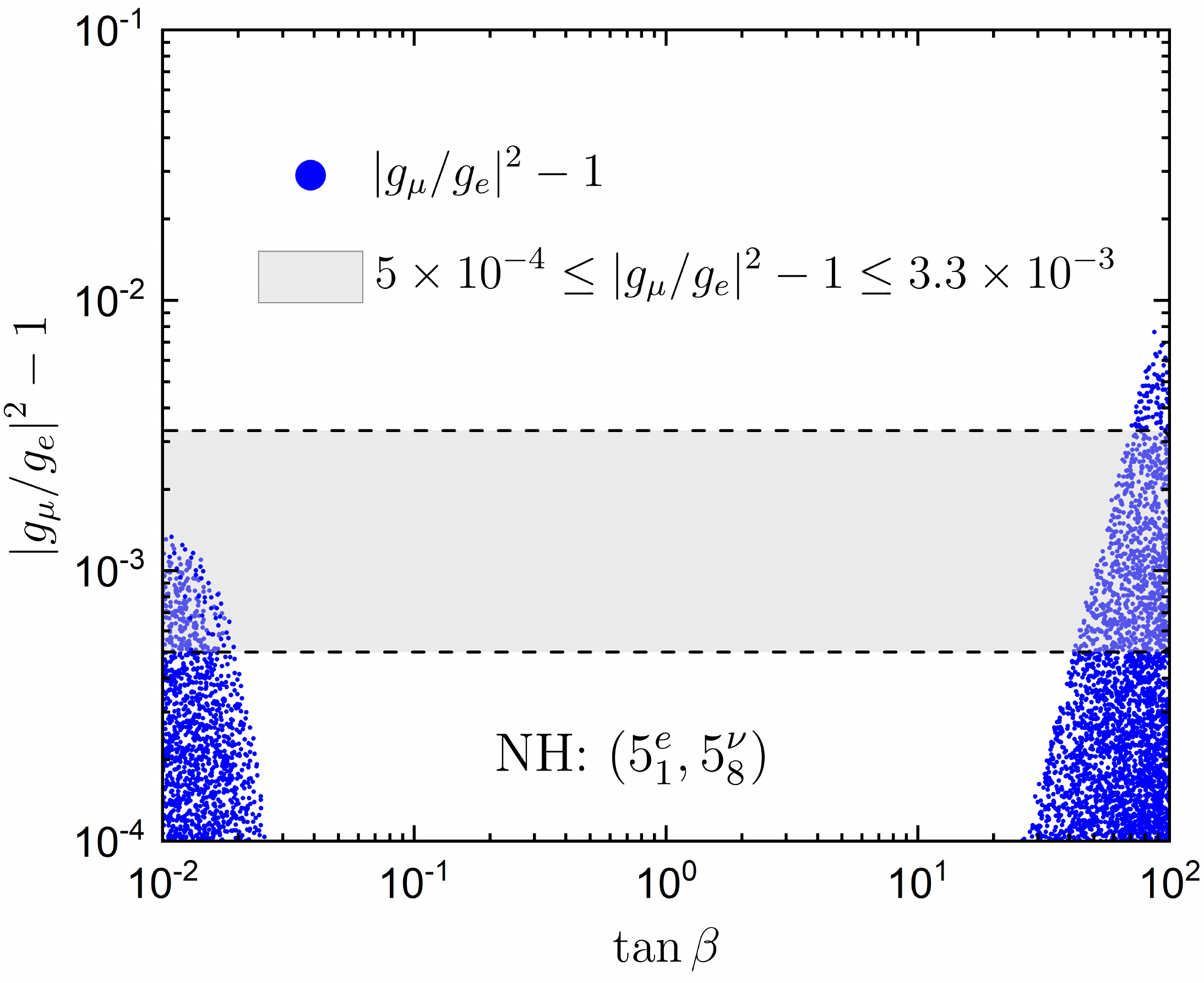}&
\includegraphics[width=0.298\textwidth]{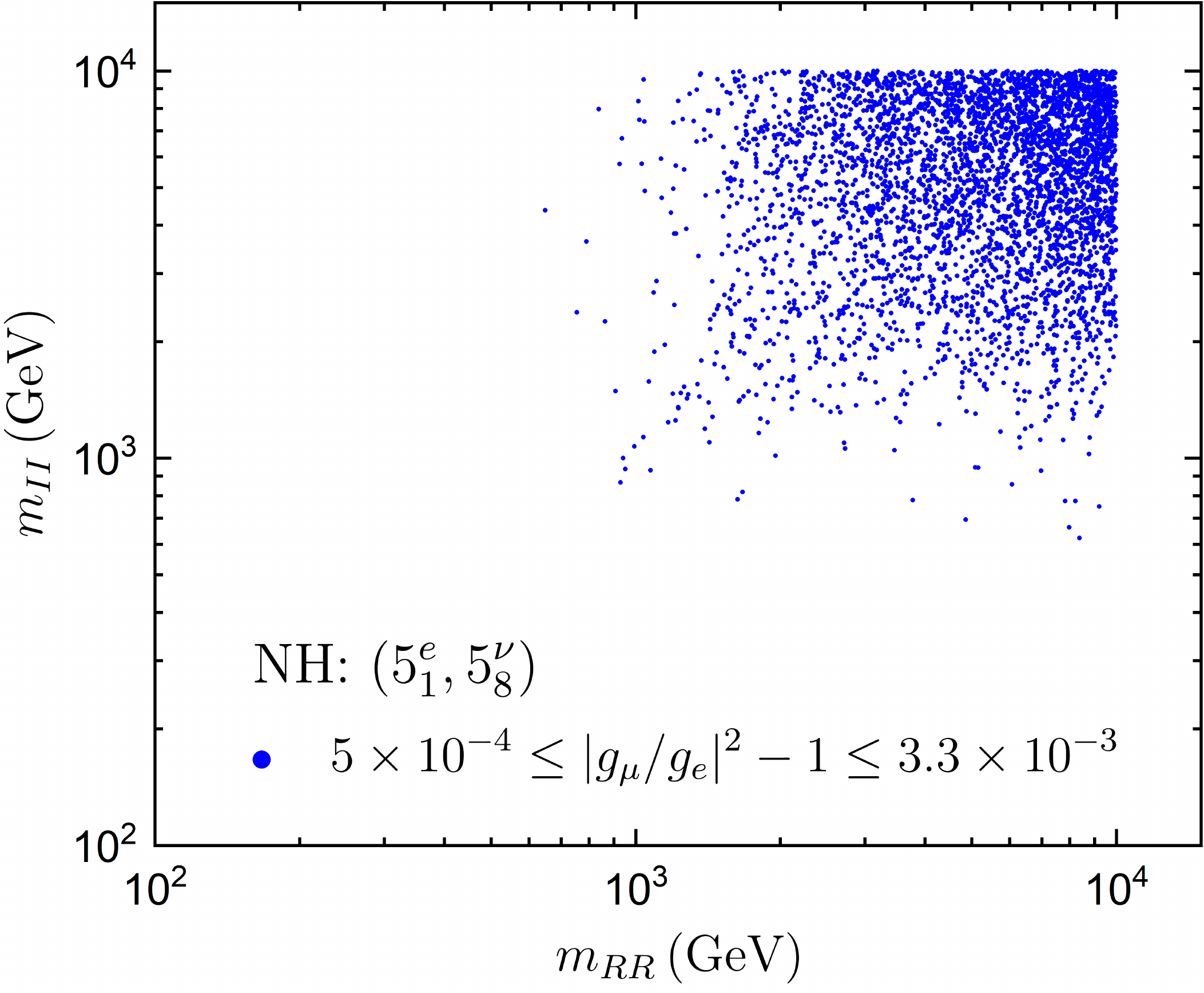} \\
\includegraphics[width=0.31\textwidth]{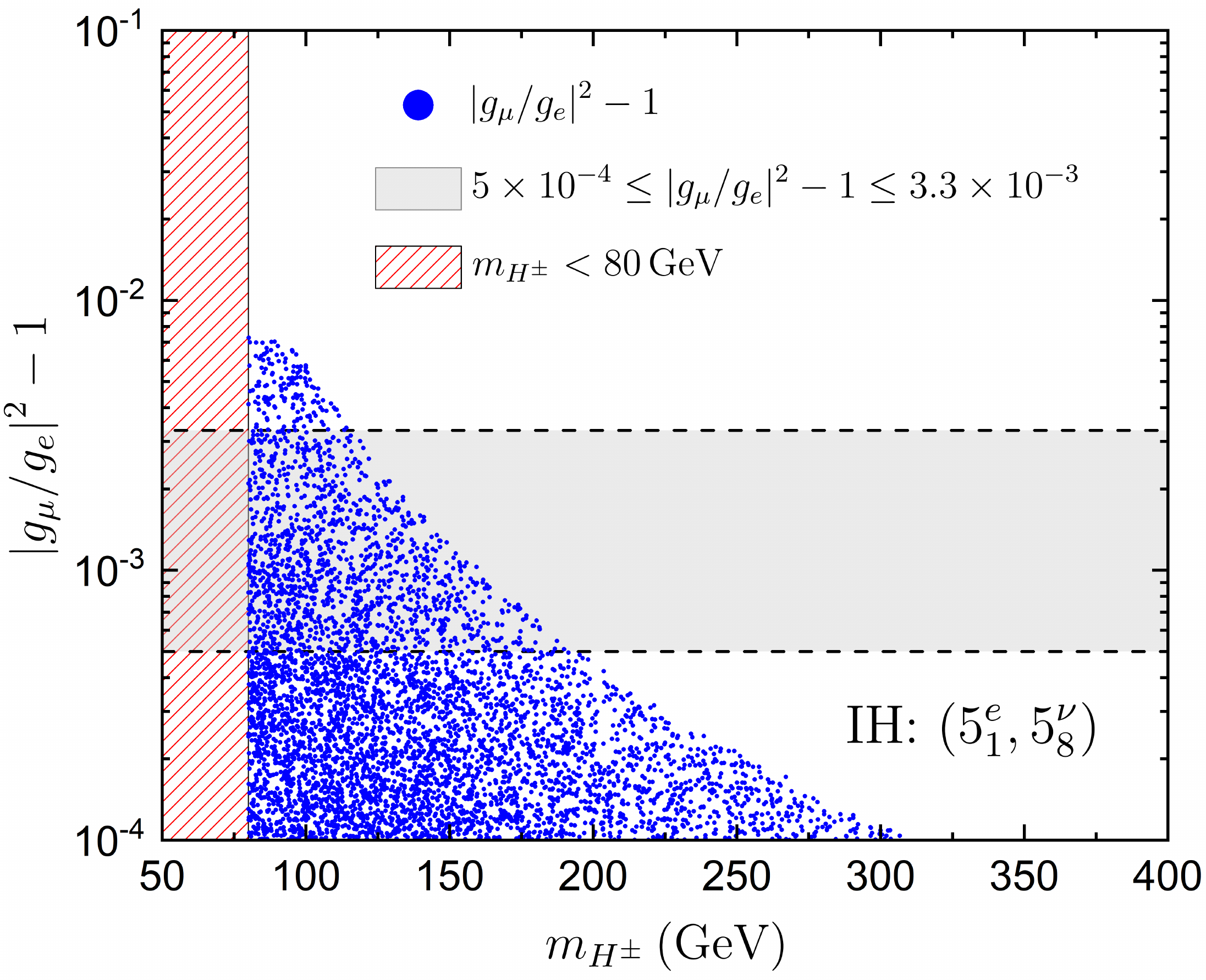} &
\includegraphics[width=0.31\textwidth]{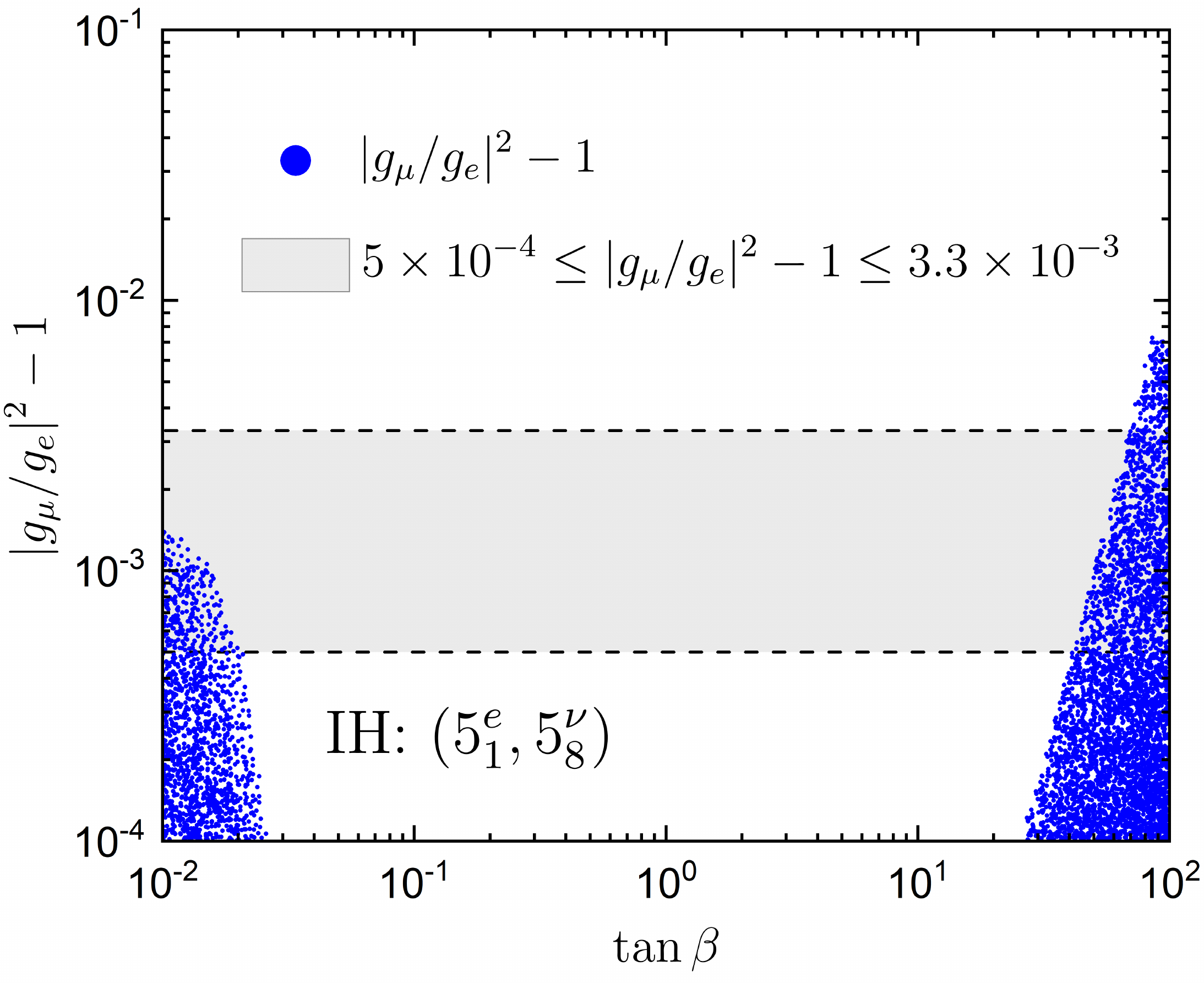}&
\includegraphics[width=0.298\textwidth]{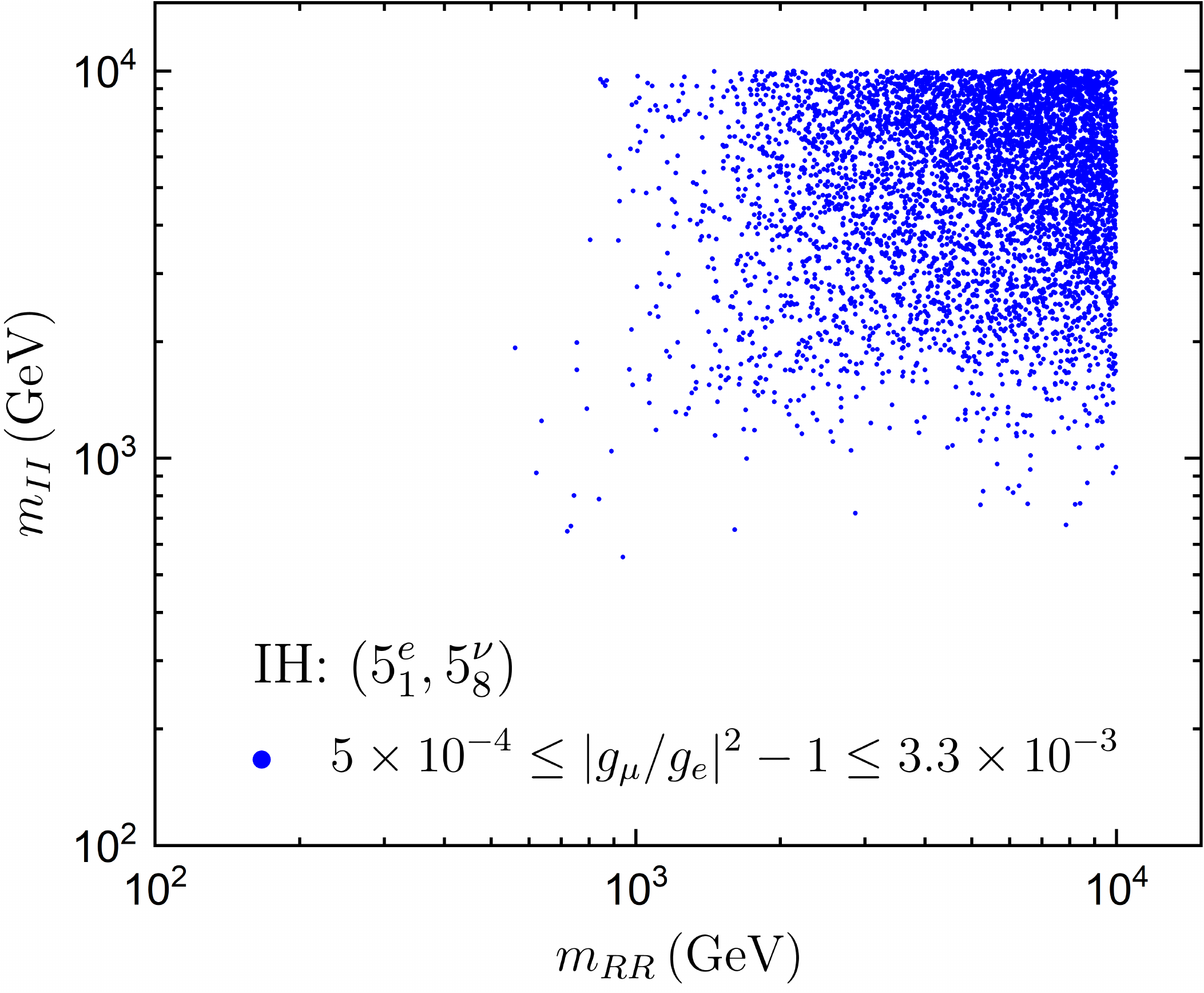}\\
\includegraphics[width=0.31\textwidth]{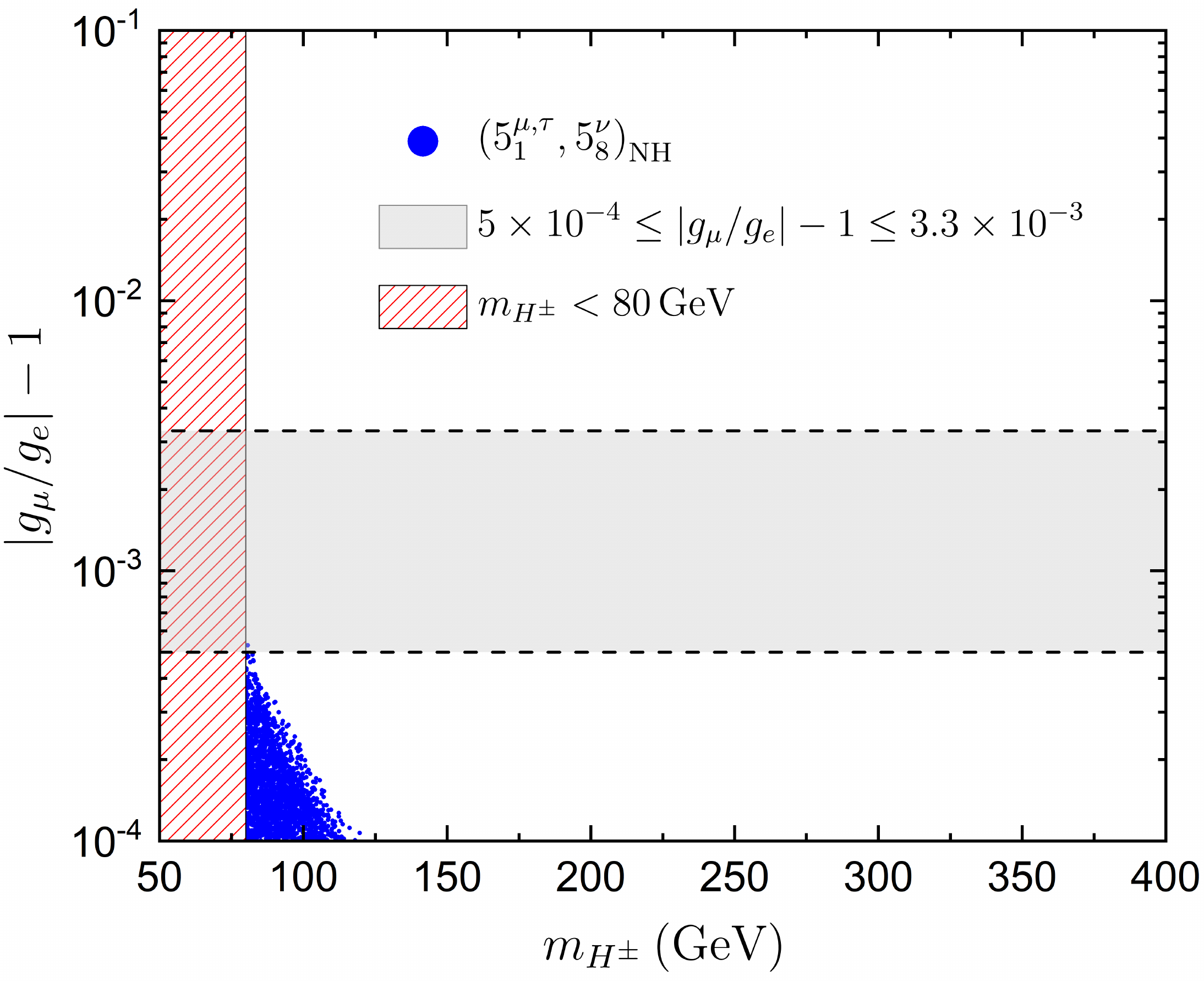} &
\includegraphics[width=0.31\textwidth]{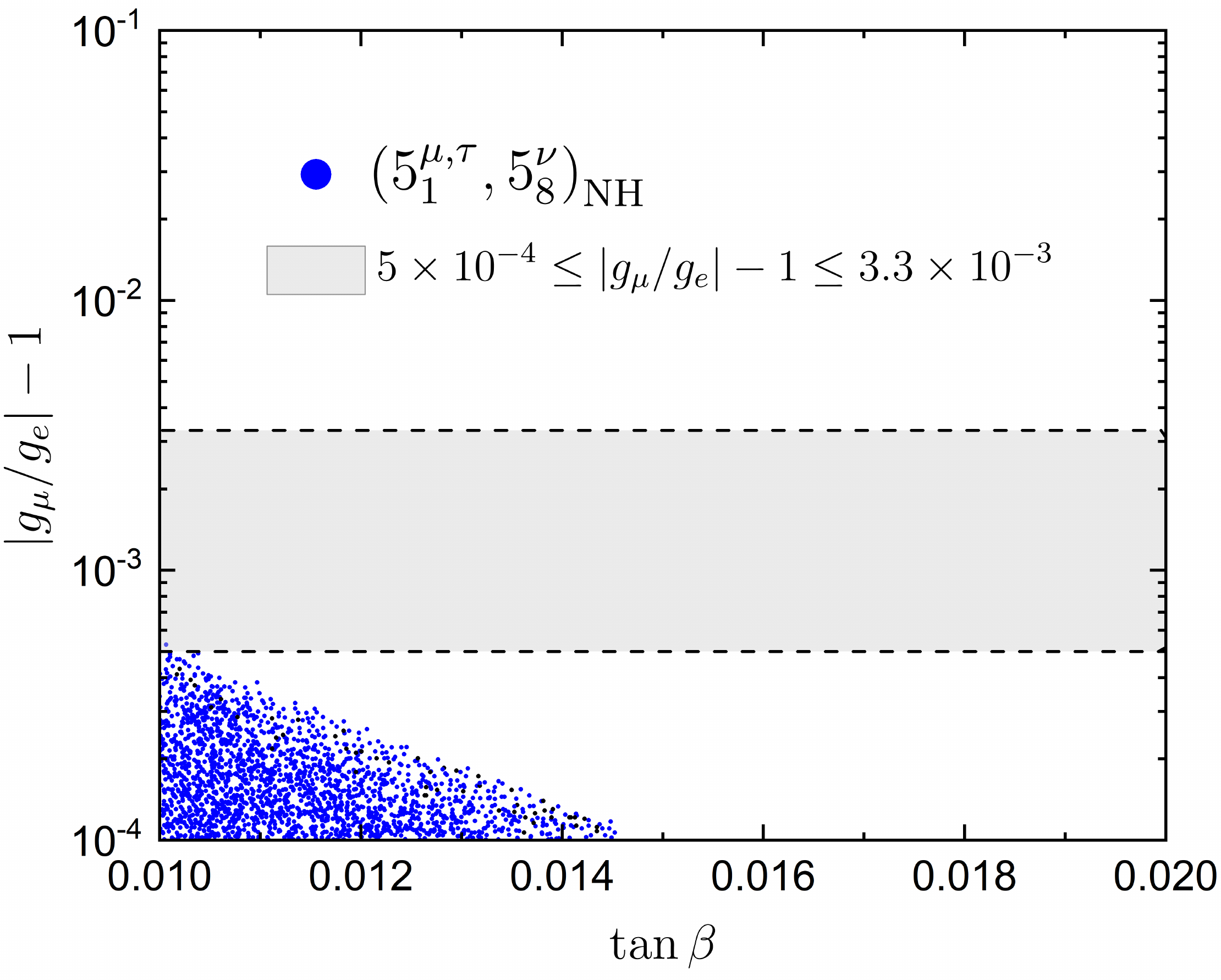}&
\includegraphics[width=0.297\textwidth]{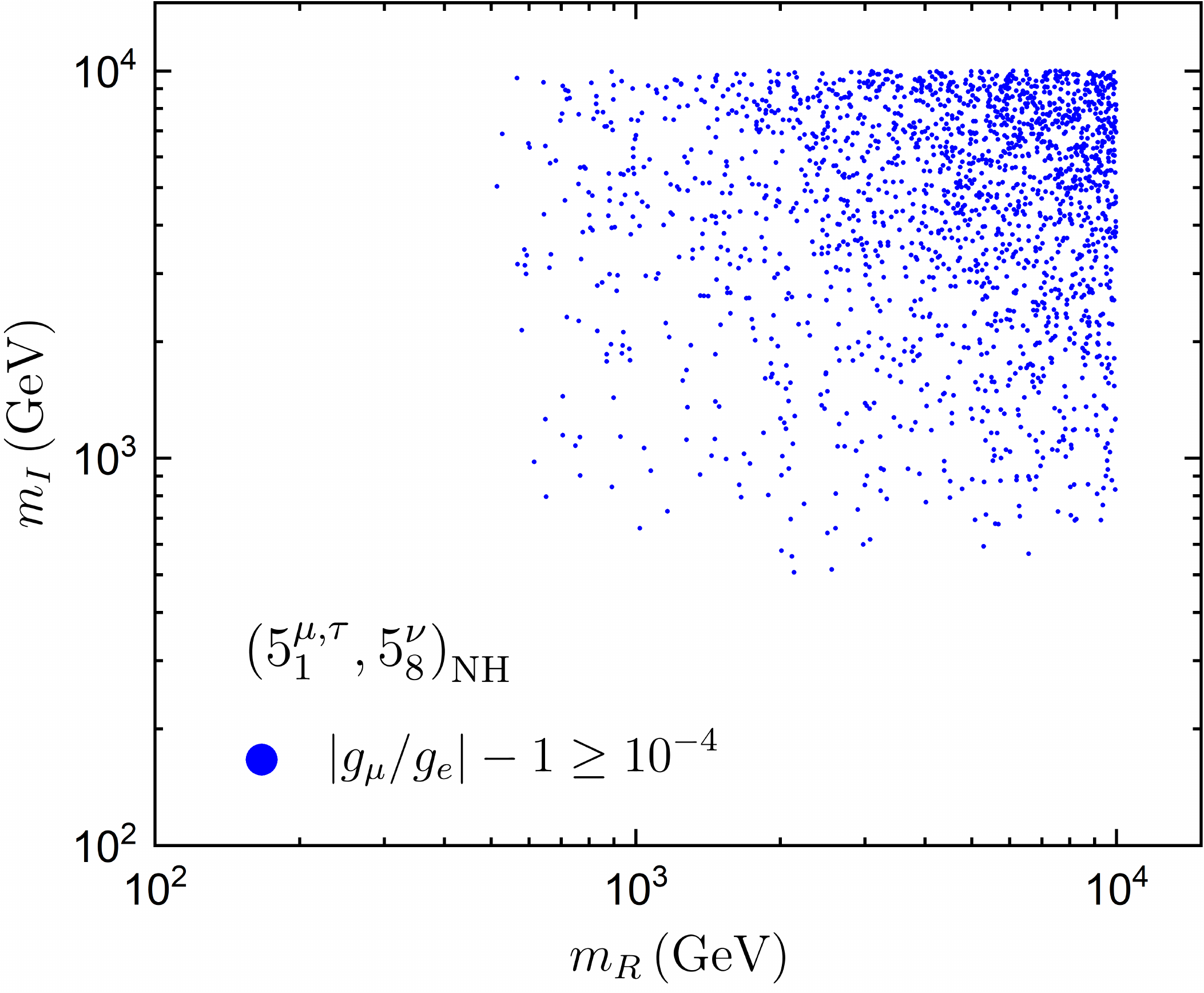} \\
\includegraphics[width=0.31\textwidth]{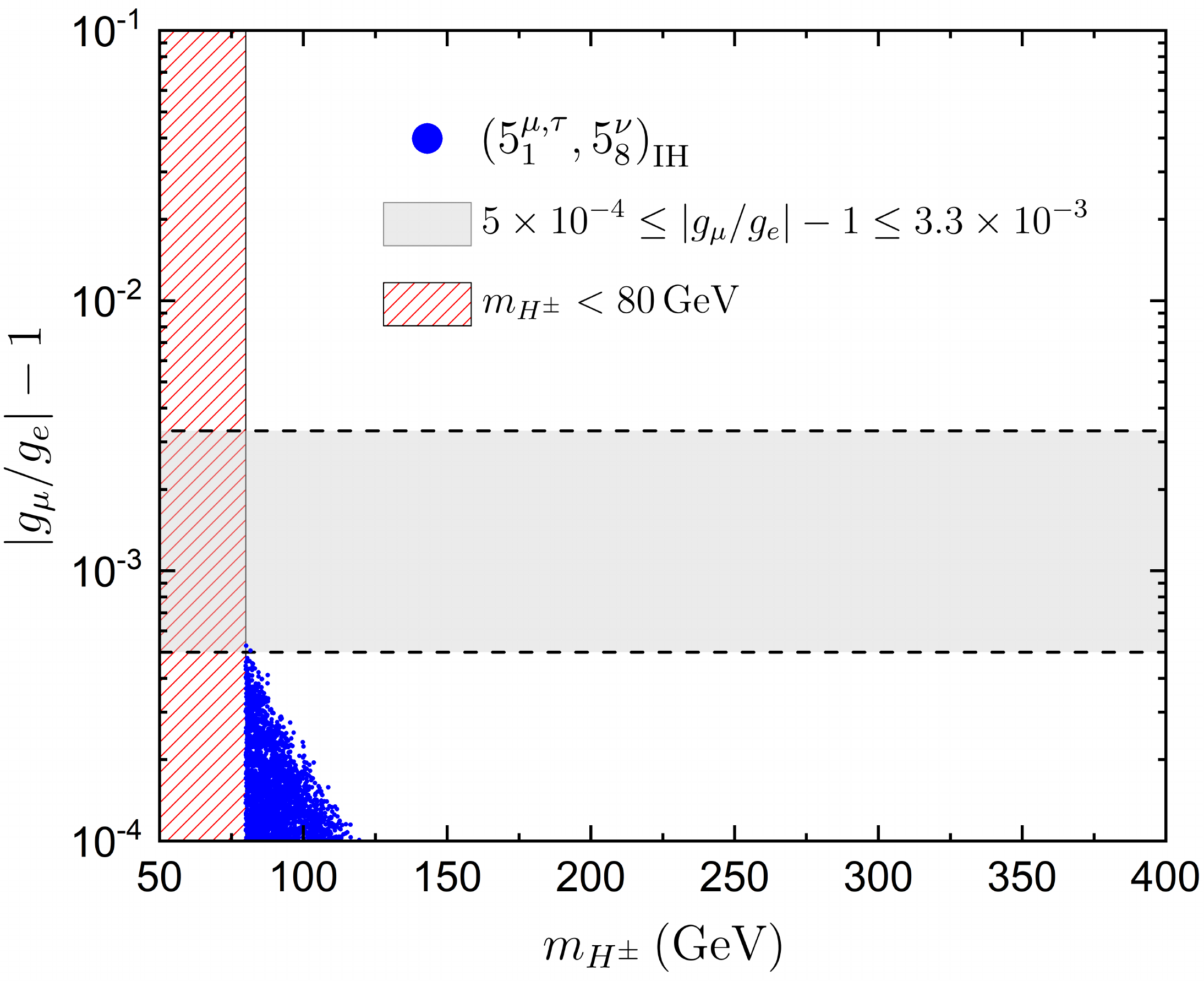} &
\includegraphics[width=0.31\textwidth]{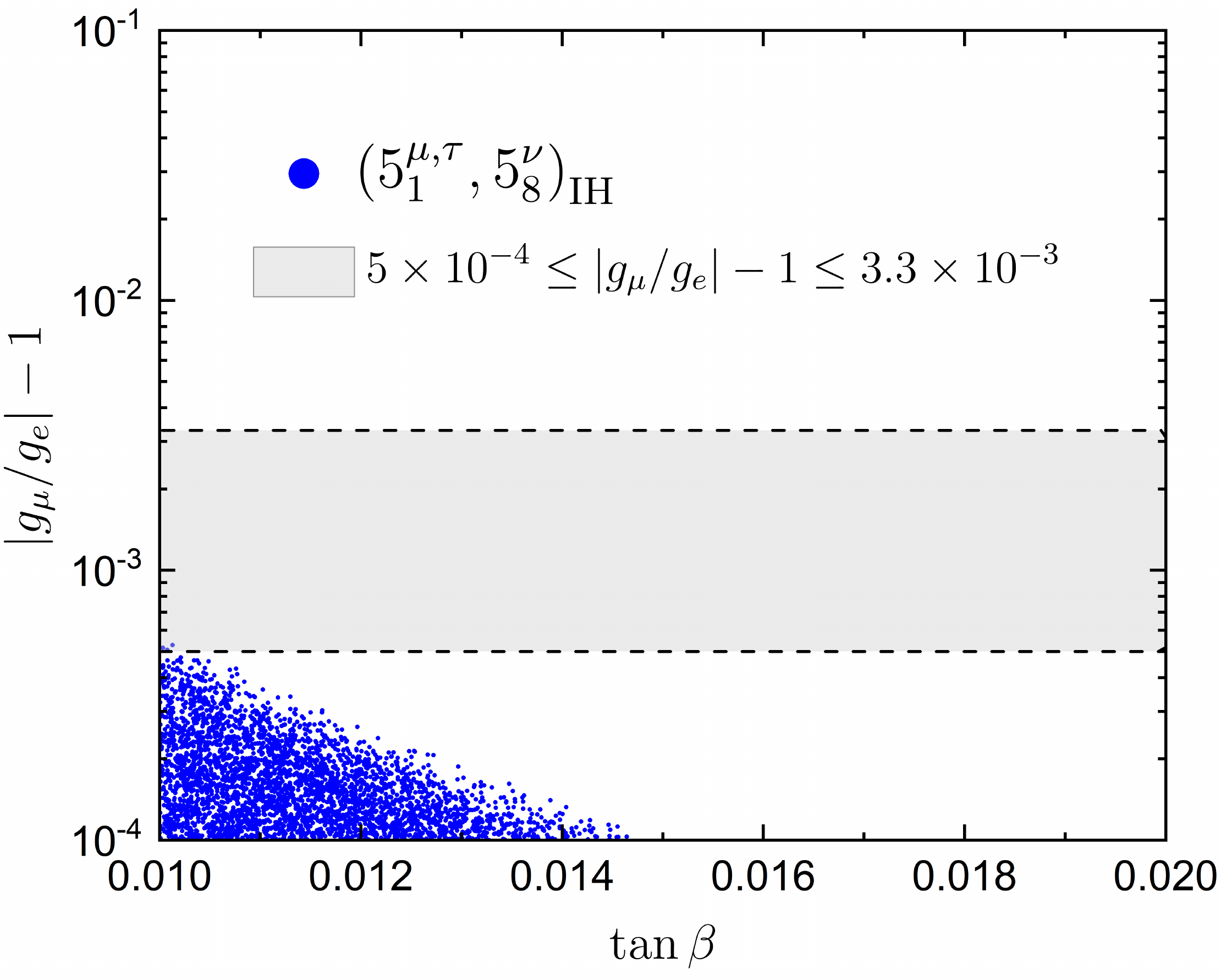}&
\includegraphics[width=0.297\textwidth]{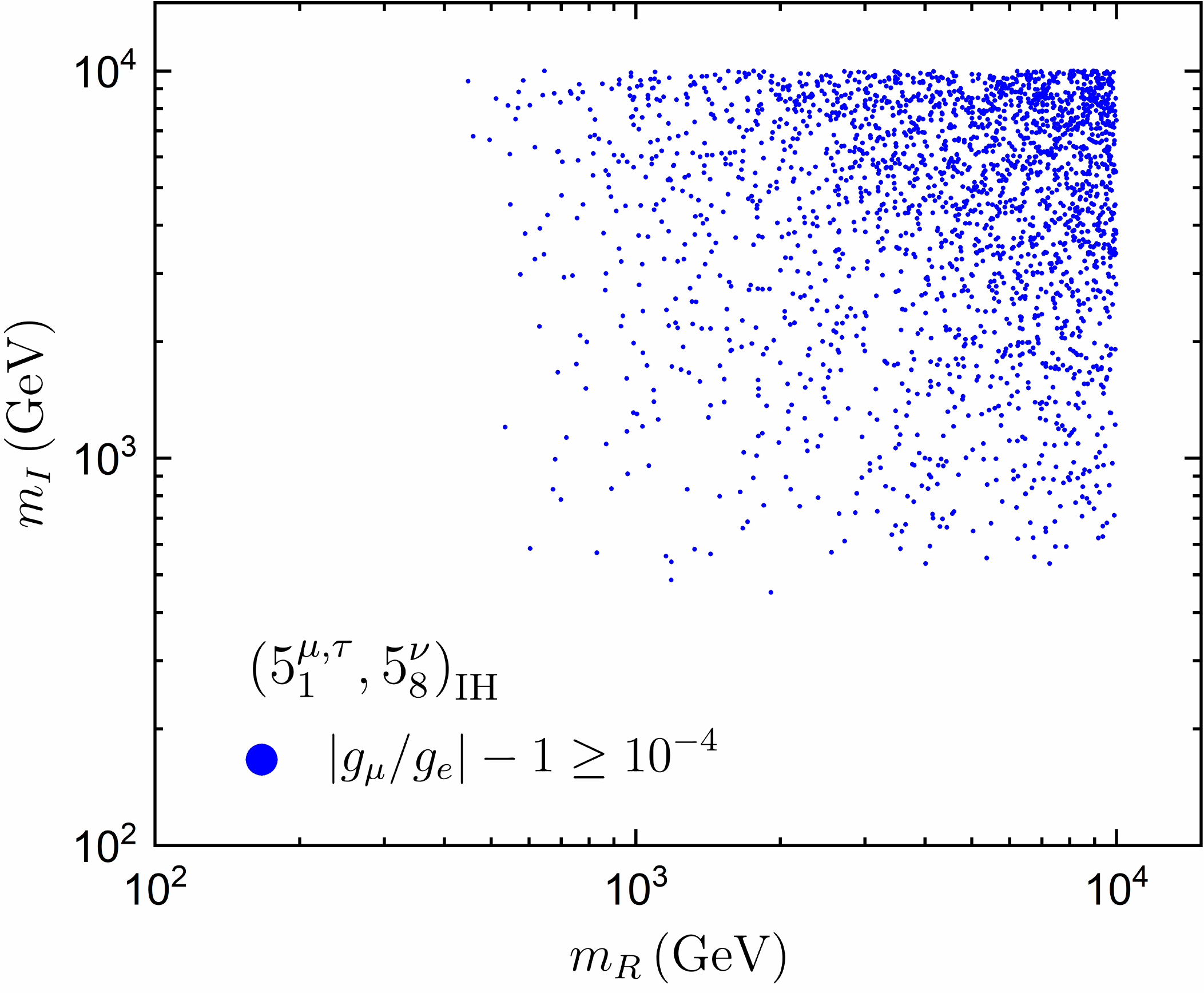}
\end{tabular}
\caption{Results for the $(5_1^e,5_8^\nu)$ and $(5_1^{\mu,\tau},5_8^\nu)$ texture pairs (upper and lower rows, respectively). In the left (middle) columns we plot $|g_\mu/g_e|-1$ as a function of $m_{H^\pm}$ ($\tan\beta$). The horizontal grey bands correspond to the constraint~\eqref{eq:expuniv1}. In the right column, the same points as in the corresponding $|g_\mu/g_e|-1$ plots are shown in the $(m_R,m_I)$-plane. All points obey the constraints \eqref{eq:expltolll} and \eqref{eq:expltolgamma}, and the mixing angles $\theta_{ij}$ lie within the $3\sigma$ ranges given in Table~\ref{datatable} for NH and IH.}
\label{fig3}
\end{figure*}
\begin{figure*}[t]
\centering
\begin{tabular}{cccc}
\includegraphics[width=0.24\textwidth]{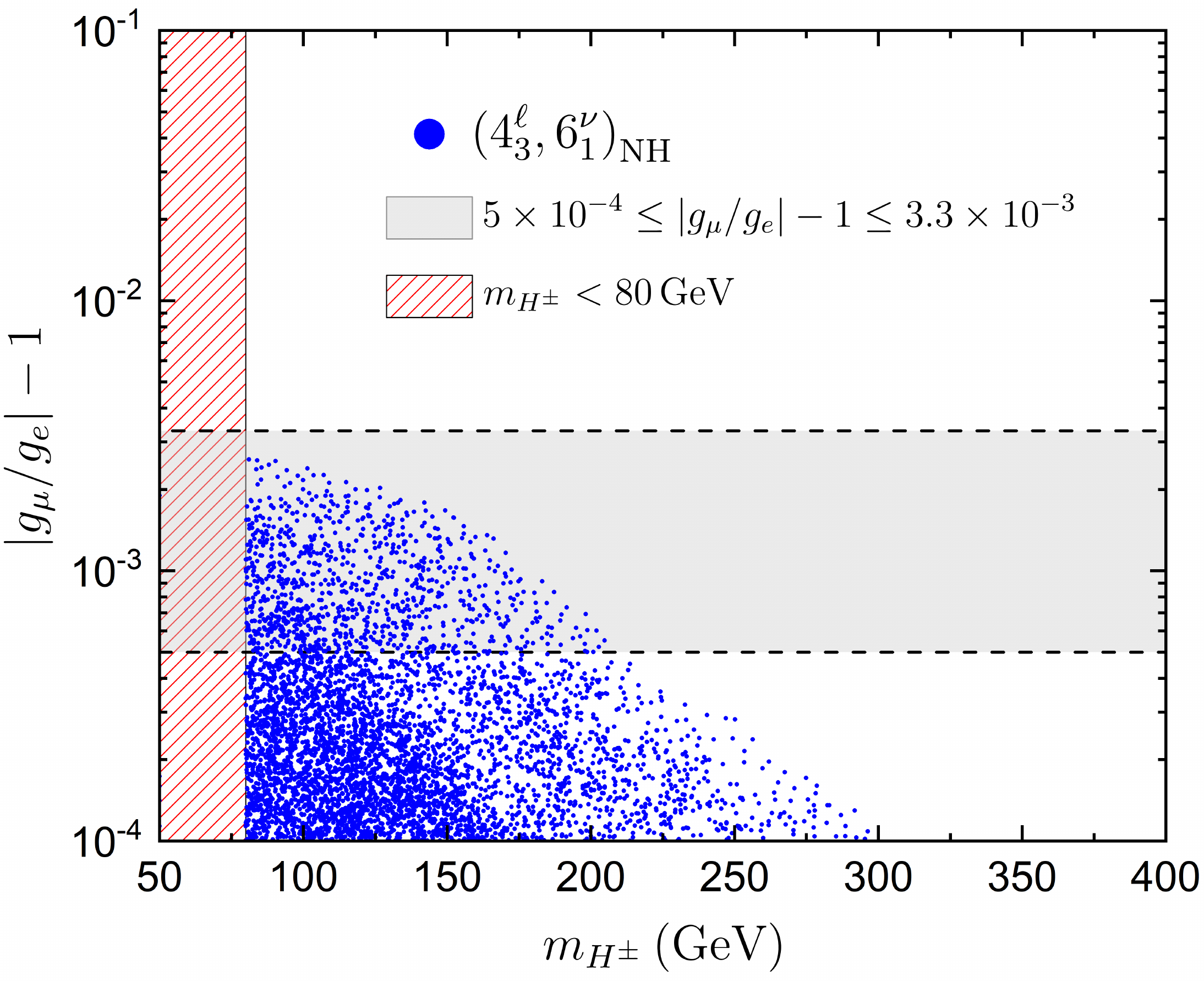} &
\includegraphics[width=0.23\textwidth]{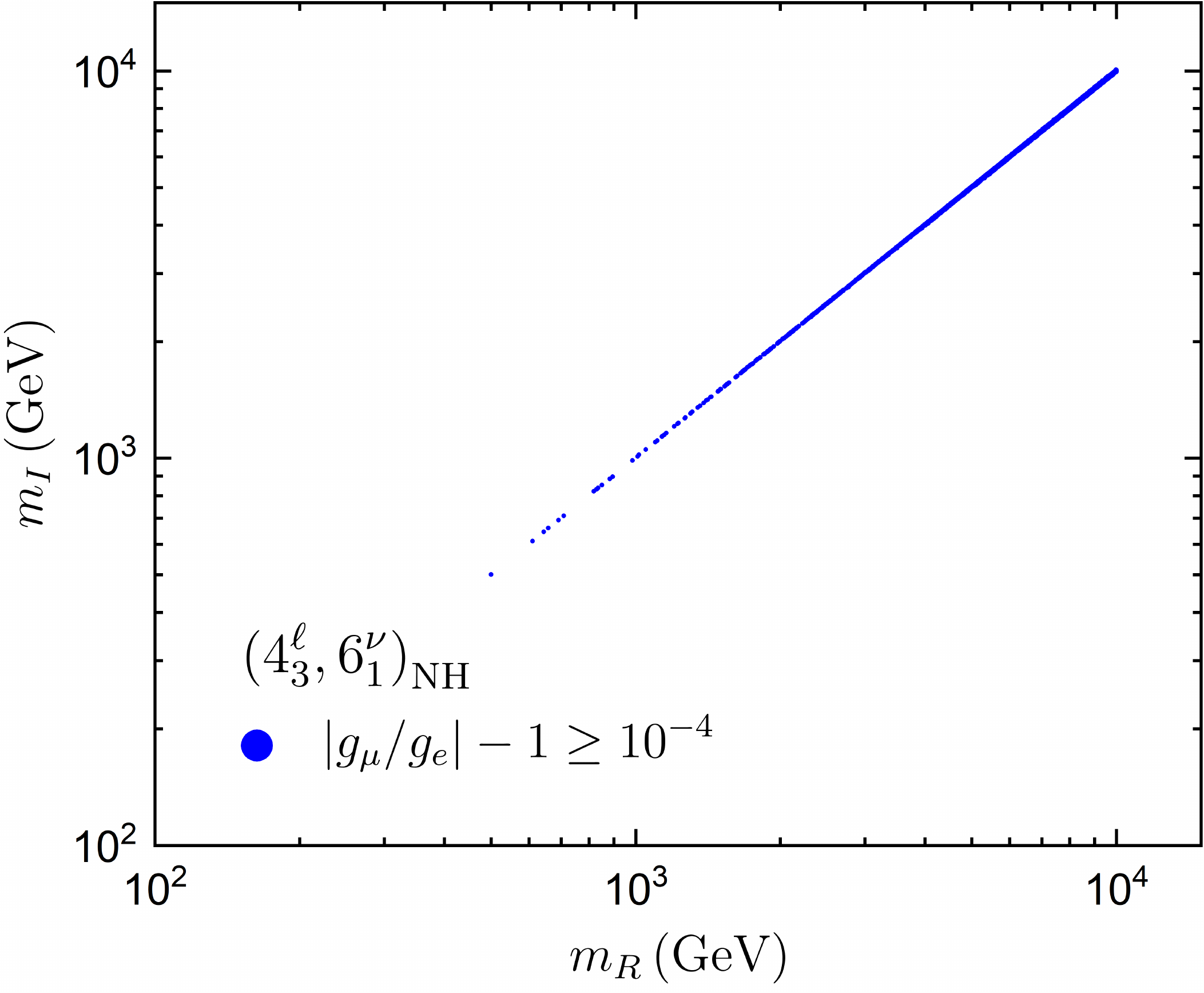} &
\includegraphics[width=0.24\textwidth]{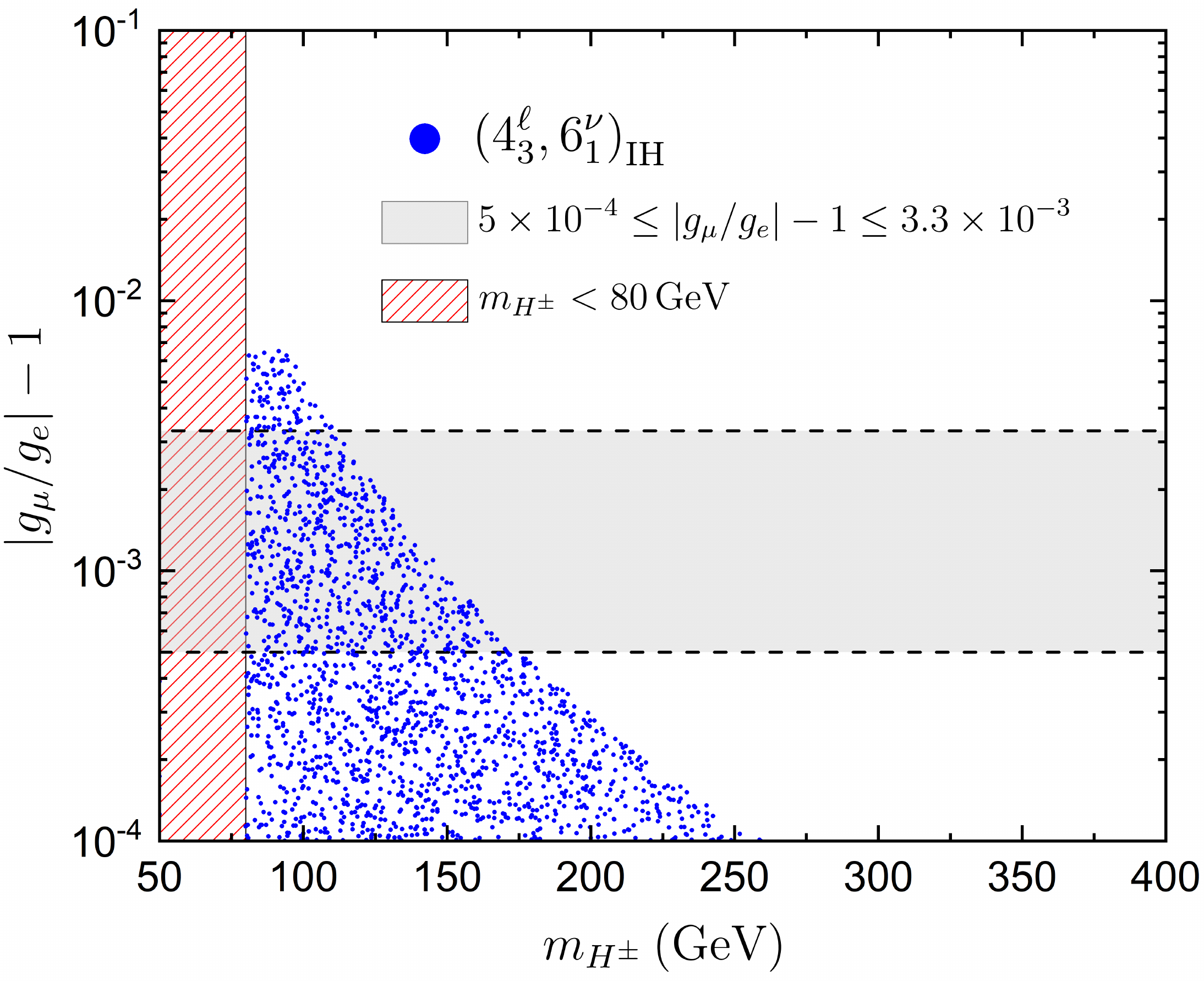} &
\includegraphics[width=0.23\textwidth]{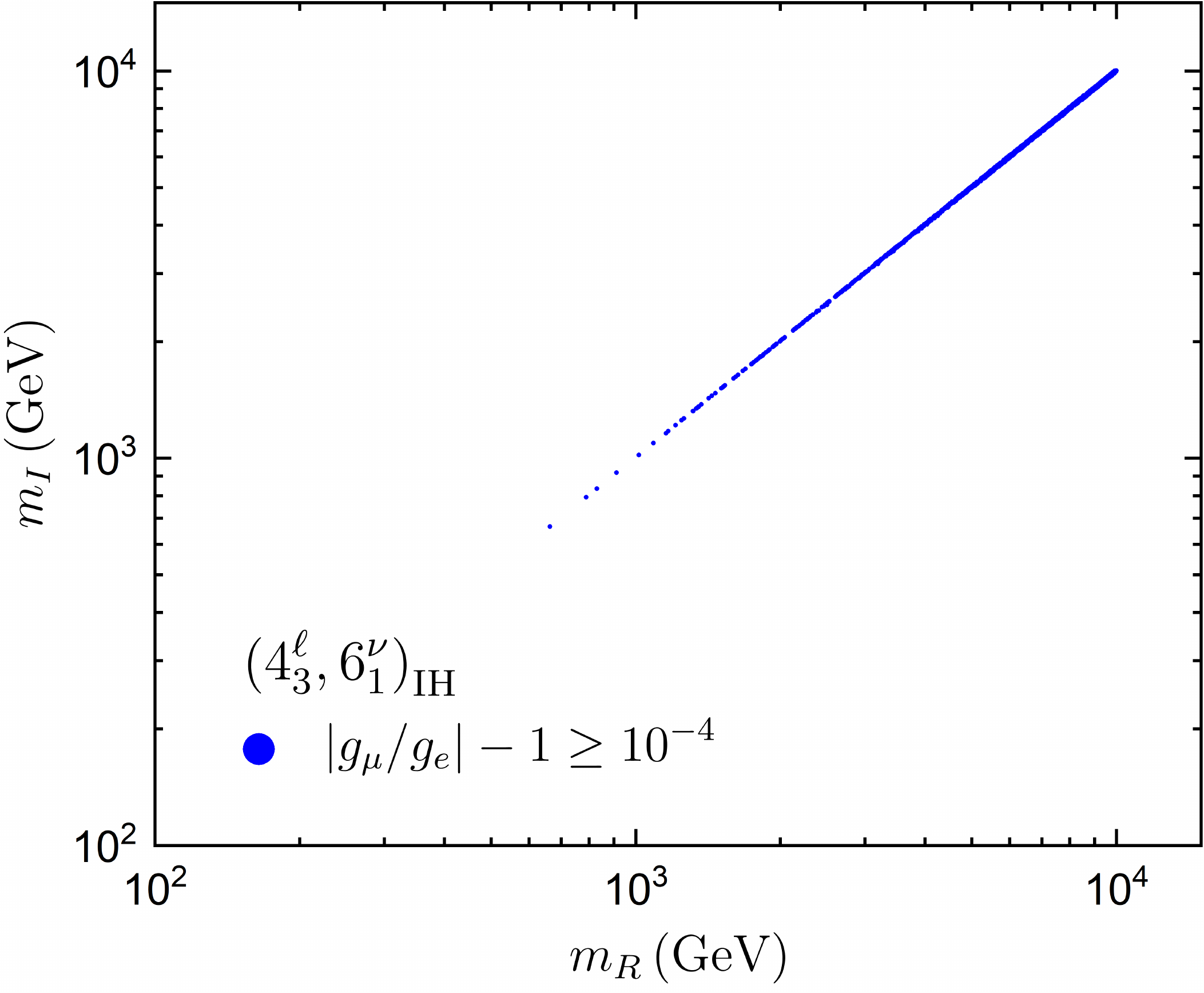}\\
\includegraphics[width=0.24\textwidth]{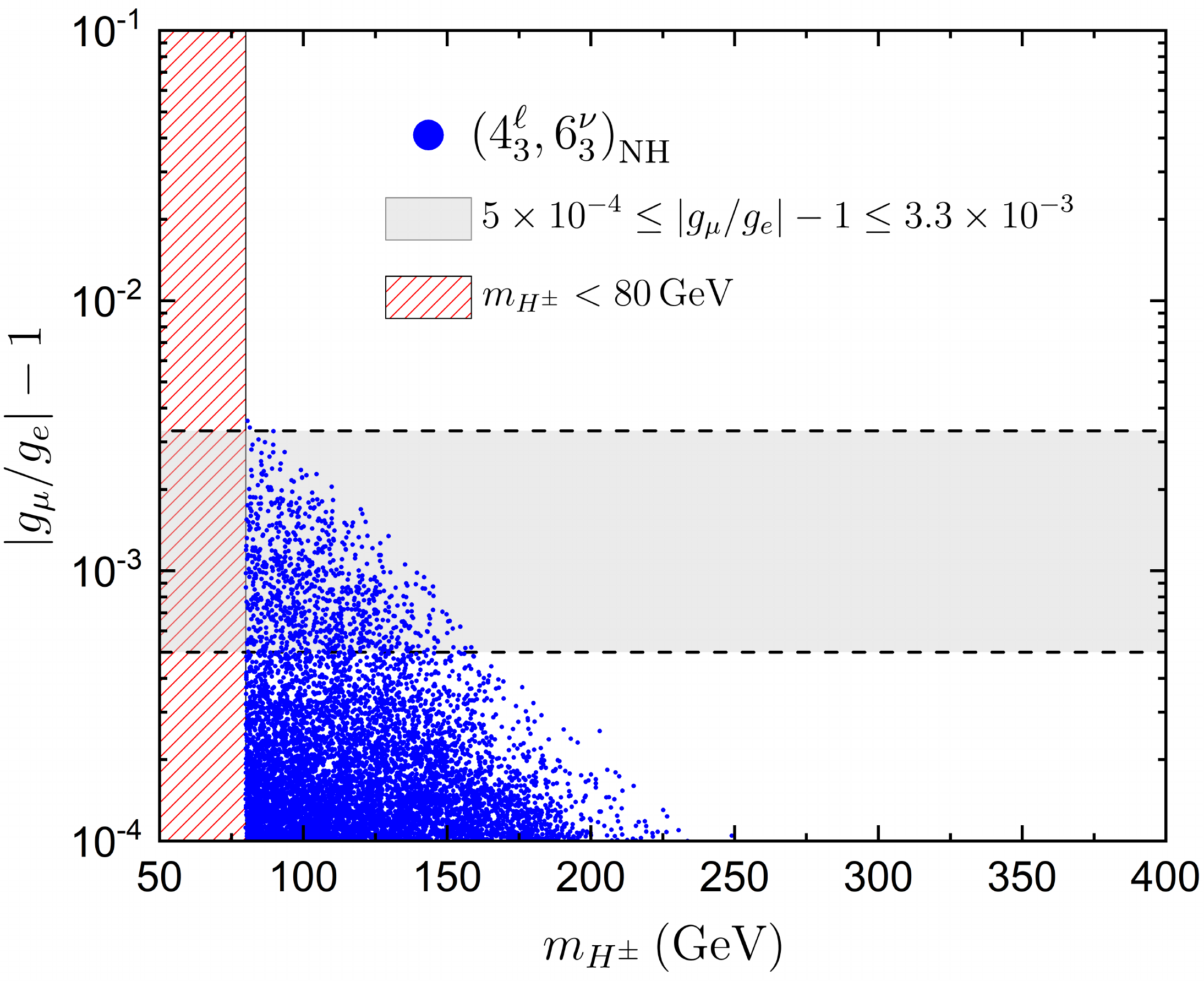} &
\includegraphics[width=0.23\textwidth]{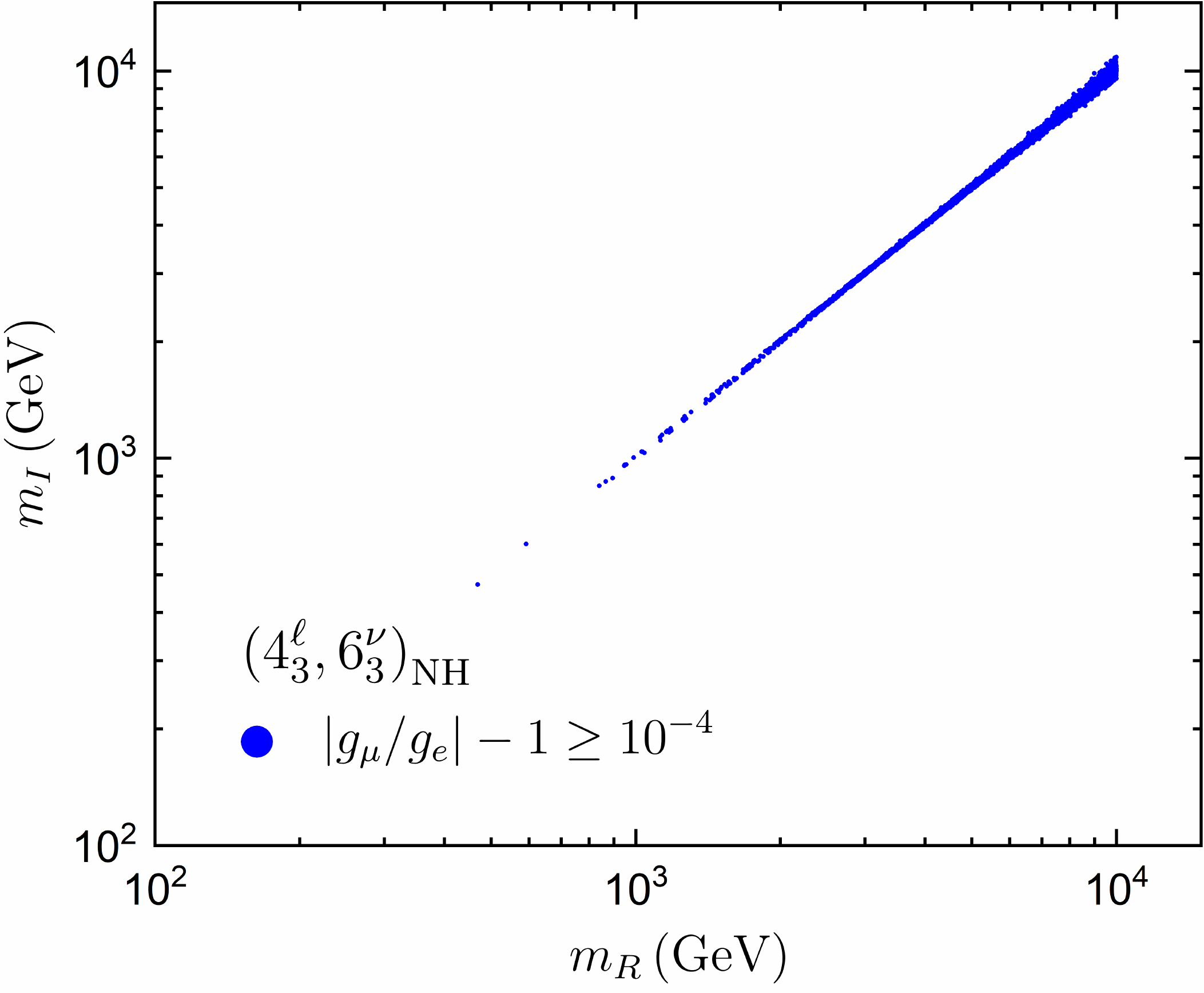} &
\includegraphics[width=0.24\textwidth]{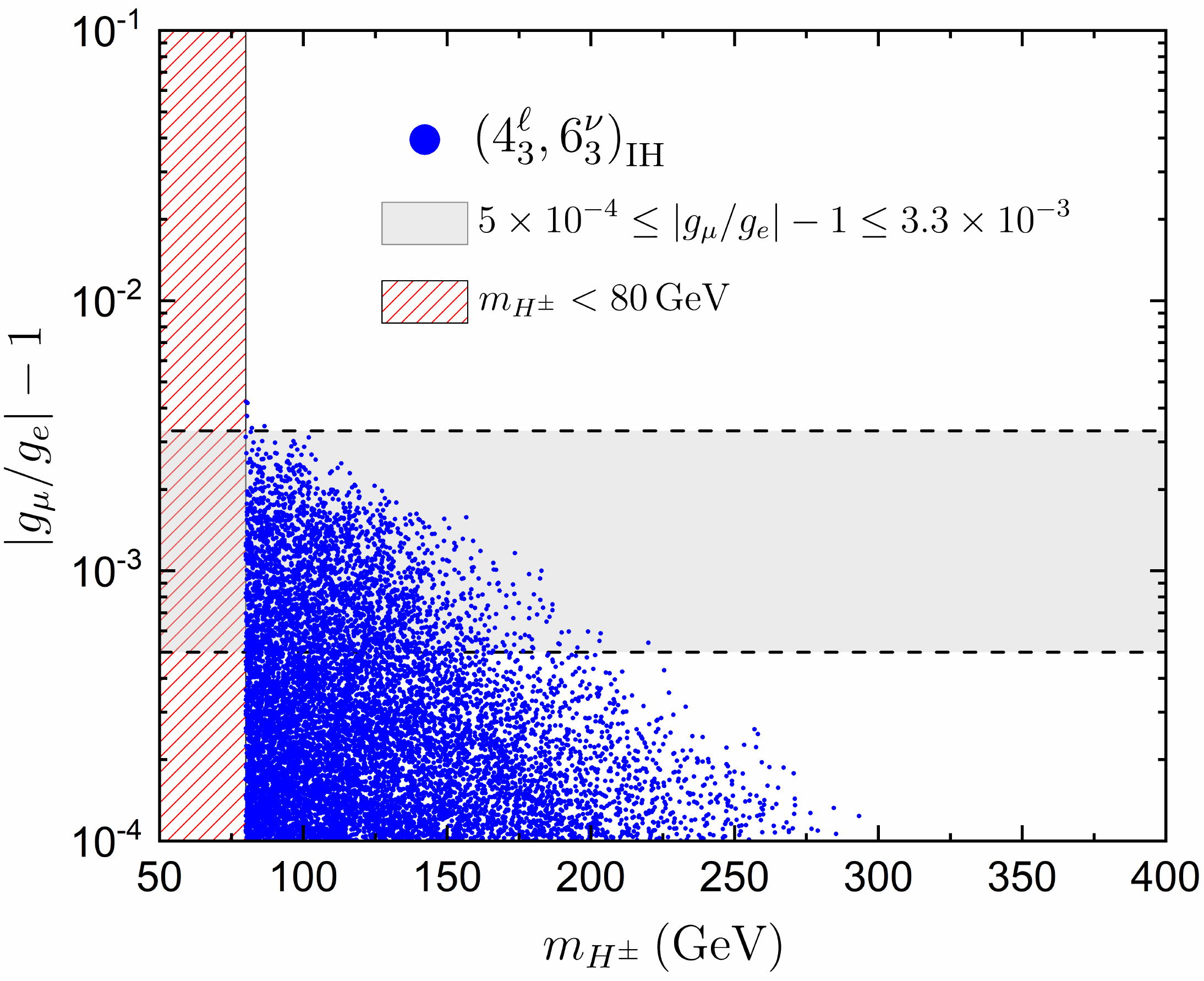} &
\includegraphics[width=0.23\textwidth]{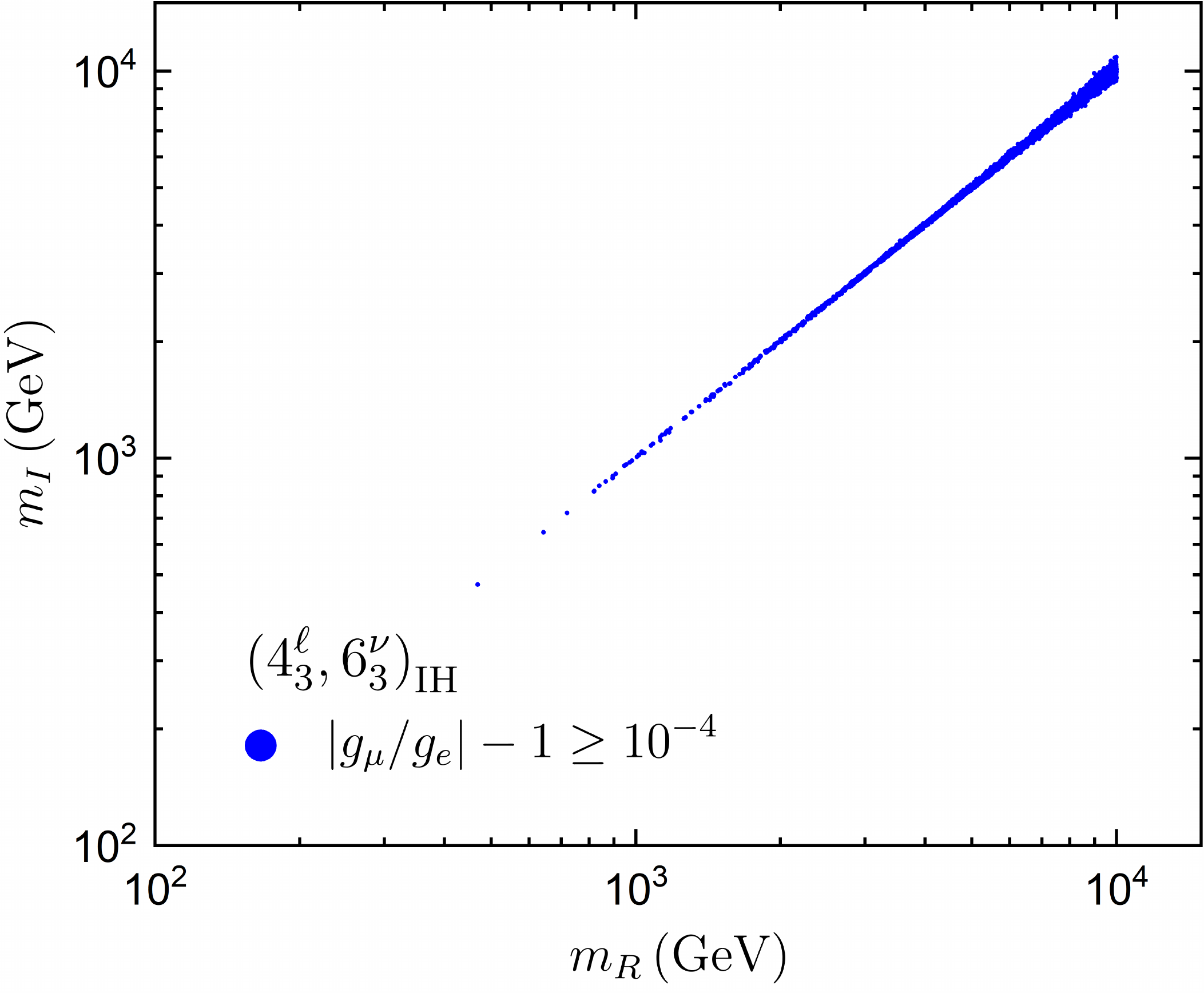}\\
\includegraphics[width=0.24\textwidth]{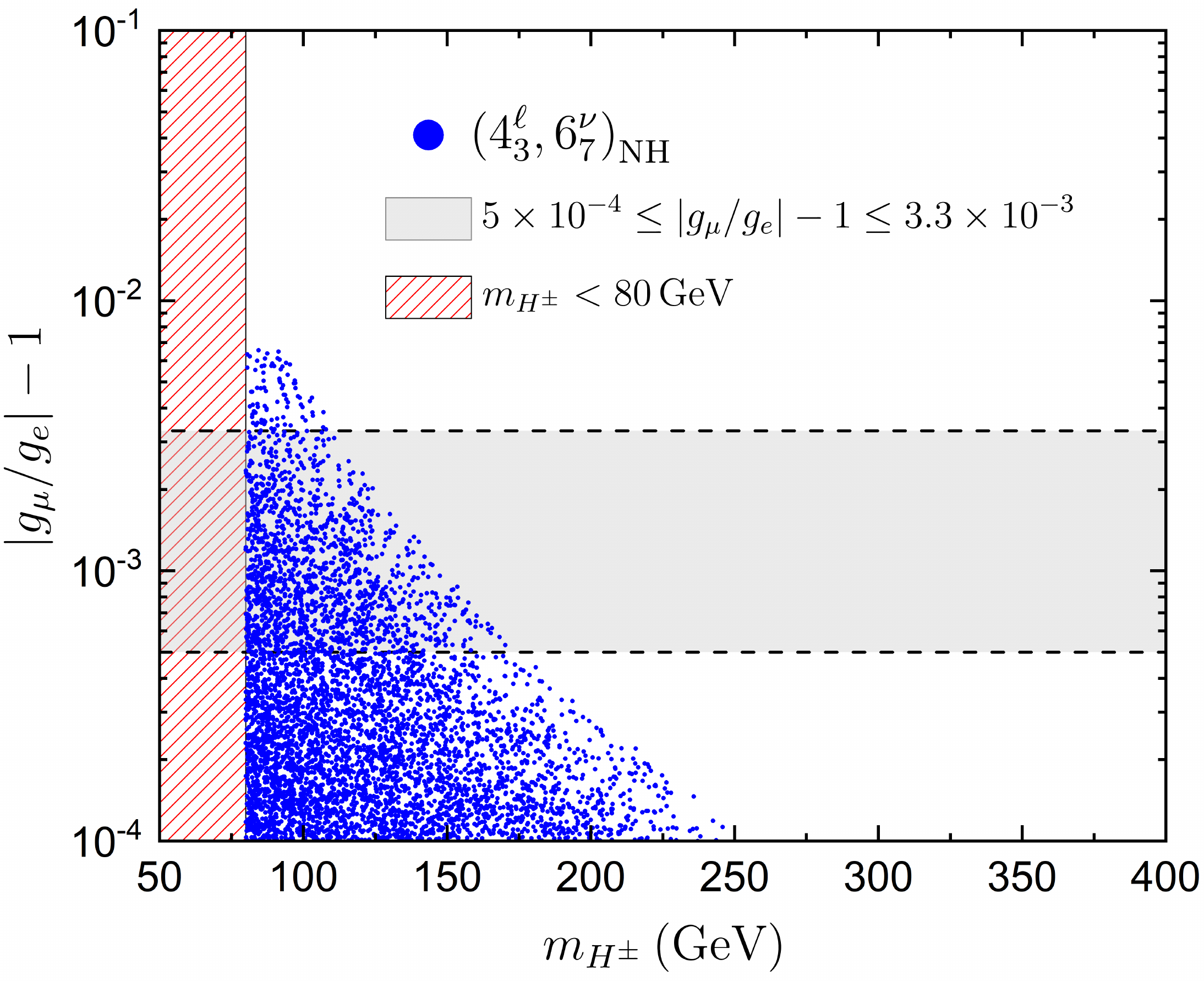} &
\includegraphics[width=0.23\textwidth]{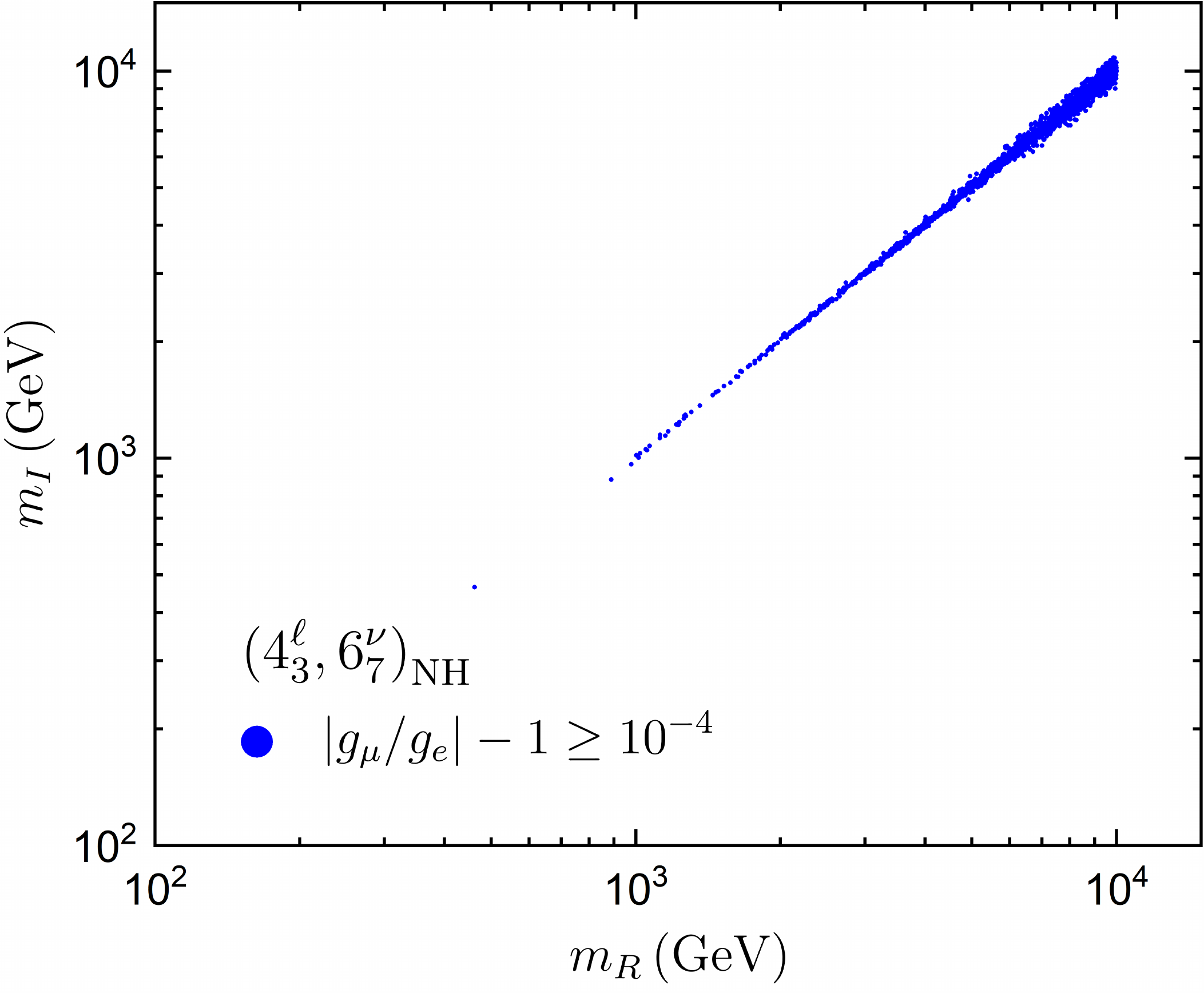} &
\includegraphics[width=0.24\textwidth]{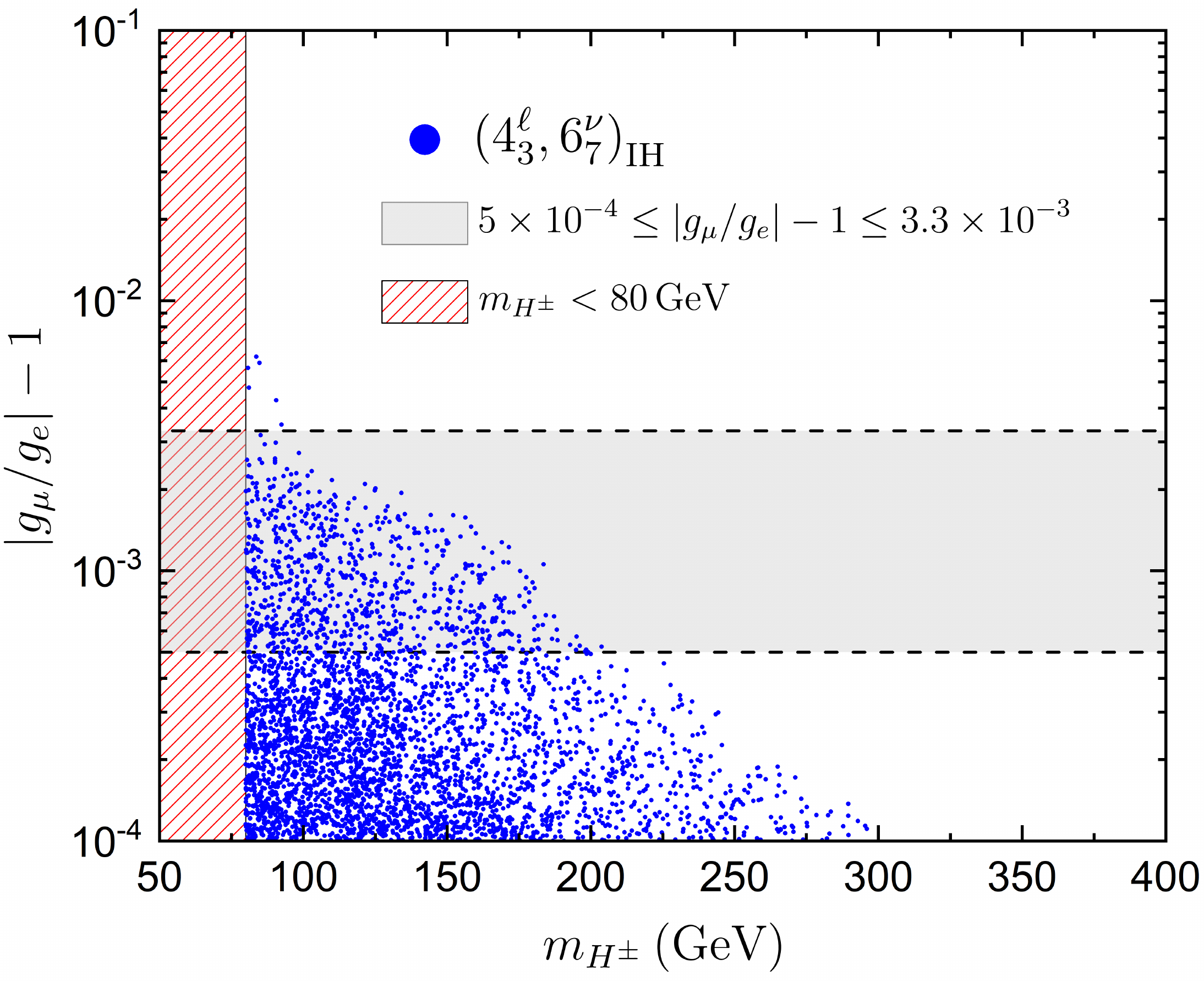} &
\includegraphics[width=0.23\textwidth]{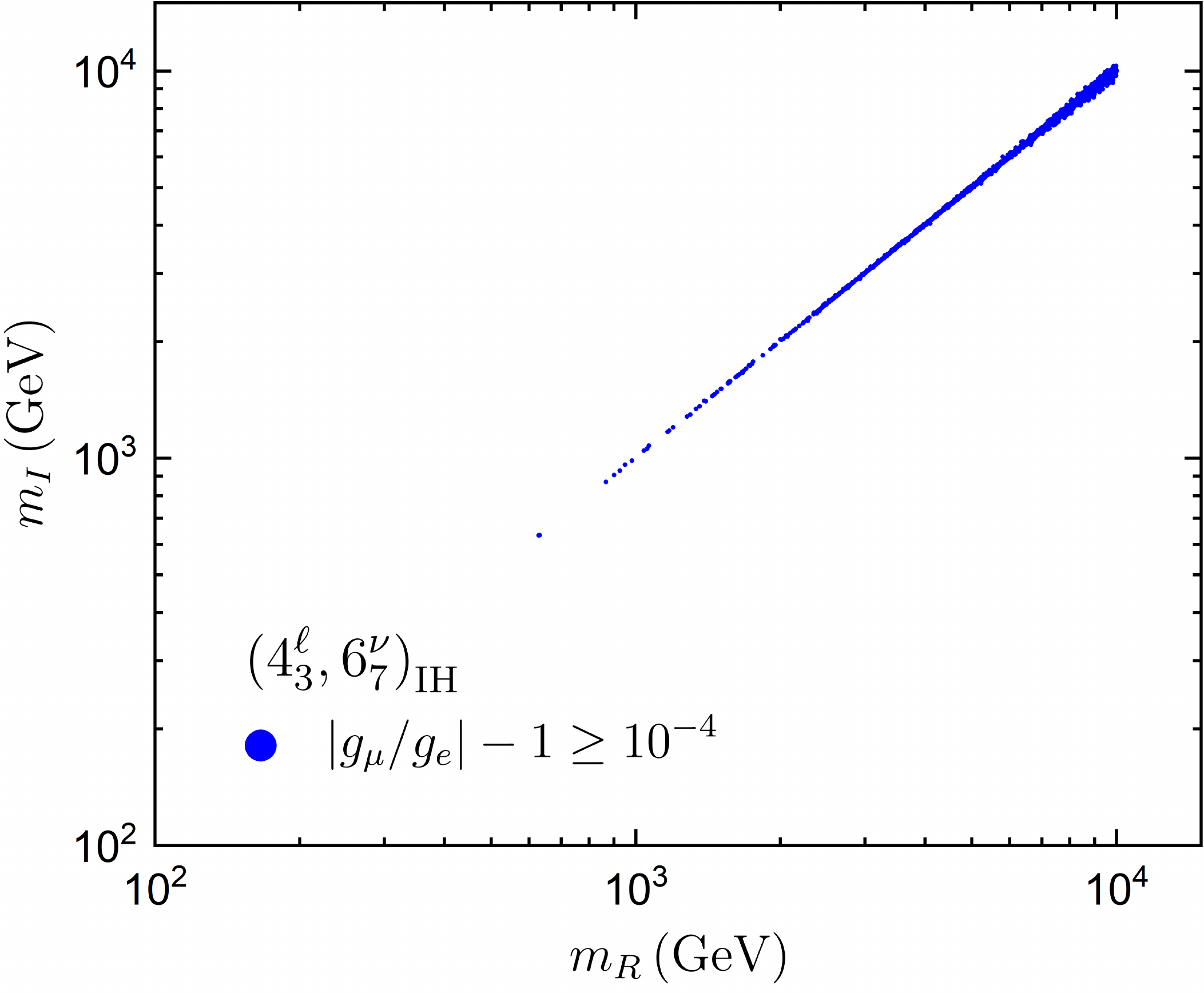}\\
\includegraphics[width=0.24\textwidth]{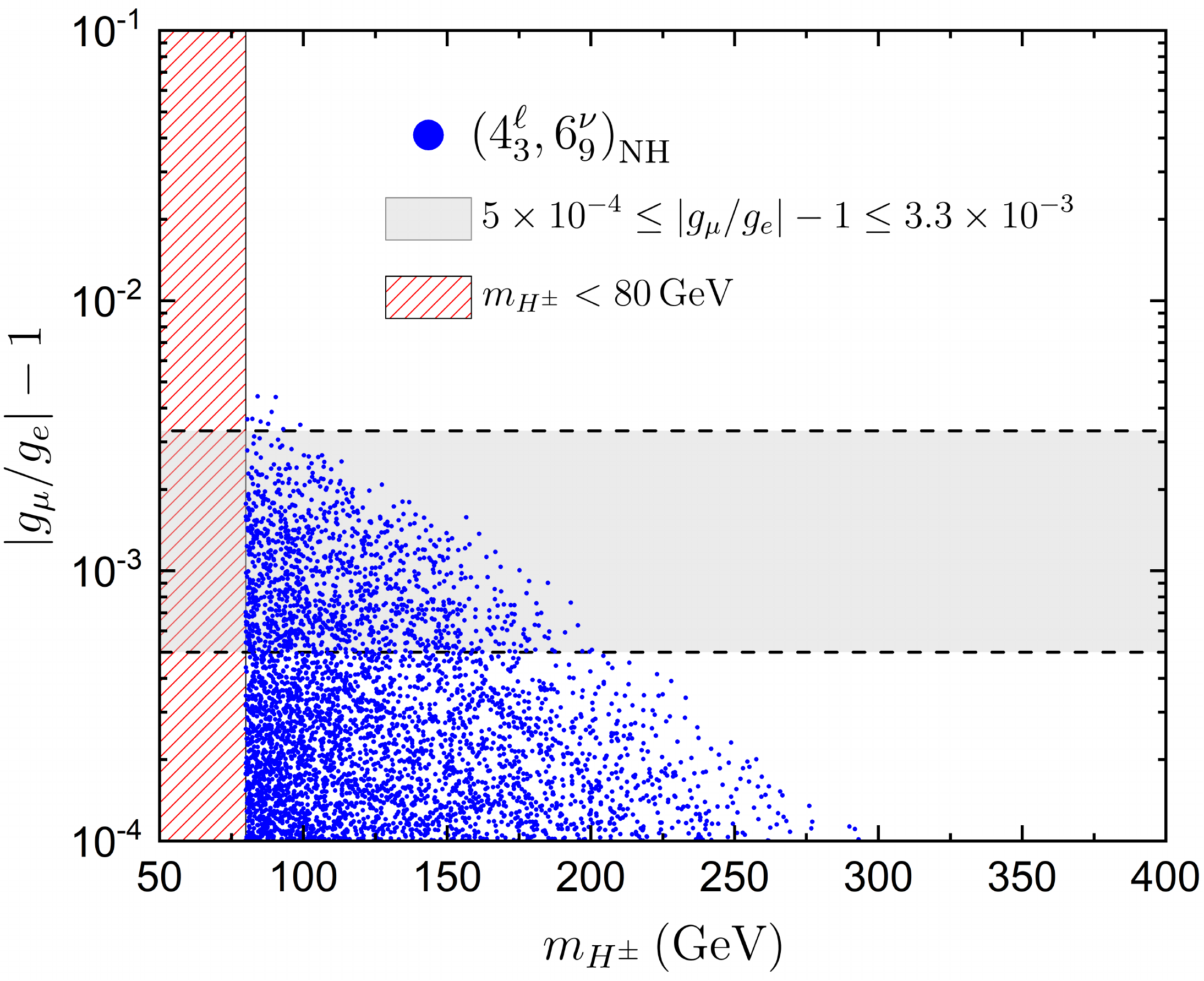} &
\includegraphics[width=0.23\textwidth]{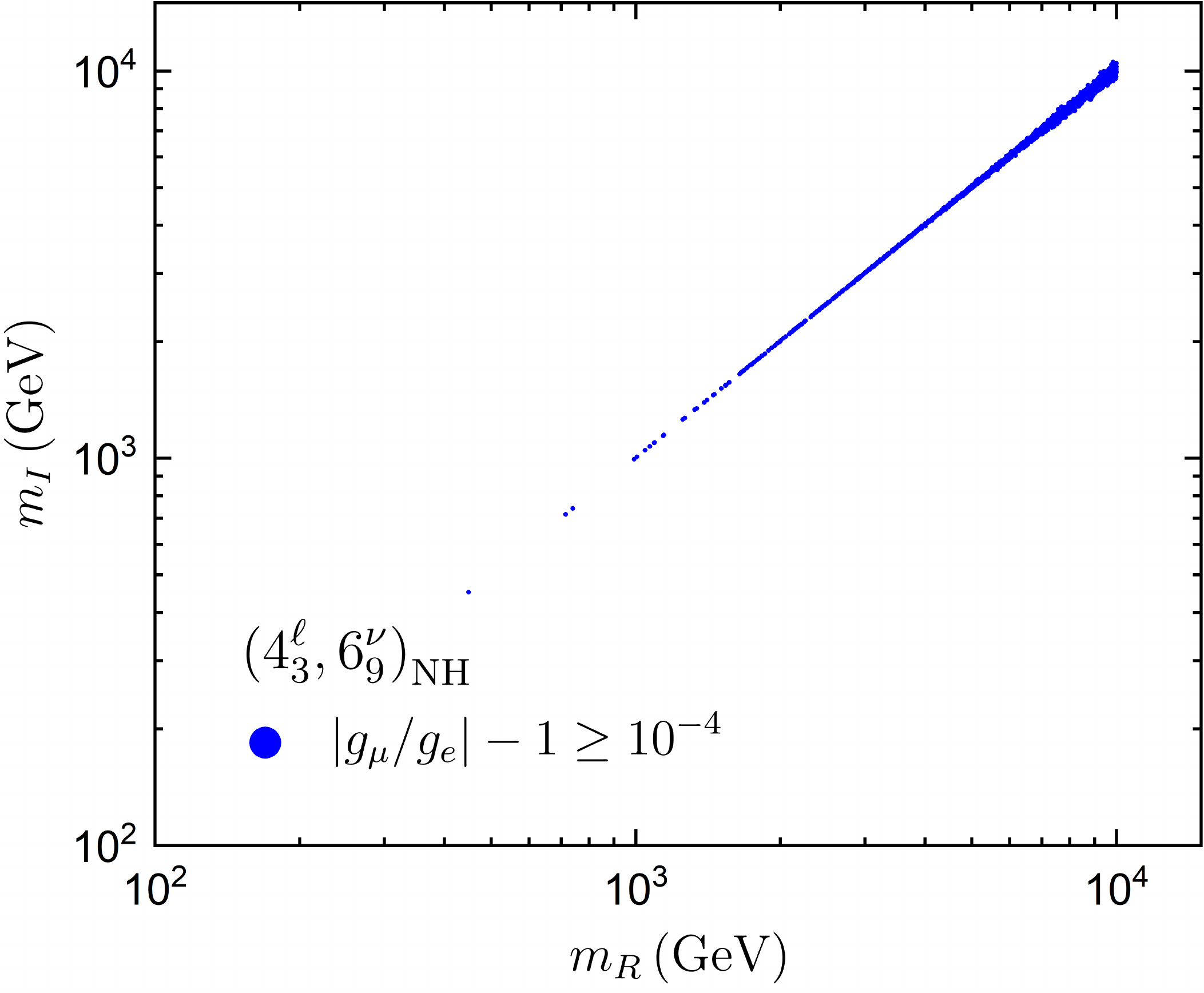} &
\includegraphics[width=0.24\textwidth]{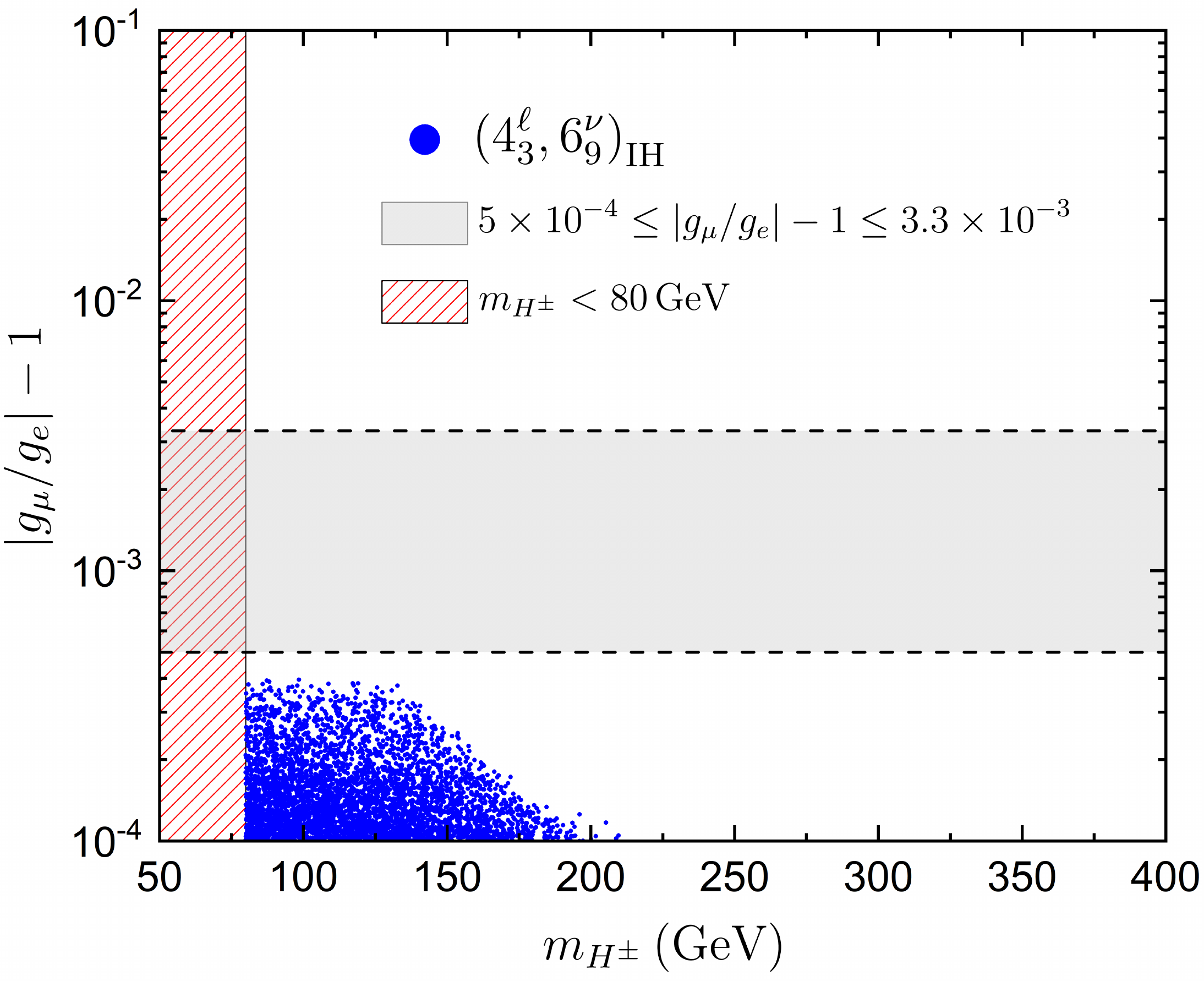} &
\includegraphics[width=0.23\textwidth]{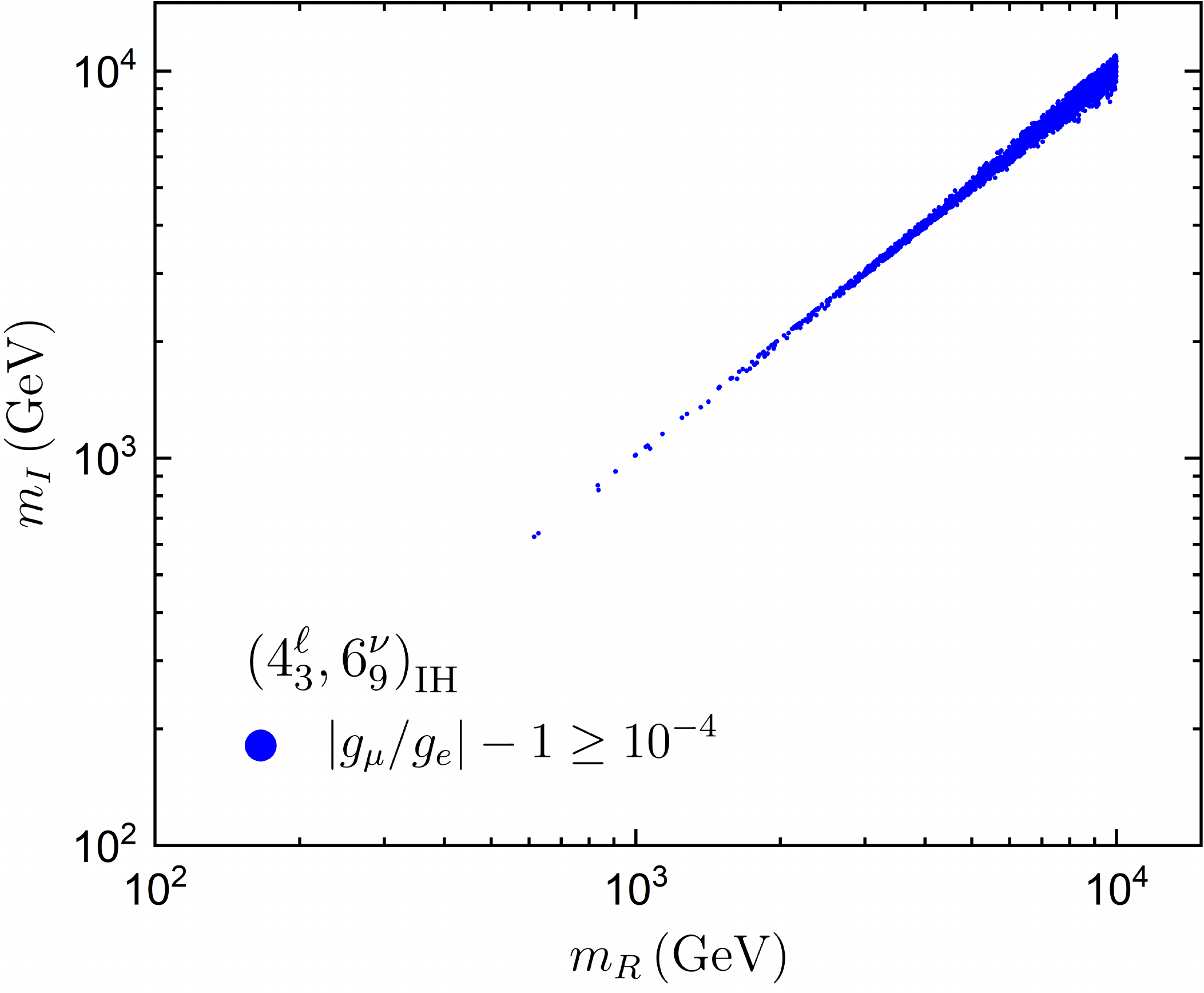}
\end{tabular}
\caption{Results for the $(4_3^\ell,6_k^\nu)$ for $k=1,3,7,9$ (from top to bottom rows). In the first (third) column we plot $|g_\mu/g_e|-1$ as a function of $m_{H^\pm}$ for NH (IH). The horizontal grey bands correspond to the experimental constraint~\eqref{eq:expuniv1}. The same points as in the corresponding $|g_\mu/g_e|-1$ plots are shown in the $(m_R,m_I)$-plane for NH (IH) in the second (forth) column. In all points the constraints \eqref{eq:expltolll} and \eqref{eq:expltolgamma} are satisfied, and the mixing angles $\theta_{ij}$ lie within the $3\sigma$ ranges given in Table~\ref{datatable} for NH or IH.}
\label{fig4}
\end{figure*}

In Fig.~\ref{fig3} we show the results for the $(5_1^e,5_8^\nu)$ and $(5_1^{\mu,\tau},5_8^\nu)$ texture pairs in the two upper and lower panels, respectively. In the left (middle) column we plot $|g_\mu/g_e|-1$ as a function of $m_{H^\pm}$ ($\tan\beta$), while in the right column the same points are shown in the $(m_R,m_I)$-plane. We conclude that the $(5_1^{\mu,\tau},5_8^\nu)$ cases are disfavored by the $|g_\mu/g_e|-1$ constraint \eqref{eq:expuniv1} (indicated by the horizontal gray bands in the plots). Instead, for $(5_1^e,5_8^\nu)$ the deviation from universality is in agreement with Eq.~\eqref{eq:expuniv1} for $80\,{\rm GeV} \lesssim m_{H^\pm} \lesssim 200~{\rm GeV}$ and $\tan\beta \lesssim 0.03$ or $\tan\beta \gtrsim 30$, for both NH and IH. Notice that for large (small) $\tan\beta$ the Yukawa couplings in $\Yl_1$ ($\Yl_2$) are enhanced, leading to an enhancement of $|g_\mu/g_e|-1$. 

Similar results are presented in Fig.~\ref{fig4} for the $(4_3^\ell,6_k^\nu)$ texture pairs given in Table~\ref{decompositions}. We do not present the results in terms of $\tan\beta$ since the behavior is similar to that of the $(5_1^{\ell},5_8^\nu)$ cases i.e., in general, there is a small and large $\tan\beta$ region. The main difference between the results in Figs.~\ref{fig3} and \ref{fig4} is evident from the comparison of the $(m_R,m_I)$ plots. While for the $(5_1^{\ell},5_8^\nu)$ texture pair all constraints are verified for non-correlated $m_{R,I}$ masses (see the left column plots in Fig.~\ref{fig3}), for the texture sets $(4_3^\ell,6_k^\nu)$ the mass tuning $m_R/m_I \simeq 1$ is needed to pass all the constraints, as shown in the second and forth column plots of Fig.~\ref{fig4}. This is easy to understand. First we notice that the matrix $\N_{e}$ defined in Eq.~\eqref{Nedef} has the following forms for the $5_1^{\ell}$ and $4_3^\ell$ textures:
\begin{eqnarray}
5_1^{e}:\;&\N_{e}\sim\begin{pmatrix}
\times &0 &0\\
0 &\times &\times\\
0 &\times & \times
\end{pmatrix},\; 
5_1^{\mu}:\;&\N_{e}\sim\begin{pmatrix}
\times &0 &\times\\
0 &\times &0\\
\times &0 & \times
\end{pmatrix},\\
5_1^{\tau}:\;&\N_{e}\sim\begin{pmatrix}
\times &\times &0\\
\times &\times &0\\
0 &0 & \times
\end{pmatrix},\;
4_3^{\ell}:\;&\N_{e}\sim\begin{pmatrix}
\times &\times &\times\\
\times &\times &\times\\
\times &\times & \times
\end{pmatrix}.
\label{Ne5158}
\end{eqnarray}
Ultimately, for a certain value of $\tan\beta$, the non-zero entries marked with a $\times$ could be expressed in terms of the charged-lepton and neutrino masses and lepton mixing angles, as illustrated for the case of the $5_1^\ell$ texture discussed in the previous section. 
\begin{table}[t]
\centering
\begin{tabular}{l@{\hskip 0.3in}c@{\hskip 0.1in}c@{\hskip 0.1in}c}
\hline \hline\\[-0.2cm]
Decay  &$5_1^e$ & $5_1^\mu$ & $5_1^\tau$ \\[0.1cm]\hline\\[-0.2cm]
$\ell_\alpha \rightarrow \ell_\beta \gamma$   & $(\tau,\mu)$ & $(\tau,e)$ & $(\mu,e)$ \\[0.2cm]
$\ell_\alpha^-\to \ell_{\beta}^-\,\ell_{\gamma}^+\,\ell_{\delta}^-$& $(\tau,\mu\mu\mu)$ & $(\tau,eee)$ & $(\mu,eee)$ \\[0.2cm]
& $(\tau,ee\mu)$ & $(\tau,\mu\mu e)$ &  \\[0.2cm]

\hline \hline
\end{tabular}
\caption{Allowed $\ell_\alpha \rightarrow \ell_\beta \gamma$ and $\ell_\alpha \rightarrow \ell_\beta \gamma$ for $5_{e,\mu,\tau}$, indicated in each case by particle flavor indices $(\alpha,\beta)$ and $(\alpha,\beta\gamma\delta)$.}
\label{tableproc}
\end{table}

Taking into account Eqs.~\eqref{3body1}, \eqref{3body2}, \eqref{2body1} and \eqref{eq:AR}, we can immediately conclude that most of the $\ell_\alpha^-\to \ell_{\beta}^-\,\ell_{\gamma}^+\,\ell_{\delta}^-$ and $\ell_\alpha \to \ell_\beta\, \gamma$ are forbidden at the one loop level for the $5_1^\ell$ textures. This is due to the coupling structure imposed by the U(1) flavor symmetry which, in the case of charged leptons, only allows mixing between two flavors. For instance, the decay $\mu \rightarrow e\gamma$ only occurs when $\Ml \sim 5_1^\tau$, since for $5_1^e$ ($5_1^\mu$) the electron (muon) is decoupled and, thus, $\mu-e$ transitions are not allowed. On the other hand, $\tau$ radiative decays are forbidden in that case since the $\tau$ is decoupled. Applying the same reasoning to the 3-body decays $\ell_\alpha^-\to \ell_{\beta}^-\,\ell_{\gamma}^+\,\ell_{\delta}^-$, we conclude that, at most, only two of these processes are allowed for each of the $5_1^\ell$ case (see Table~\ref{tableproc}). Thus, the stringent constraints coming from the $\mu$ decays, are naturally satisfied in the $5_1^{e,\mu}$ case, while for the $5_1^{\tau}$ texture the rates are suppressed by the small couplings in the $\mu-e$ sector. Summarizing, for $5_1^{\ell}$ the constraints from LFV decays are respected without requiring any special relation among the scalar masses $m_{R,I}$, as can be seen from the plots in Fig.~\ref{fig3}.
\begin{figure}[t]
\centering
\includegraphics[width=0.48\textwidth]{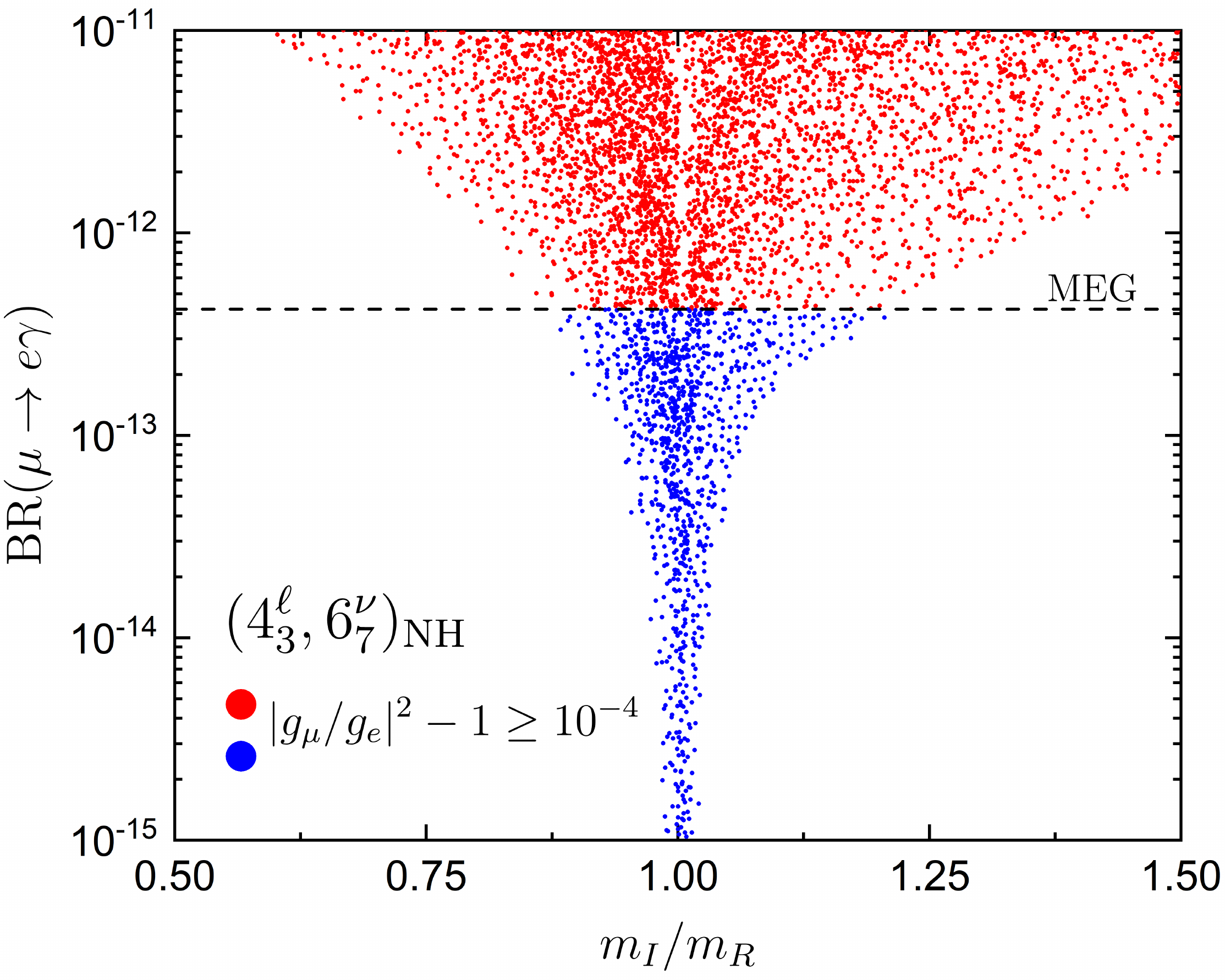} 
\caption{Dependence of $\text{Br}(\mu \to e\gamma)$ on the mass ratio $m_I/m_R$. Although the results are shown for the $(4_3^\ell,6_7^\nu)$ texture pair, for the remaining $(4_3^\ell,6_k^\nu)$ cases shown in Table~\ref{decompositions} the results are similar. In all points the mixing angles $\theta_{ij}$ lie within the $3\sigma$ ranges given in Table~\ref{datatable} and $|g_\mu/g_e|-1 \ge 10^{-4}$. The red points are excluded by the $\mu\rightarrow e \gamma$ MEG bound given in \eqref{mugamma}.}
\label{fig5}
\end{figure}

The natural suppression of LFV decays does not however occur when $\Ml \sim 4_3^\ell$. As can be seen from \eqref{Ne5158}, in these cases the couplings $\N_e$ do not exhibit any decoupling behavior and, thus, the decay rates are not naturally suppressed. In the particular case of $\mu\rightarrow e\gamma$, the terms enhanced by $m_\tau/m_\mu$ are potentially large and the experimental bound on that decay is respected only when there is a cancellation between the two terms proportional to $m_\tau/m_\mu$ in $\mathcal{A}_R$, i.e. when $m_R \simeq m_I$. This is reflected in the $(m_R,m_I)$ plots of Fig.~\ref{fig4}, where that correlation is clear. Notice that in the $5_1^\ell$ case those terms were absent since $(\N_{e})_{\mu\tau}(\N_{e})_{\tau e}=0$, which is not the case for the $4_3^\ell$ textures. For illustration, we show in Fig.~\ref{fig5} the dependence of ${\text Br}(\mu\rightarrow e \gamma)$ on the mass ratio $m_I/m_R$ for the texture pair $(4_3^\ell,6_7^\nu)$, which confirms the fact that quasi-degenerate $m_{R,I}$ masses are required to respect the MEG $\mu\rightarrow e \gamma$ bound. Notice also that, in all cases, sizable deviations from lepton universality require $m_{H^\pm} \lesssim 200-300~{\rm GeV}$.

\section{Conclusions}
\label{Conclusions}

At present, no tangible experimental evidence exists in favor of Dirac or Majorana massive neutrinos. In pure theoretical grounds, both scenarios can be implemented in a natural way from the point of view of an effective SM in which neutrino masses are suppressed by a large scale. In the case of Dirac neutrinos, once RH neutrino singlets are added, their direct couplings with the LH lepton and the SM Higgs doublets must be forbidden by a symmetry. In this way, the window is open for seesaw-like suppression of Dirac neutrino masses. This principle can be extended to the 2HDM to suppress the effective Dirac couplings to both Higgs doublets. The presence of the second Higgs doublet opens the possibility of implementing texture-zero structures in the charged-lepton and neutrino mass matrices compatible with data. In this work, we have addressed this problem by considering the maximally-restricted scenarios that can be implemented by imposing U(1) or $\mathbb{Z}_N$ Abelian symmetries in the 2HDM Lagrangian. Our approach differs from usual BGL symmetries in the sense that the number of relevant flavor parameters in the Lagrangian is the same as the number of lepton masses, mixing angles and CP phases. We stress that these are the most restrictive symmetries in the sense that the number of parameters cannot be further reduced.

Our analysis shows that from the 28 initial pairs $(\Ml,\Mnu)$ compatible with data, only 5 can be implemented in the 2HDM by imposing a U(1) flavor symmetry. Clearly, this does not preclude the possibility of implementing the remaining pairs in models with more than two Higgs doublets. For the realizable pairs we established the relation between the only complex phase parameter $\alpha$ in the Lagrangian and the Dirac CP-violating phase $\delta$ to which neutrino oscillation experiments are sensitive. In particular, for the most interesting case of $\Ml \sim 5_1^e$, we have $\delta \simeq -\alpha$. Imposing the constraints coming from lepton universality in $\tau$ decays and $\ell_\beta \rightarrow \ell_\alpha \gamma$ and $\ell_\alpha^-\to \ell_{\beta}^-\,\ell_{\gamma}^+\,\ell_{\delta}^-$ searches, we have shown that $5_1^e$ is the only case which does not require a tuning among the masses of the $R$ and $I$ scalars, due to a flavor suppression in the $\mu-e$ channel stemming from the U(1) flavor symmetry. The lepton universality constraint coming from the ratio $|g_\mu/g_e|$ selects $m_{H^+}$ to be at most $300$~GeV, with small and large $\tan\beta$ values. Nevertheless, it is worth emphasizing that these limits are loosen and wider intervals for $m_{H^+}$ and $\tan\beta$ are allowed if the universality constraint is relaxed. 

Among all the cases considered, only the pair $(5_1^e,5_8^\nu)$ leads to naturally small rates for LFV decays (in particular $\mu \rightarrow e \gamma$) without the need of cancellations between the amplitudes of $R$ and $I$ mediated contributions, only possible if $m_I \simeq m_R$. Such a condition among the scalar masses, together with a light $H^+$, does not arise naturally in the 2HDM. In fact, $m_I \simeq m_R$ can be easily accomplished in the 2HDM decoupling limit but, in general, $H^+$ is degenerate with $R$ and $I$. Therefore, all texture combinations of Table~\ref{decompositions} would be phenomenologically viable if one drops the requirement of having a non-negligible deviation from universality in $\tau$ decays. 

In conclusion, we have shown that maximally-restrictive lepton mass matrices realizable in the 2HDM with U(1) Abelian symmetries are phenomenologically viable from the point of view of lepton masses and mixing and the constraints imposed on flavour-changing processes. The results obtained in this work pose the natural question of what happens if we apply the same principle of maximally-restrictive textures to the quark sector. In this case, besides the symmetry implementation and compatibility with the observed quark masses and mixing, more severe constraints have to be checked, such as those coming from universality tests in $\tau$ and meson semileptonic decays, and from $B\rightarrow X_s \gamma$ and meson $\mu^+\mu^-$ decays. The extension of the analysis presented here to the quark sector is under preparation~\cite{future}, as well as the generalization to the case of seesaw-generated Majorana neutrino masses.

\vspace*{0.5cm}
{\bf Acknowledgements:} We thank M. Nebot and H. Ser\^{o}dio for discussions. F.R.J. thanks the CERN Theory Department for hospitality and financial support. This work was partially supported by Funda\c{c}{\~a}o para a Ci{\^e}ncia e a Tecnologia (FCT, Portugal) through the projects CFTP-FCT Unit 777 (UID/FIS/00777/2019), CERN/FIS-PAR/0004/2017 and PTDC/FIS-PAR/29436/2017, which are partly funded through POCTI (FEDER), COMPETE, QREN and EU.

\appendix
\section{Abelian symmetries in the SM with Dirac neutrinos}
\label{appdxA0}

To further motivate our analysis of texture-zero Yukawa and mass matrices in the 2HDM, let us analyze the consequences of imposing a U(1) symmetry in the SM such that the matrix $\Mx$ contains texture zeros. First we note that, in the SM, Eq.~\eqref{yukawainvariance} takes the form
\begin{equation}
\label{yukawainvariancesm}
\begin{aligned}
\Yl = \mathbf{S}_\ell \Yl \mathbf{S}_e^\dagger,\quad
\Ynu = \mathbf{S}_\ell\Ynu \mathbf{S}_\nu^\dagger ,
\end{aligned}
\end{equation}
where we have set the SM Higgs U(1) phase to zero. Unlike in the 2HDM, the lepton mass matrices will be proportional to a single Yukawa matrix, leading to the following transformation properties of the Hermitian matrices $\Hl$ and $\Hnu$ defined in Eq.~\eqref{Hdef}:
\begin{equation}
\label{hinvariance}
\begin{aligned}
\Hl = \mathbf{S}_\ell\Hl \mathbf{S}_\ell^\dagger,\quad
\Hnu = \mathbf{S}_\ell \Hnu \mathbf{S}_\ell^\dagger\,.
\end{aligned}
\end{equation}
Using Eq.~\eqref{Udef2}, and defining the unitary matrices $\mathbf{A}$ and $\mathbf{B}$ as
\begin{equation}
\label{AB}
\begin{aligned}
\mathbf{A} \equiv {\ULld} \mathbf{S}_\ell \ULl\,,\quad
\mathbf{B} \equiv {\ULnud} \mathbf{S}_\ell \ULnu\,,
\end{aligned}
\end{equation}
one obtains that the lepton mixing matrix $\U$ must satisfy the relation
\begin{equation}
\begin{aligned}
\U &= \ULld\mathbf{S}_\ell (\ULl \ULld) (\ULnu \ULnud) \mathbf{S}_\ell^\dagger \ULnu\\
&= \mathbf{A} \ULld \ULnu \mathbf{B}^\dagger = \mathbf{A} \U \mathbf{B}^\dagger\,.
\end{aligned}
\label{ABU}
\end{equation}
On the other hand, the invariance condition \eqref{hinvariance} reads
\begin{equation}
\label{commutation}
\left[\mathbf{A}, \Dl^2 \right] = 0, \quad
\left[\mathbf{B} , \Dnu^2 \right] = 0\,,
\end{equation}
which, in the case of non-degenerate charged-lepton and neutrino masses (as required by experiment), imply
\begin{equation}
\label{AB2}
\begin{aligned}
\mathbf{A} &= \text{diag}\left(e^{i A_1},e^{i A_2},e^{i A_3}\right),\\
\mathbf{B} &= \text{diag}\left(e^{i B_1},e^{i B_2},e^{i B_3}\right),
\end{aligned}
\end{equation}
where $A_i$ and $B_i$ are general phases. Together with Eq.~\eqref{ABU}, this leads to the relations
\begin{align}
\U_{ij} = e^{i (A_i - B_j)}\, \U_{ij}.
\label{pmnszeros}
\end{align}
Current experimental data (see Table~\ref{datatable}) indicate that all elements of $\U$ are nonzero. Therefore, compatibility of $\U$ with data forces the solution of the above equation to be
\begin{align}
\label{PMNSinvariance}
A_i - B_j = 0 \quad (\text{mod} ~ 2\pi), ~ \forall ~ i,j,
\end{align}
and, consequently,
\begin{equation}
\label{degeneracy}
\begin{aligned}
A_i - A_j = 0,\quad B_i - B_j = 0 \quad (\text{mod} ~ 2\pi), ~ \forall ~ i,j.
\end{aligned}
\end{equation}

Equation~\eqref{AB} can be interpreted as the diagonalization of $\mathbf{S}_\ell$. However, for fully degenerate eigenvalues, as demanded by Eq.~\eqref{degeneracy}, $\mathbf{S}_\ell$ has a unique diagonal form, up to a rephasing of the columns. This leads to the lepton mixing matrix $\U = \text{diag}\{e^{i \chi_1},e^{i \chi_2},e^{i \chi_3}\}$, which is obviously incompatible with the experimental observations.

On the other hand, one could require that $\mathbf{S}_\ell$ commutes with both $\ULl$ and $\ULnu$. Then  $\mathbf{A} = \mathbf{B} = \mathbf{S}_\ell$, implying that the charges $A_i=B_i$ are in direct correspondence with the charges of $\mathbf{S}_\ell$. Equation~\eqref{PMNSinvariance} would then yield
\begin{align}
\label{fullydegenerate}
\mathbf{S}_\ell = e^{i A_1}\text{diag}(1,1,1)\quad (\text{mod} ~ 2\pi),
\end{align}
which is the only transformation that allows for a viable mixing matrix $\U$ in the context of the SM~\cite{Low:2003dz}.

One can always set $A_1$ to zero so that $\mathbf{S}_\ell = \openone$,  $\mathbf{S}_e = \text{diag}(e^{i \beta_1},e^{i \beta_2},e^{i \beta_3})$, and $\mathbf{S}_\nu = \text{diag}(e^{i \gamma_1},e^{i \gamma_2},e^{i \gamma_3})$. The phase transformation matrices would now be
\begin{gather}
\label{chargphasesSM}
\Theta^{\ell} = \begin{pmatrix} \beta_1&\beta_2&\beta_3\\ \beta_1&\beta_2&\beta_3 \\ \beta_1&\beta_2&\beta_3 \end{pmatrix},~~~
\Theta^{\nu} = \begin{pmatrix} \gamma_1&\gamma_2&\gamma_3\\ \gamma_1&\gamma_2&\gamma_3 \\ \gamma_1&\gamma_2&\gamma_3 \end{pmatrix}\,,
\end{gather}
showing that, in the SM, imposing a single texture zero in $\Ml$ or $\Mnu$ and requiring $\U$ to be compatible with experiment implies an entire column of zeros, resulting in a massless particle. This is clearly not acceptable for charged leptons. However, for neutrinos, such a possibility is not excluded, since a massless neutrino is compatible with current data. In conclusion, in the context of the SM, one cannot impose texture zeros in the charged-lepton mass matrix, while maintaining compatibility with all experimental observations.\footnote{This is in agreement with the conclusions of Ref.~\cite{Low:2003dz}.} The most economical scenario compatible with data would be the one with textures
\begin{gather}
\label{MlMnuSMt}
\Ml = \begin{pmatrix} \times&\times&\times\\ \times&\times&\times \\ \times&\times&\times \end{pmatrix}\;,\;
\Mnu = \begin{pmatrix} 0&\times&\times\\ 0&\times&\times\\ 0&\times&\times \end{pmatrix}\!\mathcal{P}\,,
\end{gather}
where $\mathcal{P}$ is a $3\times 3$ permutation matrix (see Eq.~\eqref{Pmat}), and $\times$ denotes a general non-zero entry. In this case there are 23 independent parameters in the mass matrices, to be compared with 9 measurable quantities (5 masses, 3 mixing angles and 1 phase).

The above analysis can be trivially extended to the SM quark sector being the CKM matrix the analogue of $\U$. Since all quarks are known to be massive, neither the up nor the down quark mass matrix can have a zero eigenvalue. This means that, in the SM, no texture zeros can be imposed by Abelian flavor symmetries in the quark sector while ensuring compatibility with all experimental observations.

\section{Non-realizable texture pairs}
\label{appdxA}

In this appendix we describe in more detail some results of Section~\ref{AbComp} which were based on the application of the canonical method.

\subsection{Textures $4_1^{\ell},4_2^{\ell},4_{17}^{\nu}$} 
\label{4142417}

In the context of a 2HDM, a mass matrix texture with a row (column) full of non-zero entries can only be realized through an Abelian symmetry if it has at least two columns (rows) with an identical texture structure. Hereafter, we shall refer to such structures as identical columns (rows). To prove this statement, by means of the canonical method, let us show that a full texture row or column leads to
\begin{equation}
\beta_i-\beta_j = 0~(\text{mod}~2\pi), \quad \gamma_i - \gamma_j = 0~(\text{mod}~2\pi),\label{fullrow}
\end{equation}
or 
\begin{equation}
\alpha_i - \alpha_j = 0~(\text{mod}~2\pi),\label{fullcolumn}
\end{equation}
respectively, for some $i\neq j$. These relations, together with Eq.~\eqref{chargphases}, then lead to identical texture columns and rows, respectively.

Since $4_1^{\ell},4_2^{\ell},4_{17}^{\nu}$ have a row or a column full of non-zero entries, yet lack identical texture columns or rows, we conclude that they cannot be implemented through any continuous or discrete Abelian symmetry. 

One may wonder whether imposing consecutive symmetry transformations as those given in Eq.~\eqref{symtransf} can change the above conclusion. For instance, one could impose multiple transformations so that Eqs.~\eqref{fullrow} or~\eqref{fullcolumn} apply to different indices $i$ and $j$ in each one. Yet, the decomposition into Yukawa matrices, for the entries that remain in the final mass matrix, must be identical for all transformations. Equation~\eqref{chargphases} implies that, for $\alpha_i - \alpha_j = 0$, the entries in rows $i$ and $j$ of a full column are decomposed into $\mathbf{Y}_a$, while the remaining component into $\mathbf{Y}_{b\neq a}$. If $\alpha_1=\alpha_2=\alpha_3$ the 3 entries are decomposed into the same Yukawa matrix. Thus, imposing multiple symmetry transformations obeying either Eq.~\eqref{fullrow} or~\eqref{fullcolumn} for distinct indices $i,j$ leads to incompatible decompositions.

Symmetry transformations that allow for a full texture row or column in a mass matrix must respect Eq.~\eqref{fullrow} or~\eqref{fullcolumn}, respectively, for the same pair of indices $i,j$. Since a set of transformations with $\alpha_1=\alpha_2$ cannot generate a texture where the first and second rows are non-identical, our conclusion regarding the lack of an implementation for $4_1^{\ell},4_2^{\ell},4_{17}^{\nu}$, still holds for several symmetry transformations. The case where $\theta_1=\theta_2$ in~\eqref{symtransf} should also be considered since, out of the several transformations, only one is required to act on the scalar fields. In this case, a full texture row or column leads to $\beta_1=\beta_2=\beta_3$ and $\alpha_1=\alpha_2=\alpha_3$, respectively. Thus, the imposition of a single texture zero leads to the requirement of an entire row or column of zeros, which is not compatible with the matrices $4_1^{\ell},4_2^{\ell}$ and $4_{17}^{\nu}$.

The above analysis allows us to generalize, to any number of consecutive symmetry transformations, the conclusion that these three textures cannot be realized through any continuous or discrete Abelian symmetries in the 2HDM. We remark that this result only holds in the context of the 2HDM, since the addition of a third Higgs doublet could lift these constraints. From Table~\ref{texturepairs}, we then conclude that 11 maximally-restrictive pairs of leptonic mass matrix textures, associated to $4_1^{\ell},4_2^{\ell}$ and $4_{17}^{\nu}$, can be excluded from our model implementation perspective.

Next, we use the canonical method to eliminate the matrix pairs that are composed of textures that are individually realizable through Abelian symmetries, but for which the symmetry transformation cannot be implemented simultaneously in both matrices.

\subsection{Texture pairs $(3_2^{\ell},7_{1,3}^{\nu})$, $(6_1^{\ell},4_{1}^{\nu})$ and $(5_1^{\ell},5_{1,6}^{\nu})$}

First we note that, for each of the texture pairs $(3_2^{\ell},7_{1,3}^{\nu})$, $(6_1^{\ell},4_{1}^{\nu})$ and $(5_1^{\ell},5_{1,6}^{\nu})$, one of the textures is characterized by a column full of non-zero entries and it has two identical texture rows (generated by two identical left-handed continuous phases $\alpha_i$). Since $\alpha_i$ are common to the charged-lepton and Dirac-type neutrino sectors, Eq.~\eqref{chargphases} demands that both $\Ml$ and $\Mnu$ have two identical texture rows $i$ and $j$ when $\alpha_i = \alpha_j$ for some pair $i,j$. However, one finds that, for all these pairs, the second texture has no identical rows. As such, we conclude that these 5 pairs cannot be implemented through a continuous or discrete Abelian symmetry in a 2HDM.

Additional symmetry transformations with $\theta_1=\theta_2$ do not change this conclusion since realizing the texture with a full column of non-zero entries would require $\alpha_1=\alpha_2=\alpha_3$, exacerbating the issue found with the implementation of these mass matrix pairs. All symmetry transformations attempting at generating these pairs must respect Eq.~\eqref{fullcolumn} for the same pair $i,j$. Therefore, there is always at least one texture zero in these pairs which cannot be imposed, while allowing all of the non-zero entries. Thus, we can generalize to any number of symmetry transformations~\eqref{symtransf} the conclusion that $(3_2^{\ell},7_{1,3}^{\nu})$, $(6_1^{\ell},4_1^{\nu})$ and $(5_1^{\ell},5_{1,6}^{\nu})$ cannot be realized through any continuous or discrete Abelian symmetries in the 2HDM. Consequently, another 5 maximally-restrictive pairs of leptonic mass matrix textures can be eliminated from Table~\ref{texturepairs} for the purpose of our model implementation.

\subsection{Texture pairs $(4_3^{\ell},6_{2,5,8}^{\nu})$ and $(5_1^{\ell},5_{5}^{\nu})$}

From the application of the canonical method one can conclude that, in the 2HDM, a mass matrix texture including a column with two non-zero entries associated to non-identical rows $i$ and $j$ can only be realized by a symmetry transformation which respects an appropriate version of
\begin{equation}
\label{chargemod}
\alpha_i - \alpha_j \pm (\theta_1 - \theta_2) = 0~(\text{mod}~2\pi),~i\neq j.
\end{equation}
In each of the pairs $(4_3^{\ell},6_{2,5,8}^{\nu})$ and $(5_1^{\ell},5_{5}^{\nu})$, the constituent textures  can only be realized by transformations which obey non-compatible versions of Eq.~\eqref{chargemod}. Therefore, we conclude that these pairs cannot be realized through a continuous or discrete Abelian symmetry in the 2HDM.

Let us now consider the case of consecutive symmetry transformations. First we notice that, in the 2HDM, a texture including a column with two non-zero entries, in rows $i$ and $j$, can only be implemented through a transformation which respects either
\begin{equation}
\alpha_i - \alpha_j = 0~(\text{mod}~2\pi),~i\neq j,\label{chargemodalt}
\end{equation}
or Eq.~\eqref{chargemod}. 

Two symmetry transformations that obey Eq.~\eqref{chargemod} and Eq.~\eqref{chargemodalt}, respectively, for the same rows $i,j$ generate incompatible decompositions into Yukawa matrices and cannot be used together to implement such a texture. If the rows $i$ and $j$ are non-identical then its implementation requires at least one symmetry transformation which obeys Eq.~\eqref{chargemod} and, hence, all transformations used must do so. Furthermore, they must respect the same sign in $\pm (\theta_1-\theta_2)$, as it determines the ordering of the decomposition into Yukawa matrices, which must always be the same. This is essentially the same conclusion as for the case of a single transformation. A symmetry transformation with $\theta_1=\theta_2$ can only implement a texture with a column with two non-zero entries if it obeys Eq.~\eqref{chargemodalt} and is, therefore, not useful in this context.

We conclude that the texture pairs $(4_3^{\ell},6_{2,5,8}^{\nu})$ and $(5_1^{\ell},5_{5}^{\nu})$ cannot be realized through any continuous or discrete Abelian symmetries, in the context of the 2HDM, regardless of the number of consecutive symmetry transformations of type~\eqref{symtransf} imposed. As a result, we exclude another 4  maximally-restrictive pairs of leptonic mass matrix textures from Table~\ref{texturepairs}.

\subsection{Texture pairs $(4_3^{\ell},6_{4,6}^{\nu})$ and $(5_1^{\ell},5_{4}^{\nu})$}

From the application of the canonical method one can conclude that if a mass matrix texture $\mathbf{T}_1$ is characterized by two non-identical columns $c_1$ and $c_2$ with non-zero entries in some row $i$ and zero entries in some row $j$, then it cannot be realized through Abelian symmetries in the 2HDM, together with a texture $\mathbf{T}_2$ generated by a symmetry transformation obeying Eq.~\eqref{chargemod} for the same rows $i$ and $j$.

The above statement arises because a symmetry transformation as in Eq.~\eqref{symtransf} is unable to impose a zero in both $(\mathbf{T}_1)_{j c_1}$ and $(\mathbf{T}_1)_{j c_2}$. Despite this, it is possible, in principle, to find transformations able to generate each zero separately, which could be an ideal setup for multiple symmetry transformations generating a texture pair unattainable through a single one. However, for the texture pairs considered here, this is not the case. In the context of the statement, one of the columns $c_1/c_2$ of $\mathbf{T}_1$ already has two non-zero entries in the rows $i$ and $k\neq j$. Thus, a symmetry transformation imposing a zero in $(\mathbf{T}_1)_{j c_2}/(\mathbf{T}_1)_{j c_1}$ generates a column $c_1/c_2$ full of non-zero entries in $\mathbf{T}_1$. Such a transformation must have at least two identical charges $\alpha_i$. The only viable case is $\alpha_k = \alpha_j\; (\text{mod}~2\pi)$. For each of the three texture pairs $(4_3^{\ell},6_{4,6}^{\nu})$ and $(5_1^{\ell},5_{4}^{\nu})$, we can identify the indices $i,j,k$ and easily arrive at the conclusion that $\alpha_k = \alpha_j\; (\text{mod}~2\pi)$ is not compatible with one of the mass matrix textures in the pair. The case $\theta_1=\theta_2$ does not bear any consequence for the same arguments given in the previous section.

In conclusion, none of the pairs $(4_3^{\ell},6_{4,6}^{\nu})$ and $(5_1^{\ell},5_{4}^{\nu})$ can be realized through a continuous or discrete Abelian  in the 2HDM, regardless of the number of consecutive symmetry transformations. Another 3 of the maximally-restrictive pairs of leptonic mass matrix textures of Table~\ref{texturepairs} must be excluded from our model implementation.

In this appendix, we have determined that 23 out of the 28 maximally restrictive leptonic mass matrix texture pairs have no implementation through Abelian symmetries in the 2HDM. In the process of identifying non-realizable texture pairs, we have also discussed the possibility of imposing several consecutive symmetry transformations of the form given in Eq.~\eqref{symtransf}. It is worth emphasizing here that the same analysis can also be extended to the realizable pairs, namely, $(4_3^{\ell},6_{1,3,7,9}^{\nu})$ and $(5_1^{\ell},5_{8}^{\nu})$. We find that no other decompositions (besides those given in Section~\ref{decompyukawa}) arise from the application of such consecutive transformations.

Finally, we stress that all our conclusions were drawn in the context of the 2HDM. The addition of more Higgs doublets would relax most of the constraints, even for the case of a single symmetry transformation.

\begin{table*}[!ht]
\centering
\begin{tabular}{@{}l@{\hskip 0.4in}l@{\hskip 0.6in}l@{}}
\hline \hline \vspace{-0.2cm}\\
Texture & \multicolumn{2}{l}{Defining conditions}\\ \hline \vspace{-0.5cm}\\ \\
$4_3^\ell ~:~ \begin{pmatrix} 0&0&a_1\\0&a_2&b_1\\ a_3&b_2&0 \end{pmatrix}$ & \multicolumn{2}{l}{$a_1^2=\dfrac{b_2^2\,\Delta}{\Delta-\Sigma}\;\;,\;\;
a_2^2=\dfrac{\Delta-\Sigma}{a_3^2b_2^2}\;\;,\;\;b_1^2=a_3^2+\dfrac{a_3^4}{b_2^2}+\dfrac{\chi}{b_2^2}-\dfrac{a_3^2\,{\rm T}}{b_2^2}-\dfrac{b_2^2\,\Delta}{\Delta-\Sigma}>0$}\,\\
& \multicolumn{2}{l}{$\Sigma=a_3^2\left[\chi+(a_3^2+b_2^2)(a_3^2+b_2^2-{\rm T})\right]<\Delta$}\\
\\
& \multicolumn{2}{l}{$\Delta=m_e^2 m_\mu^2 m_\tau^2\;,\;\chi=m_e^2 m_\mu^2 +m_e^2 m_\tau^2+m_\mu^2 m_\tau^2\;,\;{\rm T}=m_e^2+m_\mu^2+m_\tau^2$}
\\\vspace{-0.1cm} \\ \hline \vspace{-0.2cm}\\
$5_1^\ell ~:~ \begin{pmatrix} 0&0&a_1\\0&b_1&0\\ a_2&0&b_2 \end{pmatrix}$ & \multicolumn{2}{l}{$a_1^2=\dfrac{m_{\ell_2}^2m_{\ell_3}^2}{a_2^2}\;\;,\;\;b_1^2=m_{\ell_1}^2\;\;,\;\;b_2^2=\dfrac{(a_2^2-m_{\ell_2}^2)(m_{\ell_3}^2-a_2^2)}{a_2^2}\;\;{\rm with}\;\; m_{\ell_2}<a_2<m_{\ell_3}$}\\\vspace{-0.1cm} \\ \hline \vspace{-0.2cm}\\
& NH & IH \vspace{-0.3cm} \\ \\ \hline \\ 
$5_8^\nu ~:~ \begin{pmatrix} 0&y_1&x_1\\ 0&0&y_2 \\ 0&x_2e^{i\alpha}&0 \end{pmatrix}$& $y_1^2=\dfrac{(\Delta_{21}-x_2^2)(\Delta_{31}-x_2^2)}{y_2^2-x_2^2}$ & $y_1^2=\dfrac{(x_2^2-\Delta_{31})(\Delta_{21}+\Delta_{31}-x_2^2)}{y_2^2-x_2^2}$\\
\vspace{0.2cm} & $x_1^2=\dfrac{(\Delta_{31}-y_2^2)(y_2^2-\Delta_{21})}{y_2^2-x_2^2}$ & $x_1^2=\dfrac{(y_2^2-\Delta_{31})(\Delta_{21}+\Delta_{31}-y_2^2)}{y_2^2-x_2^2}$\\
\vspace{0.2cm}& $i)\,\Delta_{21} < y_2^2< \Delta_{31}\,,\,x_2^2 < \Delta_{21}$ & $i)\,\Delta_{31} < y_2^2 < \Delta_{21}+\Delta_{31}\,,\,x_2^2 < \Delta_{31}$\\ 
& $ii)\,\Delta_{21} < x_2^2< \Delta_{31}\,,\,y_2^2 < \Delta_{21}$ & $ii)\,\Delta_{31} < x_2^2 < \Delta_{21}+\Delta_{31}\,,\,y_2^2 < \Delta_{31}$ \\\vspace{-0.1cm} \\ \hline \vspace{-0.2cm}\\
$6_1^\nu ~:~ \begin{pmatrix} 0&0&0\\ 0&0&x_2 \\ 0&x_1&ye^{i\alpha} \end{pmatrix}$ & $x_2^2=\dfrac{\Delta_{21}\Delta_{31}}{x_1^2}$ & $x_2^2=\dfrac{\Delta_{31}(\Delta_{21}+\Delta_{31})}{x_1^2}$\\
& $y^2=\dfrac{(x_1^2-\Delta_{21})(\Delta_{31}-x_1^2)}{x_1^2}$ & $y^2=\dfrac{(x_1^2-\Delta_{31})(\Delta_{31}+\Delta_{21}-x_1^2)}{x_1^2}$\\
& $\Delta_{21} < x_1^2 <\Delta_{31}$ & $\Delta_{31} < x_1^2 <\Delta_{31}+\Delta_{21}$\\\vspace{-0.1cm} \\ \hline \vspace{-0.2cm}\\
$6_3^\nu ~:~ \begin{pmatrix} 0&0&x_1\\ 0&0&ye^{i\alpha} \\ 0&x_2& 0\end{pmatrix}$ & $i)\,x_2^2=\Delta_{21}\,,\,y^2=\Delta_{31}-x_1^2$ & $ i)\,x_2^2=\Delta_{31}\,,\,y^2=\Delta_{21}+\Delta_{31}-x_1^2$\\
\vspace{0.2cm}& $x_1^2<\Delta_{31}$ & $x_1^2<\Delta_{31}+\Delta_{21}$\\
\vspace{0.2cm}&$ii)\,x_2^2=\Delta_{31}\,,\,y^2=\Delta_{21}-x_1^2$ & $ ii)\,x_2^2=\Delta_{21}+\Delta_{31}\,,\,y^2=\Delta_{31}-x_1^2$\\
& $x_1^2<\Delta_{21}$ & $x_1^2<\Delta_{31}$\\\vspace{-0.1cm} \\ \hline \vspace{-0.2cm}\\
$6_7^\nu ~:~ \begin{pmatrix} 0&0&x_2\\ 0&ye^{i\alpha}&0 \\ 0&x_1& 0\end{pmatrix}$ & $i)\,x_2^2=\Delta_{21}\,,\,y^2=\Delta_{31}-x_1^2$ & $ i)\,x_2^2=\Delta_{31}\,,\,y^2=\Delta_{21}+\Delta_{31}-x_1^2$\\
\vspace{0.2cm}& $x_1^2<\Delta_{31}$ & $x_1^2<\Delta_{31}+\Delta_{21}$\\
\vspace{0.2cm}&$ii)\,x_2^2=\Delta_{31}\,,\,y^2=\Delta_{21}-x_1^2$ & $ ii)\,x_2^2=\Delta_{21}+\Delta_{31}\,,\,y^2=\Delta_{31}-x_1^2$\\
& $x_1^2<\Delta_{21}$ & $x_1^2<\Delta_{31}$\\\vspace{-0.1cm} \\ \hline \vspace{-0.2cm}\\
$6_9^\nu ~:~ \begin{pmatrix} 0&x_1&ye^{i\alpha}\\ 0&0&x_2 \\ 0&0&0 \end{pmatrix}$ & $x_2^2=\dfrac{\Delta_{21}\Delta_{31}}{x_1^2}$ & $x_2^2=\dfrac{\Delta_{31}(\Delta_{21}+\Delta_{31})}{x_1^2}$\\
& $y^2=\dfrac{(x_1^2-\Delta_{21})(\Delta_{31}-x_1^2)}{x_1^2}$ & $y^2=\dfrac{(x_1^2-\Delta_{31})(\Delta_{31}+\Delta_{21}-x_1^2)}{x_1^2}$\\
& $\Delta_{21} < x_1^2 <\Delta_{31}$ & $\Delta_{31} < x_1^2 <\Delta_{31}+\Delta_{21}$
\\[0.2cm]
\hline \hline 
\end{tabular}
\caption{Defining conditions for the charged-lepton and neutrino Yukawa textures given in Table~\ref{decompositions}. For the $4_3^\ell$ ($5_1^\ell$) texture we write $a_1^2$, $a_2^2$ and $b_1^2$ ($a_1^2$, $b_1^2$ and $b_2^2$) in terms of $a_3^2$ and $b_2^2$ ($a_2^2$) and the charged-lepton masses. In $5_1^\ell$, the state $\ell_1$ can be identified with $e$, $\mu$ or $\tau$ leading, respectively, to the cases $5_1^e$, $5_1^\mu$ and $5_1^\tau$ discussed in Section~\ref{Phenom}. For the $6_k^\nu$ ($5_8^\nu$) textures we write $x_2^2$ and $y$ ($x_1^2$ and $y_1^2$) in terms of $x_1^2$ ($x_2^2$ and $y_2^2$) and the neutrino mass-squared differences $\Delta_{21}$ and $\Delta_{31}$. The Yukawa couplings in $\Y_1^{\ell,\nu}$ ($\Y_2^{\ell,\nu}$) entering Eq.~\eqref{lagrangian} are determined by dividing $a_i$ and $x_i$ ($b_i$ and $y_i$) by $v\cos\beta$ ($v\sin\beta$). For simplicity, we use the notation $\Delta_{21}\equiv \dmsol$ and $\Delta_{31}\equiv \dmatm>0$ ($\dmatm$ corresponds to $|\dmatm|$ in the IH case).}
\label{definingconditions}
\end{table*}

\vspace{5mm}
\section{Relations for $\Ml$ and $\Mnu$ parameters}
\label{appdxB}

We have seen that for the maximally-restrictive textures the total number of parameters in the mass matrices is nine, equalling that of lepton mass and mixing parameters, namely three charged-lepton and two neutrino masses, three mixing angles and one CP-violating phase. For the mass matrices $\Ml$ and $\Mnu$, the number of free parameters not fixed by requiring the lepton masses to be compatible with experiment is two (one) for the $4_3^\ell$ and $5_8^\nu$ ($5_1^\ell$ and $6_k^\nu$) textures. Those free parameters are determined by requiring compatibility with the measured values of the mixing angles and CP phase. In Table~\ref{definingconditions}, we provide the defining relations of the parameters that can be written as functions of the masses and free parameters for the textures realizable with Abelian symmetries in the 2HDM (see Table~\ref{decompositions}). The choice of the free parameters in the mass-defining relations is not unique but, obviously, the results do not depend on it. We also give the ranges within which the free parameters can vary. For each texture combination known to be compatible with data, the results shown in Section~\ref{Phenom} are obtained as follows.

The free parameters are varied within the ranges given in Table~\ref{definingconditions}, considering all possible cases of lepton mass ordering and both the NH and IH cases. This gives rise to possible different scenarios labelled as $i)$ and $ii)$. Notice that, since the masses do not depend on the phase $\alpha$, we consider $\alpha \in [0, 2\pi]$ . For each set of input parameters, the lepton mixing angles $\theta_{ij}$ and the Dirac CP-violating phase $\delta$ in Eq.~\eqref{Uparam} are determined by computing $\U$ via Eqs.~\eqref{Hdiag} and \eqref{Udef2}. For the texture $5_1^\ell$, $\U_\ell$ is determined by Eqs.~\eqref{Hl5158a}-\eqref{Ul51tau}. In the case $4_3^\ell$, an analytical form can also be obtained for $\U_\ell$, but since the relations are more involved it is not presented here. As for $\U_\nu$, we have in the $5_8^\nu$ case $\U_\nu={\rm \bf K}_\alpha\U_\nu^\prime$, where ${\rm \bf K}_\alpha={\rm diag}(e^{-i\alpha},^{-i\alpha},1)$ and
\begin{widetext}
\begin{align}
{\rm NH:}\;\U_\nu^\prime=
\begin{pmatrix}
-\dfrac{x_2 y_2}{\sqrt{\Delta_{21}\Delta_{31}}} &\pm\sqrt{\dfrac{(x_2^2-\Delta_{21}) (y_2^2-\Delta_{21})}{\Delta_{21}\Delta_-}}  & \sqrt{\dfrac{(\Delta_{31}-x_2^2) (\Delta_{31}-y_2^2)}{\Delta_{31}\Delta_-}}\\ 
x_2\sqrt{\dfrac{(y_2^2-\Delta_{21}) (\Delta_{31}-y_2^2)}{\Delta_{21}\Delta_{31}(y_2^2-x_2^2)}} 
&y_2\sqrt{\dfrac{(\Delta_{21}-x_2^2) (\Delta_{31}-y_2^2)}{\Delta_{21}\Delta_-(y_2^2-x_2^2)}}  
&y_2\sqrt{\dfrac{(y_2^2-\Delta_{21}) (\Delta_{31}-x_2^2)}{\Delta_{31}\Delta_-(y_2^2-x_2^2)}}  \\
y_2\sqrt{\dfrac{(x_2^2-\Delta_{21}) (\Delta_{31}-y_2^2)}{\Delta_{21}\Delta_{31}(y_2^2-x_2^2)}} 
&-x_2\sqrt{\dfrac{(y_2^2-\Delta_{21}) (\Delta_{31}-x_2^2)}{\Delta_{21}\Delta_-(y_2^2-x_2^2)}} 
&x_2\sqrt{\dfrac{(\Delta_{21}-x_2^2) (\Delta_{31}-y_2^2)}{\Delta_{31}\Delta_-(y_2^2-x_2^2)}} 
\end{pmatrix}\,,
\label{UnupNH}
\end{align}
\begin{align}
{\rm IH:}\;\U_\nu^\prime=\begin{pmatrix}
\mp\sqrt{\dfrac{(y_2^2-\Delta_{31}) (\Delta_{31}-x_2^2)}{\Delta_{21}\Delta_{31}}}
&\sqrt{\dfrac{(\Delta_+-x_2^2) (\Delta_+-y_2^2)}{\Delta_{21}\Delta_+}}
& -\dfrac{x_2 y_2}{\sqrt{\Delta_{21}\Delta_+}}\\ 
y_2\sqrt{\dfrac{(\Delta_{31}-x_2^2) (\Delta_+-y_2^2)}{\Delta_{21}\Delta_{31}(y_2^2-x_2^2)}} 
&y_2\sqrt{\dfrac{(y_2^2-\Delta_{31}) (\Delta_+-x_2^2)}{\Delta_{21}\Delta_+(y_2^2-x_2^2)}}  
&x_2\sqrt{\dfrac{(y_2^2-\Delta_{31}) (\Delta_+-y_2^2)}{\Delta_{31}\Delta_+(y_2^2-x_2^2)}}  \\
-x_2\sqrt{\dfrac{(y_2^2-\Delta_{31}) (\Delta_+-x_2^2)}{\Delta_{21}\Delta_+(y_2^2-x_2^2)}}  
&x_2\sqrt{\dfrac{(\Delta_{31}-x_2^2) (\Delta_+-y_2^2)}{y_2^2-x_2^2}} 
&y_2\sqrt{\dfrac{(\Delta_{31}-x_2^2) (\Delta_+-x_2^2)}{\Delta_{21}\Delta_+(y_2^2-x_2^2)}}  
\end{pmatrix}\,,
\label{UnupIH}
\end{align}
\end{widetext}
for NH and IH neutrino mass spectra, respectively. In the above equations and in Table~\ref{definingconditions}, we have used the notation $\Delta_\pm=\Delta_{31}\pm \Delta_{21}$. Moreover, the $-$(+) sign in $\U_\nu^\prime$ corresponds to the case $i$ ($ii$) of the validity ranges for $x_2$ and $y_2$. 

The numerical procedure starts by varying the free parameters in their validity ranges. For each input set $(a_i,b_i,x_i,y_i,\alpha)$, the lepton mixing matrix $\U$ is obtained and compatibility with the $3\sigma$ ranges for $\theta_{ij}$ given in Table~\ref{datatable} is checked. If compatibility is found, the phase $\delta$ is plotted as a function of $\alpha$. This procedure fixes all mass parameters $a_i$, $b_i$, $x_i$, $y_i$ and $\alpha$ in the mass matrices $\Ml$ and $\Mnu$. The Yukawa couplings in $\Y_1^{\ell,\nu}$ ($\Y_2^{\ell,\nu}$) are determined dividing the corresponding mass-matrix elements by $v\cos\beta$ ($v\sin\beta$). Thus, for a given $\tan\beta$, the couplings $\N_e$ are completely determined in terms of lepton masses and mixing parameters, as illustrated in Section~\ref{Phenom} for the case $(5_1^e,5_8^\nu)$. Finally, the  $\ell_\beta \rightarrow \ell_\alpha \gamma$ and $\ell_\alpha^-\to \ell_{\beta}^-\,\ell_{\gamma}^+\,\ell_{\delta}^-$ BRs, as well as the value of $|g_e/g_\mu|-1$, are computed.

\end{document}